\documentclass[aps,prx,reprint,amsmath,amssymb,floatfix,longbibliography]{revtex4-2}

\usepackage{graphicx}% Include figure files
\usepackage{dcolumn}% Align table columns on decimal point
\usepackage{bm}% bold math
\usepackage[english]{babel}
\usepackage{soul,multirow,booktabs,comment}
\usepackage{array}
\usepackage{tabularx}
\newcommand{\my}[1]{\textcolor{black}{#1}}
\newcommand{\myy}[1]{\textcolor{black}{#1}}

\def\mys{\color{black}}
\def\mye{\color{black}}
\usepackage{hyperref}
\hypersetup{
	colorlinks = true,
	linkcolor = blue,
	citecolor = blue,
	urlcolor = blue,
	filecolor = blue
}

\begin{document}

% Use the \preprint command to place your local institutional rephttps://www.overleaf.com/project/5e3d7920017c950001954321ort
% number in the upper righthand corner of the title page in preprint mode.
% Multiple \preprint commands are allowed.
% Use the 'preprintnumbers' class option to override journal defaults
% to display numbers if necessary
%\preprint{}

\title{{Laplacian-Level} {Quantum} {Hydrodynamic} {Theory} for {Plasmonics}}

\author{Henrikh M. Baghramyan$^{1}$}
\thanks{These authors contributed equally to this work}
\author{Fabio Della Sala$^{1,2}$}
\thanks{These authors contributed equally to this work}
\author{Cristian Cirac\`{i}$^{1}$}
 \email{cristian.ciraci@iit.it}
% \footnote[*]{These authors contributed equally to this work}.

\affiliation{$^{1}$Center for Biomolecular Nanotechnologies, Istituto Italiano di Tecnologia, Via Barsanti 14, 73010 Arnesano (LE), Italy}
\affiliation{$^{2}$Institute for Microelectronics and Microsystems (CNR-IMM), Via Monteroni, Campus Unisalento, 73100 Lecce, Italy}

\date{\today}

\begin{abstract}
An accurate description of the optical response of subwavelength metallic particles and nanogap structures is a key problem of plasmonics.  Quantum hydrodynamic theory (QHT) has emerged  as a powerful method to calculate the optical response of metallic nanoparticles (NPs) since it takes into account nonlocality and spill-out effects. 
%quantum properties of electron response. 
%The conventional QHT framework allowing a correct calculation of plasmon resonance, spill-out and retardation effects.true
Nevertheless, the absorption spectra of metallic NPs \my{obtained with conventional QHT, i.e., incorporating Thomas-Fermi (TF) and von Weizs\"acker (vW) kinetic energy (KE) contributions, can be} affected by several spurious resonances at energies higher than the main \my{localized surface plasmon (LSP)}. 
\my{These peaks are not present in} reference time-dependent density-functional-theory \my{(TD-DFT) spectra, where, instead, only a broad shoulder exists}.
Moreover, we show here that these peaks \my{ incorrectly reduce the LSP peak intensity and} have a strong dependence on the simulation domain size so that \my{a proper calculation of QHT absorption spectra  can be} problematic. In this article, we introduce a \my{more general} QHT method accounting for KE contributions depending on the Laplacian of the electronic density \my{($q$)}, thus, beyond the gradient-only dependence of \my{the TFvW functional}. 
\my{We show that employing a KE functional with a term proportional to $q^2$ results in an absorption spectrum free of spurious peaks, with LSP resonance of correct intensity and numerically stable Bennett state.
Finally, we present a novel Laplacian-level KE energy functional that is very accurate for the description of the optical properties of NPs with different sizes as well as for dimers.}
Thus, the Laplacian-level QHT represents a novel, efficient, and accurate platform to study plasmonic systems.
\end{abstract}


\maketitle
%%%%%%%%%%%%%%%%%%%%%%%%%%%%%%%%%%%%%%%%%%%%%%%%%%%%%%%%%%%%%
\section{Introduction}\label{sec01}
%%%%%%%%%%%%%%%%%%%%%%%%%%%%%%%%%%%%%%%%%%%%%%%%%%%%%%%%%%%%
Metal nanoparticles (NPs) play a crucial role in the enhancement of the optical field due to plasmonic effects \cite{maradudin}, which make them an ideal platform for nonlinear optics \cite{kauranen,ren}, hot-electron enhancement for photovoltaics \cite{brongersma,goykhman}, surface-enhanced Raman scattering \cite{xu}, and imaging \cite{kawata}. When it comes to the nanoscale, nonlocal and quantum effects play a crucial role in light-matter interaction \cite{weiner2017}.
Among theoretical approaches \cite{bohren98,quinten,morton11,esteban12,luo13,yan15dip,raza15,varas16,chris17}, time-dependent density-functional theory (TD-DFT) \cite{rungegross84,ullrich} stands out since it allows to accurately resolve the optical response of plasmonic structures at the nanoscale, including both quantum and atomistic effects \cite{morton11,zhang14,barbry2015,varas16,roy17,zhang18al,fds19}. However, TD-DFT is computationally expensive since all occupied orbitals need to be evaluated.

Another approach would be to treat the electron system semiclassically: a fluid, characterized by the macroscopic local quantities, such as the electron-density $n\left(\textbf{\textrm{r}},t\right)$ and the electron velocity field $\textbf{\textrm{v}}\left(\textbf{\textrm{r}},t\right)$ \cite{madelung,bloch,jensen,ying74}, but at the same time considering quantum effects through energy functionals of the electron-density fluctuations. This approach is known as hydrodynamic theory (HT). 
%An essential initial step in the formulation of HT is the choice of the functionals that account for quantum and nonlocal effects.
The HT is part of a larger class of methods based on the orbital-free (OF) \cite{parrbook,wang1998orbital,ofdft_book}  description of quantum electronic systems dating back to the works of Thomas \cite{thomas} and Fermi \cite{fermi}. Although the interest in OF-DFT methods has gradually decreased in favor of Kohn-Sham (KS) orbital-based methods, the last decades have witnessed a reinvigorated interest due to the ideal scaling of computational resources with respect to the size of the electronic system offered by the OF-DFT approach \cite{gavini07}.
Most of the research efforts in this field, however, have been devoted to static properties \cite{xia2012,shinCarter2014,constantin2018,witt2018} and, more recently, also to response properties with the time-dependent OF-DFT \cite{domps1998time,Neuhauser:2011ch,xiang2016,zhang2017}.
In both cases, the central quantity that controls the accuracy of these methods is the noninteracting kinetic energy (KE) functional.

The most simple KE functional is the Thomas-Fermi (TF) functional, which accounts for the Pauli exclusion principle for a homogeneous system of noninteracting electrons \cite{parrbook}, and it yields the electron quantum pressure $p\left(\textbf{\textrm{r}},t\right) \propto n\left(\textbf{\textrm{r}},t\right)^{5/3}$ \cite{ashcroft} that accounts for the nonlocal electron response.
It has been demonstrated that TF-HT is able to provide surprisingly accurate predictions that match well experiments with noble metal NPs, such as Au \cite{ciraci_science} and Ag \cite{raza}, both qualitatively and quantitatively.
Nevertheless, for alkali metals or aluminum, the TF-HT predicts a blueshift of the localized surface-plasmon resonance with respect to the classical Mie resonance \cite{teperik}, in contradiction with the redshift from the experiments \cite{reiners} and TD-DFT calculations \cite{li13}.
The origin of this difference lies in neglecting the spill-out of the plasmon-induced charges at the NP surface \cite{teperik}. In fact, the TF-HT approach  employs (with some recent exception \cite{david14}) a spatially uniform electronic density inside the NP and zero outside (i.e., hard-wall boundary) \cite{raza15}.

\begin{figure}[!hbt]
\centering
\resizebox{0.95\columnwidth}{!}{\includegraphics[width=0.3\textwidth,angle=0]{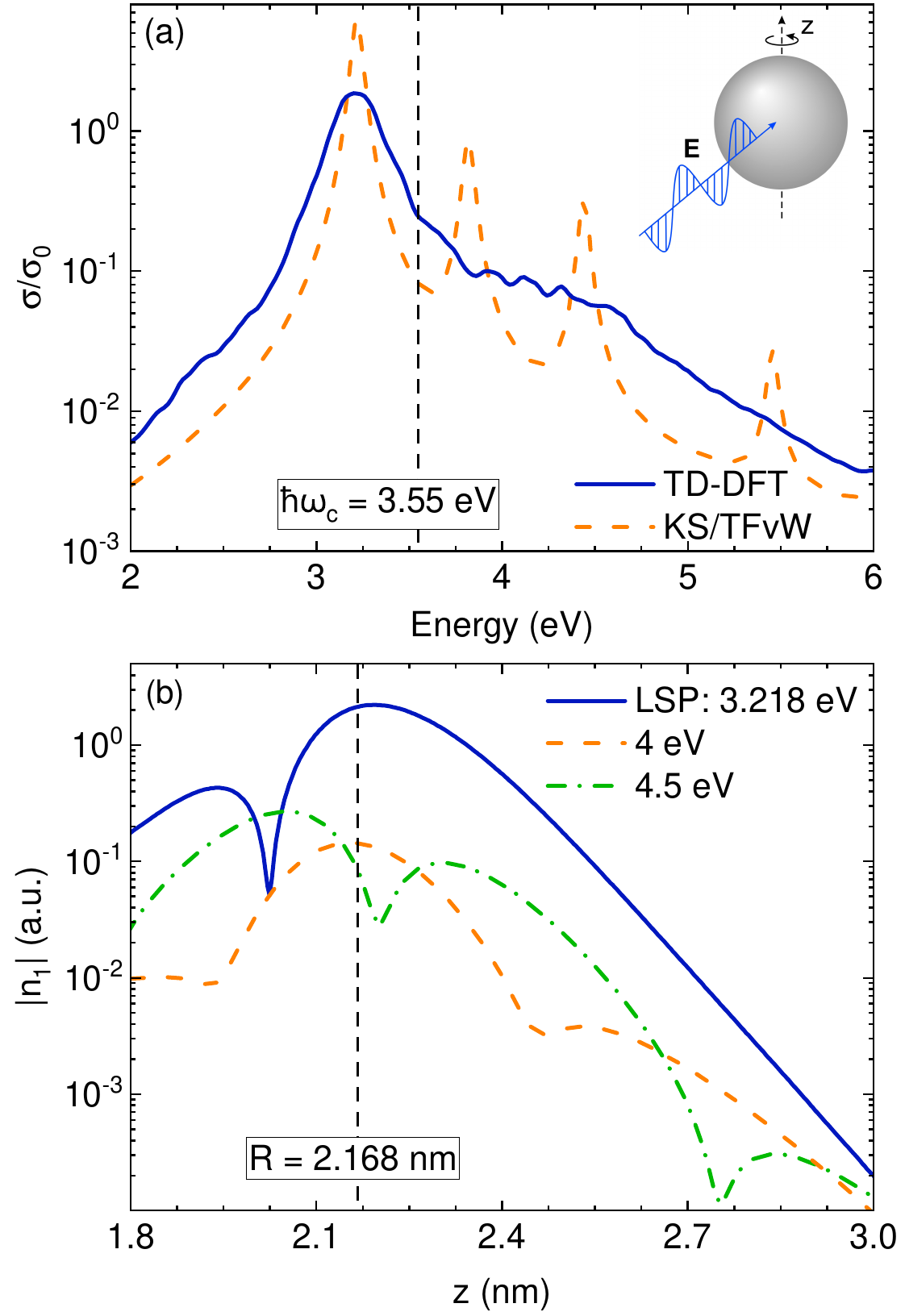}}
\caption{\small (a) Normalized absorption cross section (in logarithmic scale) for a Na jellium sphere with radius $R=2.168\,$ nm (and $N_e$=1074 electrons) as obtained from TD-DFT and \my{QHT} (using KS density and TFvW functional);
\my{$\hbar\omega_c$ is the critical frequency, see Eq. \eqref{eq:crit}} (b) Induced charge density $n_1$ in atomic units (a.u.) at \my{different energies excitation}  \my{as calculated from QHT with the KS density and the TFvW functional}. The inset in (a) schematically illustrates the nanosphere interaction with the incident plane wave. See Appendix \ref{appabs} for definitions and details on the absorption spectra calculation.}
\label{f00}
\end{figure}

To properly address spill-out effects, the spatial dependence of electron density as well as a correction to the KE functional, in order to describe the density variation effects, must be introduced.
The simplest functional that depends on the gradient of the density is the von Weizs\"acker (vW) functional \cite{vW,parrbook}.
The TF-HT with a fraction ($\lambda$, with $0<\lambda\le1$) of the vW correction (i.e., the TF$\lambda$vW KE functional) is usually referred to in the literature as the quantum hydrodynamic theory (QHT) since the vW functional does not have a classical counterpart.
The QHT has been largely used in plasma physics \cite{manfredi05,plasmabook,shukla12,michta2015,moldabekov2018,bonitz19}, \my{magnetoplasmonics \cite{zaremba94,zaremba99,tsozaremba}}, plasmonic response properties of metal NPs of different geometries \cite{banerjee2008,toscano,ciraciQHT,ding17,ding18,khalidMatr,khalid2019shell}, as well as for surfaces \cite{zhang14plas,yan15prb,palade16} and strongly coupled plasmonic structures \cite{ciraci2017visc,ciraciQEmmit,khalidMatr}.
It has been shown that the QHT can predict plasmon resonance, spill-out, and retardation effects in noble and simple metal NPs, matching very well with TD-DFT calculations \cite{khalidMatr,ciraciQHT}.
There are also other works on the development of the QHT that consider the viscous contribution of electron fluid \cite{ciraci2017visc,deceglia2018} and formulation of HT for nonlinear phenomena \cite{bergara1996,crouseilles2008,khalid2020}.

However, it is important to highlight that the QHT results depend on the approximation made for the KE functional (e.g., $\lambda$ parameter) as well as on the electronic density, which is an input quantity. The input electronic density can be obtained from a preceding OF-DFT calculation using the same KE functional used for the response, i.e., the self-consistent QHT (SC-QHT) approach of Ref. \onlinecite{toscano}. Other approaches use the exact KS density \cite{ciraciQHT} or, more efficiently, a model density \cite{banerjee2008,ciraciQHT}
\my{
that reproduces the decay of the exact KS density.}
%We recall, however, that in the sc-QHT approach, $\lambda=1/9$ was employed, leading to an input density decaying three times faster than the KS density, thus strongly underestimating spill-out effects \cite{ciraciQHT}.

Although QHT can describe different quantum effects relevant in plasmonics, it is not unaffected by drawbacks:

(i) Various
\mys 
TF-HT \cite{bennet,eguiluz1975influence,schwartz82} and QHT \cite{tsozaremba,toscano,ciraciQHT,palade16,ding17,ding18} investigations for NPs \cite{toscano,ciraciQHT}, rods \cite{ding17,ding18}, and surfaces or slabs \cite{bennet,eguiluz1975influence,schwartz82,tsozaremba,palade16} show the presence of one (or even more)
\mye 
additional resonances above the main plasmon peak and below the plasma frequency ($\omega_p$). These resonances originate from the spatial variation of the electronic density as first pointed out by Bennett \cite{bennet}.
No Bennett states are observed when using hard-wall boundary conditions, both in the QHT \cite{moradi,kupre2020,moradibook} and in the TF-HT \cite{ruppin73,fuchs81,raza11,moradibook}. In these approaches, several peaks (volume plasmons) occur due to nonlocality but only at frequencies larger than $\omega_p$. 
Such Bennett states are thus peculiar to \my{models} with nonuniform density.
\my{In TD-DFT a single Bennett state has been computed for jellium-surfaces \cite{tsuei90,tsuei91,liebbook97}, at about 0.8$\omega_p$ in the case of sodium. Instead, for} large jellium spheres, only a shoulder above the main plasmon peak is present \cite{varas16} due to the interaction between single-particle transitions and surface modes \cite{beck87,brack93,yanno93,rein96,lieb93}.

\my{To better illustrate this point, we anticipate in Fig.~\ref{f00}-(a) the absorption spectra of one of the systems investigated in this work, i.e. a sodium jellium nanosphere. All calculations in this work will be focused on sodium (Wigner-Seitz radius $r_s = 4$ a.u., with  plasma frequency $\omega_p$=5.89 eV) which
is commonly investigated as a model metallic system.   Fig.~\ref{f00}-(a) reports a direct comparison of the QHT with the TFvW functional (KS/TFvW denotes using $\lambda=1$ and the KS density as the input density) and reference TD-DFT.}
TD-DFT can be considered as a reference for QHT because the latter method can be directly derived from TD-DFT equations (and for the two-electron case, the methods coincide) \cite{harbola98,ciraci2017visc,bonitz19}. 
Although the energy position of the main peak, the localized surface plasmon (LSP), is very well reproduced, additional peaks are
present in the QHT spectrum, which is not the case for the reference TD-DFT spectrum.
The results in Fig.~\ref{f00}-(a) represent the current state-of-the-art of QHT calculations. Clearly, the presence of the other peaks 
strongly limits the QHT accuracy and applicability.
\my{Note that also the number and the position of the Bennett states strongly depend on the input density as well as on the $\lambda$ parameter \cite{eguiluz1975influence,schwartz82,palade16}}.

%For example, the sc-QHT
%approach 
%of Ref. \onlinecite{toscano} yields 
%only one Bennett state is present for metal NPs\cite{toscano}, but it to so only using the  %(incorrect) self-consistent density with $\lambda=1/9$.

(ii)  \my{The QHT absorption spectrum of metal nanoparticle is characterized by the} critical frequency \cite{tsozaremba,ciraciQHT,yan15prb}: 
\mys
\begin{equation}
\hbar\omega_c=\frac{\hbar^2}{m_e}\frac{\kappa^2}{8}\sqrt{\lambda}=|\mu|\frac{\sqrt{\lambda}}{\lambda_g}    
\label{eq:crit}
\end{equation}
where $\kappa$ is the exponential decay constant of the ground-state density, 
$\lambda_g$ is the vW fraction used for the ground-state-density calculation ($\lambda_g$=1 for exact KS density), 
$\lambda$ is the vW fraction used in the QHT response calculation, 
and $\mu=\lambda_g{(\hbar^2\kappa^2)}/{(8 m_e)} $ is the chemical potential \cite{ciraciQHT}.
For the KS calculations of the large Na jellium nanosphere, we have $\kappa\approx1.05$ a.u. 
and $\mu\approx$ 3.75 eV \cite{ciraciQHT}.
For energies above $\hbar\omega_c$,
\mye
the induced density (i.e., the first-order change of the electronic density due to the excitation)  has both an oscillating and exponentially decaying behavior, as shown in Fig.~\ref{f00}-(b), which is problematic to treat numerically.
We show in this article that the energy position of all the peaks above the critical frequency strongly changes with the computational domain size so that a numerically converging QHT spectrum is challenging to obtain.
\mys
The critical frequency can be artificially increased using an input density that decays faster (i.e., $\kappa>1.05$ or $\lambda_g < 1$).
In Ref. \cite{toscano}, for example, the  SC-QHT approach with $\lambda_g=\lambda=1/9$ (and thus, $\hbar\omega_c=3|\mu|$) was employed \myy{(i.e. the second-order gradient expansion \cite{brack76})}, leading to an input density decaying three times faster than the KS density, and thus strongly underestimating the spill-out effects \cite{ciraciQHT}.
In the present work, we focus only on the more physical case of correct input density.
\mye

iii) The TF$\lambda$vW functional is known to be quite a rough approximation of the exact KE, and different limitations of this functional have been shown in different contexts, e.g., lack of dynamical corrections \cite{Neuhauser:2011ch,yan15prb,moldabekov2018,palade18} and incorrect response for homogeneous electron gas  \cite{wang2002orbital,wang1998orbital,fabio01}. Thus, the great accuracy of QHT calculations with the TF$\lambda$vW functional obtained in some cases \cite{ciraciQHT,toscano} should be related to some error cancellation and, therefore, cannot have general validity.

\mys
In order to overcome these limitations, in this article, we extend the QHT approach to Laplacian-level KE functionals \cite{ho73,brack76,perdew2007,kara09,laricchia14,cancio16,seino18,golub18,fabio01,const19}.   
Laplacian-level KE functionals have been investigated in the past for ground-state properties, but with limited success \cite{perdew2007,kara09,laricchia14,cancio16}.  
Only recently, Laplacian-level functionals performing well for semiconductors and metals in the framework of OF-DFT have been introduced \cite{fabio01,const19}: The Pauli-Gaussian second-order and Laplacian (PGSL) functional has an improved Lindhard response \cite{fabio01}, which is an important property for the description of metallic systems.
\mye
Laplacian-level KE functionals are much simpler than fully 
nonlocal functionals based on the Lindhard response in the reciprocal space \cite{wang2002orbital,wang1998orbital,palade18,Neuhauser:2011ch} and can be easily applied to finite systems \cite{fabio01}.
\mys 
While Laplacian-level functionals have been applied for the ground-state properties, their application for optical properties is completely unexplored.
In this work, we introduce the Laplacian-level QHT linear-response equations in the frequency domain.
We carry out a general form of the QHT equations that holds for any arbitrary Laplacian-level functional, boosting the QHT potential in an unprecedented manner.
%Note, in fact, that the QHT was so far only limited to the TF$\lambda$vW KE functionals.
\mye
%We clearly show that the calculation of the 
%QHT absorption spectra is corrupted by additional peaks which strongly depend on the computational domain-size.
%by doing simulations on Na jellium nanospheres in computational
%domains of different sizes.
%First, we analyze the limits of We show that QHT in the limit of gradient-level correction can give additional nonphysical solutions which can interfere with the localized surface plasmon (LSP) resonance.
\my{We perform calculation for Na jellium nanospheres (up to 6000 electrons)}
and demonstrate that in the QHT-PGSL approach, only the main LSP peak appears in the \my{lower part of the} absorption spectrum, which is stable to the changes of computational domain size \my{as well as on the input density}. In fact, in QHT-PGSL, the induced density decays in the same way for all frequencies, and no critical frequencies exist anymore.
%Moreover, as we will show later, in QHT-PGSL, the intensity of the main plasmon peak is more accurate than in the QHT-TFvW approach.  

\mys
Finally, we go beyond the PGSL approximation and introduce the plasmonic tailored PGSLN functional, which gives very accurate
plasmon energy, peak intensity, and Feibelman $d$ parameter \cite{feibelmann82}, as well as
a single numerically stable Bennett state.
We present a detailed comparison of the different KE functional for QHT, and
we clearly demonstrate that the QHT-PGSLN approach is the most accurate and numerically stable method to treat plasmonics nanosystems.
\mye
%Finally, we report results for nanospheres with different number of electrons and a dimer of Na jellium sphere.

The article is organized as follows: in Sec. \ref{sec03}, we introduce the equations governing the Laplacian-level QHT, which also contains the conventional QHT-TFvW approach as a special case. In Sec. \ref{sec:asy} we discuss theoretically the properties
of the induced density in the tail region in spherical systems, showing that the QHT-PGSL has an unexpected and completely different behavior with respect to the conventional QHT. In Sec. \ref{sec04}, we provide numerical details of our implementation which can efficiently describe systems with spherical and cylindrical symmetry. In Sec. \ref{sec05}, we compare in detail the absorption spectra of Na jellium nanospheres from TD-DFT, QHT-TFvW, QHT-PGS and QHT-PGSL, showing their different dependence on the computational domain size as well as their oscillator strength. In Sec. \ref{sec06}, we describe the numerical results of the induced density decay for Na jellium nanospheres which confirms the theoretical prediction of Section \ref{sec:asy}. \my{In Sec. \ref{sec:pgsln}, we present the derivation and the results for the  PGSLN functional, which can be tuned to have a Bennett state at the correct energy.} In Sec. \ref{sec07}, \my{we benchmark the energy, the oscillator strength and the Feibelman $d$ parameter} as a function of the particle size. In Sec. \ref{sec08}, we present the results for spherical dimers. Finally, the conclusion and future perspectives are drawn in Sec. \ref{sec09}.

%%%%%%%%%%%%%%%%%%%%%%%%%%%%%%%%%%%%%%%%%%%%%%%%%%%%%%%%%%%%%%%%%%%%
%\section{Limits of the conventional QHT}\label{sec0002}
%%%%%%%%%%%%%%%%%%%%%%%%%%%%%%%%%%%%%%%%%%%%%%%%%%%%%%%%
%Although the QHT predicts very well the plasmon resonance, when compared to %more sophisticated TD-DFT approaches, the presence of additional peaks is a 
%major shortcoming. These peaks have in fact an energy higher than %$\hbar\omega_c$, and thus can be hardly treated in an efficient numerical %scheme.
%These features are not limited to spherical NPs, but could happen in %other geometries or materials since: in fact, for $\omega_c$ the identical %expression was obtained for a jellium sphere \cite{ciraciQHT} and slab %\cite{yan}. Thus, in general, one could have for LSP $\omega_{lsp}\simeq %\omega_c$ or even  $\omega_{lsp}>\omega_c$. In such cases the QHT will not %accurately predict the LSP energy.

%%%%%%%%%%%%%%%%%%%%%%%%%%%%%%%%%%%%%%%%%%%%%%%%%%%%%%%%%%%%%%%%%%%%%55555
\section{Laplacian-level functionals in quantum hydrodynamics}\label{sec03}
The linearized QHT response \cite{ying74,stott} is governed by the following equations \cite{ciraciQHT,toscano} for the electric field $\textrm{\textbf{E}}$ and polarization vector $\textrm{\textbf{P}}$:
\begin{subequations}
    \begin{gather}
        \nabla\times\nabla\times\textrm{\textbf{E}} - \frac{\omega^2}{c^2}\textrm{\textbf{E}} = \omega^2\mu_0 \textrm{\textbf{P}}, \label{fq01a} \\
        \frac{en_0}{m_e}\nabla\left(\frac{\delta G\left[n\right]}{\delta n}\right)_1 + \left(\omega^2 + i\gamma\omega\right)\textrm{\textbf{P}} = -\varepsilon_0\omega_p^2\textrm{\textbf{E}}, \label{fq01b}
    \end{gather}
\label{fq01}
\end{subequations}

\noindent where $c$ is the speed of light, $\varepsilon_0$ and $\mu_0$ are the vacuum permittivity and permeability, $m_e$ and $e$ are the electron mass and charge (in absolute value), $\gamma$ is the phenomenological damping rate, and $\omega_p\left(\textbf{r}\right) = \sqrt{e^2n_0\left(\textbf{r}\right)/\left(m_e\epsilon_0\right)}$ is the plasma frequency with $n_0\left(\textbf{r}\right)$ being the ground-state (equilibrium) electron density. $\left(\frac{\delta G\left[n\right]}{\delta n}\right)_1$ is the first-order term for the potential associated with the energy functional $G\left[n\right]$ given by
\begin{equation}
    G\left[n\right] = T_s\left[n\right] + E_{\textrm{XC}}^{\textrm{LDA}}\left[n\right],
\label{fq02}
\end{equation}
where $E_{\textrm{XC}}^{\textrm{LDA}}\left[n\right]$ is the exchange-correlation (XC) energy functional in the local density approximation (LDA), while $T_s$ is the noninteracting KE functional.

\mys
In general, the exact energy functional can be written as 
\begin{equation}
T_s[n]=\int \left[\tau^{\rm TF}(n)+\tau^{\rm vW}\left(n,w\right)\right]\textit{d}^3\textrm{\textbf{r}} + C_s[n]+ C_d[n,\omega],
\end{equation}
where
\begin{equation}
\tau^{\rm TF}\left(n\right)=\left(E_h a_0^2\right) \frac{3}{10} \left(3\pi^2\right)^{2/3}  n^{5/3}
\label{eq:tf}
\end{equation}
is the TF kinetic energy density (a simple local function
of the electronic density), and
\begin{equation}
\tau^{\rm vW}\left(n,w\right)=\left(E_h a_0^2\right) 
\frac{w}{ 8 n},
\label{eq:vw}
\end{equation}
is the vW term which depends on both $n$ and on the
squared gradient of the density 
$w=\nabla n \cdot \nabla n$. In Eqs. (\ref{eq:tf} and \ref{eq:vw}), $E_h = \hbar^2/(m_e a_0^2)$ is the Hartree energy, and $a_0$ is the Bohr radius. Finally, $C_s$ and $C_d$ represent the generic density functionals for static and dynamic corrections, respectively. 
\mye
Although some schemes have been proposed \cite{Neuhauser:2011ch, ciraci2017visc,palade18}, the first-principles derivation of dynamic corrections presents fundamental challenges, especially for finite-size systems.   

In this article, we consider only static corrections. In particular, at the Laplacian-level of theory, the KE has the form:
\begin{equation}
    T_s\left[n\right] = \int\tau\left(n,w,q\right)\textit{d}^3\textrm{\textbf{r}},
\label{fq03}
\end{equation}
where the Laplacian of the density is $q=\nabla^2 n$, which is a new ingredient in addition to $w$. 
\mys
The function $\tau\left(n,w,q\right)$ can be approximated in several ways \cite{ho73,brack76,perdew2007,kara09,laricchia14,fabio01}.
%The oldest example is the fourth-order gradient expansion %(GE4) with has been shown to have many limitations\cite{ho73, %Later other functionals has been developed [], but only %recently with success [].
In the PGSL functional \cite{fabio01},
\mye 
the function  $\tau\left(n,w,q\right)$ is approximated as the sum of the vW \cite{vW}, Pauli-Gaussian (PG$\alpha$), and Laplacian (L$\beta$) terms \cite{fabio01}
\begin{equation}
    \tau\left(n,w,q\right) = \tau^{\textrm{vW}}\left(n,w\right) + \tau^{\textrm{PG}\alpha}\left(n,w\right) + \tau^{\textrm{L}\beta}\left(n,q\right),
\label{fq04}
\end{equation}
where
\begin{subequations}
    \begin{gather}
%        \tau^{\textrm{vW}}\left(n,w\right) = A n^{-1}w, \label{fq05a} \\
        \tau^{\textrm{PG}\alpha}\left(n,w\right) = \tau^{\rm TF}\left(n\right) e^{-\alpha C n^{-8/3}w}, \label{fq05d} \\
        \tau^{\textrm{L}\beta}\left(n,q\right)  =\beta \tau^{\rm TF}\left(n\right) q_r^2= \beta D n^{-5/3}q^2, \label{fq05c}
    \end{gather}
\label{fq05}
\end{subequations}
\noindent with the coefficients being 
%$A = E_h a_0^2 /8$, 
%$B = 3\left(3\pi^2\right)^{2/3}E_h a_0^2/10$,
$C = \left(3\pi^2\right)^{-2/3}/4$ and 
$D = 3\left(3\pi^2\right)^{-2/3} E_h a_0^2/160$.   

\mys
In Eq. \eqref{fq05c}, we also introduce the (adimensional) reduced Laplacian \cite{laricchia14,fabioijqc,fabio01}, i.e., $q_r=3 q/(40 \tau^{TF})$, which is largely used for the development of KE functionals.
\mye

It is useful to identify the following cases:
\begin{itemize}
\item $\alpha = 0$, $\beta = 0$. Equations \eqref{fq02}~-~\eqref{fq05} reduce to the models employed in previous works \cite{yan15prb,ciraciQHT,khalidMatr}; i.e.,
\my{
$T_s\left[n\right]$ is approximated as the sum of TF and vW  functionals; and we will indicate this case as TFvW (i.e., TF$\lambda$vW with $\lambda=1$).}
%Note that here we use the full vW term, i.e., $\lambda=1$. Other implementations in literature use $\lambda=1/9$, but as shown in Ref. \citenum{ciraciQHT} $\lambda=1$ is required to better reproduce TD-DFT results.

\item $\alpha \neq 0$, $\beta = 0$. It corresponds to the case where the QHT is improved with the addition of the PG$\alpha$ functional. 
We refer to this case as QHT-PG$\alpha$.
\item $\alpha \neq 0$, $\beta \neq 0$. This is the more complex case in which the Laplacian-level correction L$\beta$ is included in the energy functional. 
This case is be referred to as QHT-PG$\alpha$L$\beta$.
\end{itemize}

The parameters $\alpha$ and $\beta$ can be determined in a nonempirical way by imposing exact asymptotic solutions. In particular, we set $\alpha=40/27$ in order to satisfy second-order gradient expansion \cite{kirzhnits,fabio01} and use PGS for $\textrm{PG}40/27$.
Moreover, we follow the results of Ref. \citenum{fabio01} and fix $\beta=0.25$ such that the overall correction functional PGSL0.25 accurately reproduces the linear-response function of a noninteracting homogeneous electron gas at both small and large wave vectors \cite{fabio01}. For brevity, we use the acronym PGSL for PGSL0.25. %but for other

%More in general, in this work, we will consider functionals of the type:
%\begin{equation}
%     \tau\left(n,w,q\right) = \tau^{0}\left(n\right) + \tau^1\left(n,w\right) +  %\tau^2\left(n,q\right) 
%\end{equation}
%i.e. without terms involing products of $w$ and $q$.

In order to calculate the potential, we take the functional derivative of $T_s\left[n\right]$ \cite{fabioijqc} and obtain
% %
% \begin{equation}
%     \begin{split}
%         \frac{\delta T_s}{\delta n} = \tau_n + w\tau_{nnq} - 2w\tau_{nw} + \tau_{nq}q - 2q\tau_w \\
%         + 2\tau_{nqq} \nabla n \cdot\nabla q - 2\tau_{ww}\nabla n \cdot\nabla w  + \tau_{qq}\nabla^2q,
%     \end{split}
% \label{fq06}
% \end{equation}
%

% {\color{red}  }

\begin{equation}
    \begin{split}
\frac{\delta T_s}{\delta n} & = \, \tau_{n}+w\left(\tau_{n n q}-2 \tau_{n w}\right)+\left(\tau_{q n}-2 \tau_{w}\right) q \\
& +2\left(\tau_{n q q}-{\color{black} \tau_{w q} }\right) \left(\nabla n \cdot \nabla q\right) \\
& +2\left({\color{black} \tau_{n w q} }-\tau_{w w}\right) \left(\nabla n \cdot \nabla w\right) \\
& + {\color{black} 2\tau_{w q q} \left(\nabla w \cdot \nabla q\right) + \tau_{w w q} |\nabla w|^{2} } \\
& +{\color{black} \tau_{w q} \nabla^{2} w } +\tau_{q q} \nabla^{2} q +{\color{black} \tau_{q q q} }|\nabla q|^{2},
    \end{split}
\label{fq06}
\end{equation}

\noindent where the subscripts  $i = n, w, q$ denote the corresponding partial derivatives.
The detailed derivation of Eq.~\eqref{fq06} is given in Sec.~I of the Supplementary Material (SM) \cite{SM};
\my{a similar derivation can be found in Ref. \cite{kara09}.}

\my{
While the kinetic potential in Eq.~\eqref{fq06} is the key quantity for self-consistent OF-DFT calculations,
it is not used in the QHT linear response, where, instead, the second-order
functional derivative (never investigated so far) is required. In particular, 
}
the first-order term of the potential $\left(\frac{\delta T_s}{\delta n}\right)_1$ is required and it can be obtained using a perturbation approach where the perturbed density is taken as $n = n_0 + n_1$, with $n_1 = \frac{1}{e}\nabla\cdot\textrm{\textbf{P}}$ being the electron density perturbation. After some tedious algebra and neglecting higher-order terms, we obtain the following expression for the linear potential (see Secs.~I and II of the Supplemental Material for the full derivation \cite{SM}):
\begin{equation}
        \left(\frac{\delta T_s}{\delta n}\right)_1 = 
          \left(\frac{\delta T_s^{\textrm{I}}}{\delta n}\right)_1
        + \left(\frac{\delta T_s^{\textrm{II}}}{\delta n}\right)_1 \\ 
        + \left(\frac{\delta T_s^{\textrm{III}}}{\delta n}\right)_1,
        \label{fq07}
\end{equation}
where
\begin{widetext}
    \begin{subequations}
        \begin{align}
            \left(\frac{\delta T_s^{\textrm{I}}}{\delta n}\right)_1 = & \, \tau_{nn}^{(0)} n_1 \label{eqf:1}\\
            %%%%%%%%%%%%%%%%%%%%%
            \left(\frac{\delta T_s^{\textrm{II}}}{\delta n}\right)_1 = & - 2\tau_{nnw}^{(0)}\left|\nabla n_0\right|^2n_1 - 2\tau_{nw}^{(0)}\left[n_1 \nabla^2 n_0 + \nabla n_0 \cdot \nabla n_1\right] - 2\tau_w^{(0)} \nabla^2 n_1  \label{eqf:2}\\  
            & -2\tau_{ww}^{(0)} \left[2\left(\nabla n_0 \cdot \nabla n_1\right)\nabla^2 n_0 + \nabla \left(\left|\nabla n_0\right|^2 \right) \cdot \nabla n_1 + 2\nabla n_0 \cdot \nabla \left(\nabla n_0 \cdot \nabla n_1\right)\right]  \label{eqf:3} \\ 
                    & - 4\tau_{www}^{(0)} \left(\nabla n_0 \cdot \nabla n_1 \right)\left[\nabla n_0 \cdot \nabla \left(\left|\nabla n_0\right|^2 \right) \right] \label{eqf:4} \\
                    & -2\tau_{nww}^{(0)} \left[2\left(\nabla n_0 \cdot \nabla n_1 \right)\left|\nabla n_0\right|^2 + \left\{\nabla n_0 \cdot \nabla \left(\left|\nabla n_0\right|^2 \right) \right\}n_1\right], \label{eqf:5} \\
                    %%%%%%%%%%%%%%%
           \left(\frac{\delta T_s^{\textrm{III}}}{\delta n}\right)_1 = & \,
           \tau_{nnnq}^{(0)}\left|\nabla n_0\right|^2  n_1 + \tau_{nnq}^{(0)} \left[2\nabla n_0 \cdot \nabla n_1 + n_1 \nabla^2 n_0 \right] + 2\tau_{nq}^{(0)} \nabla^2 n_1  \label{eqf:6} \\
                    & + \tau_{nnqq}^{(0)} \left[\left|\nabla n_0\right|^2 \nabla^2 n_1 + 2 n_1\left\{\nabla n_0 \cdot \nabla\left(\nabla^2 n_0\right)\right\}\right]  \label{eqf:7} \\
                    & + \tau_{nqq}^{(0)} \left[\nabla^2 n_0 \nabla^2 n_1 + 2 \nabla n_1 \cdot \nabla\left(\nabla^2 n_0\right) + 2\nabla n_0 \cdot \nabla \left(\nabla^2 n_1\right) + \nabla^2\left(\nabla^2n_0\right)n_1\right] \label{eqf:8} \\
                    & +  \tau_{qq}^{(0)} \nabla^2\left(\nabla^2n_1\right) \label{eqf:9}  \\
              & {\color{black} + \tau_{q q q}^{(0)}\left[2\left(\nabla\left(\nabla^{2} n_{0}\right) \cdot \nabla\left(\nabla^{2} n_{1}\right)\right)+\nabla^{2}\left(\nabla^{2} n_{0}\right) \nabla^{2} n_{1}\right] } \label{eqf:10} \\
                    & {\color{black} +\tau_{n q q q}^{(0)}\left[2\left\{\nabla n_{0} \cdot \nabla\left(\nabla^{2} n_{0}\right)\right\} \nabla^{2} n_{1}+\left|\nabla\left(\nabla^{2} n_{0}\right)\right|^{2} n_{1}\right] } \label{eqf:11} \\
                    & {\color{black} +\tau_{q q q q}^{(0)}\left|\nabla\left(\nabla^{2} n_{0}\right)\right|^{2} \nabla^{2} n_{1} } \label{eqf:12}.
            \end{align}
            \label{fq08}
    \end{subequations}
\end{widetext}
The superscript $(0)$ indicates that the function is evaluated at $n = n_0$.
\mys
The terms are grouped so that  $\left(\frac{\delta T_s^{\textrm{I}}}{\delta n}\right)_1$ includes
only derivatives of $\tau$ with respect to $n$,
$\left(\frac{\delta T_s^{\textrm{II}}}{\delta n}\right)_1$  includes derivatives  of $\tau$ with respect to $w$, and
finally, $\left(\frac{\delta T_s^{\textrm{III}}}{\delta n}\right)_1$ includes derivatives of $\tau$ with respect to $q$.
Equation (\ref{fq08}) thus represents a novel and a quite general expression 
for the QHT first-order potential with increasing
complexity.

The term  $\left(\frac{\delta T_s^{\textrm{I}}}{\delta n}\right)_1$ is the only one
included in the TF-HT model, which is a local model.
In the case of the TFvW functional, only the  terms (\ref{eqf:1}) and (\ref{eqf:2}) survive. 
With the PGS functional, all terms in $\left(\frac{\delta T_s^{\textrm{II}}}{\delta n}\right)_1$ are included,
whereas PGSL includes all terms but \eqref{eqf:10}-\eqref{eqf:12}, as third- and fourth-order derivatives of $\tau$
with respect to $q$ are not present in Eq. \eqref{fq04}.
The terms in Eqs. \eqref{eqf:10}-\eqref{eqf:12} are present in the functional described
in Sec. \ref{sec:pgsln}, where the Laplacian term does not have a simple quadratic dependence on $q$.    
%in Eq.~\eqref{fq07} survives; $\left(\frac{\delta T_s^{\textrm{TFvW}}}{\delta n}\right)_1$ then %corresponds to the TFvW (with $\lambda = 1$) functional used in previous works %\cite{ciraciQHT,yan15prb,toscano}. 
%Clearly, $\left(\frac{\delta T_s^{\textrm{PG}\alpha}}{\delta n}\right)_1$ and $\left(\frac{\delta %T_s^{\textrm{L}\beta}}{\delta n}\right)_1$ are the contribution associated with the PG and L %corrections respectively.
% cases value of $\beta$ will be indicated.

Despite its apparent complexity, Eq. \eqref{fq08} can be implemented in finite-element codes.
Moreover, we note that Eq. \eqref{fq08} is not
the most general expression for a Laplacian-level KE functional:
When $\tau$ includes terms with products  of $w$ and $q$, additional terms are present, which will be investigated elsewhere. \myy{Such a product is present in the fourth-order gradient expansion \cite{ho73}}.
\mye 

Finally, we recall that the first-order term $\left(\frac{\delta E_{\textrm{XC}}^{\textrm{LDA}}\left[n\right]}{\delta n}\right)_1$ for the XC potential can be obtained via Perdew-Zunger LDA parametrization \cite{perdew}, a and its full expression can be found in Ref. \citenum{ciraciQHT}. 

%%%%%%%%%%%%%%%%%%%%%%%%%%%%%%%%%%%%%%%%%%%%%%%%%%%%%%%%%%
\newcommand{\RR}{\mathbf{r}}
\section{Asymptotic analysis}\label{sec:asy}
%%%%%%%%%%%%%%%%%%%%%%%%%%%%%%%%%%%%%%%%%%%%%%%%%%%%
\my{As discussed in Ref. \citenum{ciraciQHT} the tail of the ground-state density plays a fundamental role in the determination of the QHT solutions.
In this section we will summarize and generalize the derivation in Ref. \citenum{ciraciQHT} to Laplacian-level functional.}
We \my{start by taking} the divergence of Eq.~(\ref{fq01b}), and we use the quasistatic approximation (so that $\varepsilon_0 \nabla\cdot{\bf E}=\nabla\cdot{\bf P}=e n_1$), obtaining
\begin{equation}
\begin{split}
\nabla\cdot \frac{{e{n_0}}}{{{m_e}}} & \nabla {\left( {\frac{{\delta G}}{{\delta n}}} \right)_1} = \\
& - {{\omega ^2}  } e n_1 - \frac{e^2}{m_e} \left( \frac{e}{\varepsilon_0} n_0 n_1 + \nabla n_0 \cdot {\bf E} \right ).
\end{split}
\label{eq:asy1}
\end{equation}
\indent To obtain the asymptotic form of Eq.~(\ref{eq:asy1}), we assume that \cite{ciraciQHT}
\begin{eqnarray}
 n_0(\RR)&\rightarrow& {A_0}\exp(-\kappa r),  \\ 
 n_1(\RR)&\rightarrow& {B_0}\exp(-\nu \kappa r) \cos(\theta),  
\end{eqnarray}
where $\kappa>0$ is the decay constant of the ground-state density, and $\nu\kappa$ is
the decay constant of the (dipole excited) induced density.

The right-hand side (rhs) of Eq.~(\ref{eq:asy1}) is asymptotically vanishing, and it decays as
\begin{equation}
    -{\omega ^2}  e n_1 +  \frac{3\,e^2\, \kappa\, d_1}{4\pi\epsilon_0 m_e}\frac{n_0\cos(\theta)}{r^3},  
\end{equation}
where $d_1$ is the dipole moment of $n_1$ (see Ref. \citenum{ciraciQHT}).

For the left-hand side (lhs), \my{we firstly note terms like $\tau^{\rm TF}$,  $\tau^{\rm PG\alpha}$ and the XC term will vanish exponentially \cite{ciraciQHT}. Thus in the case of the PGSL functional we need to consider only $\tau^{\rm vW}$ and the new term $\tau^{\rm L\beta}$.}
\mys
For spherical systems, we have that the lhs of Eq. \eqref{eq:asy1} can be written as
\begin{equation}
  \nabla n_0(r)  \nabla  \left(\frac{\delta T_s^{\rm L\beta}}{\delta n}\right)_1=\sum_{n=0}^6 F_k[r,n_0(r)] \frac{d^k n_1(r)}{dr^k},   
  \label{eq:polynom}
\end{equation}
where $n_1({\bf r})=n_1(r)\cos(\theta)$ and $F_k$ are functions reported in Sec. III of the Supplemental Material \cite{SM}. Note for the PGSL functional, Eq. \eqref{eq:polynom} involves derivatives of $n_1$ up to the sixth order.
After some \my{algebra} (see Sec. III of the Supplemental Material \cite{SM}),  we obtain
\mye
\begin{widetext}
  \begin{eqnarray}
%    \nabla \frac{e n_0}{m_e}  \nabla  \left(\frac{\delta E_{xc}}{\delta n}\right)_1
%  &\rightarrow& 0  \label{eq:aterm0}\\   
  \nabla n_0  \nabla  \left(\frac{\delta T_s^{\textrm{vW}}}{\delta n}\right)_1
  &\rightarrow&
  \left (  -\frac{\nu^4}{4} + \frac{\nu^3}{2} -\frac{\nu^2}{4} 
  %+ O(\frac{1}{r})
  \right) \kappa^4 n_1  \label{eq:aterm1}\, ,\\
  %
%   \nabla \frac{e n_0}{m_e}  \nabla  \left(\frac{\delta T_s^{PG\alpha}}{\delta n}\right)_1
%   &\rightarrow&  0 \label{eq:aterm2} \\
   %
   \nabla n_0  \nabla  \left(\frac{\delta T_s^{\textrm{L}\beta}}{\delta n}\right)_1
   &\rightarrow&
   %\left (
   \beta \frac{\sqrt[3]{3}} {\pi^{4/3}}
   \left (
            \frac {
            %         \left(
            243  {{\nu}}^{6}      
          -1377  {{\nu}}^{5}
            +2025 {{\nu}}^{4}
            +765  {{\nu}}^{3}
            -3885 {{\nu}}^{2}
            +2865 {{\nu}}
            -650
       %\right)
              }{19\,440 }
%    
%     \right)
            \kappa^6
            %+ O\left (\frac{1}{r}\right)
            \right )
            \frac{n_1}{n_0^{2/3}}  \label{eq:aterm3} \, .
\end{eqnarray}
\end{widetext}
\my{Eq. \eqref{eq:aterm1} has already been derived in Ref. \cite{ciraciQHT}, whereas Eq. \eqref{eq:aterm3} is a key finding of the present work.}
\my{We recall that the} terms with $E_{\textrm{XC}}^{\textrm{LDA}}$ and $T_s^{\textrm{PG}\alpha}$ decay exponentially faster than $n_1$, and  that Eqs.~\eqref{eq:aterm1} and \eqref{eq:aterm3} represent only the leading terms in the asymptotic region.
%and the with the slowest $r$ power.

\my{ When $\beta=0$, the PGSL functional is asymptotically equivalent to the vW functional:} When $\hbar\omega$ is higher than critical energy see Eq. \eqref{eq:crit}, 
%\begin{equation}
%\hbar\omega_c = \hbar\frac{\kappa^2}{8}\sqrt{\frac{E_h a_0^2}{m_e}} \label{eq:wc}
%\end{equation}
the asymptotic decay is complex valued and oscillating. Otherwise, the asymptotic decay is exponential, and $\nu$ depends on $\omega$.

When $\beta>0$, we find, interestingly, that the L$\beta$ term gives an exponentially increasing contribution due to the division
by $n_0^{2/3}$, which dominates over the term in Eq.~(\ref{eq:aterm1}) as well as the term on the rhs.
Thus, the asymptotic solution does not depend on $\omega$, as in the conventional
QHT approach with the TFvW functional, but it is related to the solutions of the sixth-degree
polynomial in $\nu$ in Eq.~(\ref{eq:aterm3}), which are
\begin{equation}
\begin{split}
-1.320,& \; +0.543, \; +2/3,         \\
+1.123,& \; +5/3, \; +2.987.  
% \frac{5}{3}\kappa, \frac{2}{3}\kappa, \frac{5}{6}\pm \frac{1}{6}\sqrt{85\pm 8\sqrt{105}}
\end{split}
\label{eq:nusol}
\end{equation}
Only for those values of $\nu$, the lhs term vanishes asymptotically, as does the rhs.
Some of these solutions are not possible or unstable, i.e., those 
with $\nu \le 2/3$, as the term $n_1/n_0^{2/3}$ will not decay asymptotically.
The other three values of $\nu$ give the right asymptotic solution, but a high-order analytical analysis or a full numerical solution
is required to select the actual value of $\nu$.
Interestingly, all these solutions have  $\nu>1$, which is another
difference with respect to the QHT approach with the TFvW functional \cite{ciraciQHT}, where $\nu<1$.

%%%%%%%%%%%%%%%%%%%%%%%%%%%%%%%%%%%%%%%%%%%%%%%%%%%%%%%%%%%%%%%%%%%%
\section{Numerical implementation}\label{sec04}
%%%%%%%%%%%%%%%%%%%%%%%%%%%%%%%%%%%%%%%%%%%%%%%%%%%%%%%%%%%%%%%%%%%%
The system of Eqs.~\eqref{fq01} with Eq.~\eqref{fq02} and expressions \eqref{fq08} is solved for a plane-wave excitation using a commercial implementation of the finite-element method (FEM) \cite{comsol}. 
\begin{figure}[!htb]
\centering
\resizebox{0.85\columnwidth}{!}{\includegraphics[width=0.3\textwidth,angle=0]{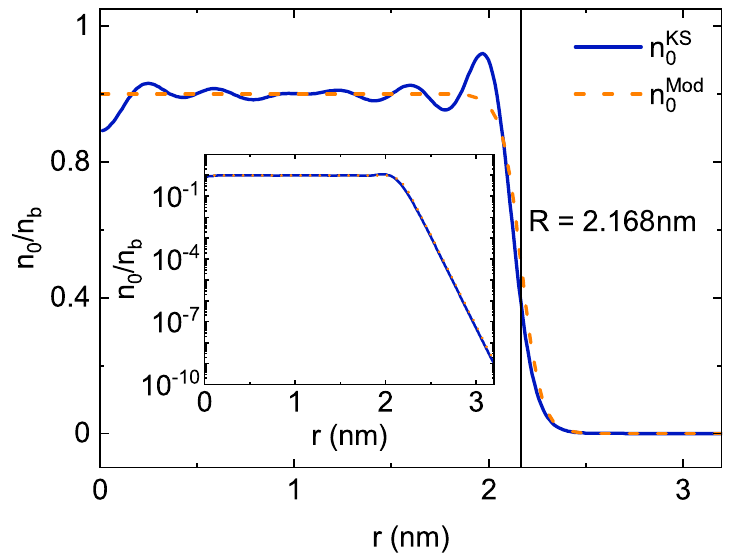}}
\caption{\small KS $\left(n_0^{\textrm{KS}}\right)$ and model $\left(n_0^{\textrm{Mod}}\right)$ ground-state densities for a Na jellium sphere with $N_e = 1074$ electrons. The inset shows the variation of densities in the logarithmic scale. Values are normalized to the bulk density $n_b=(4/3\pi r_s^3)^{-1}$.}
\label{f02}
\end{figure}

In order to easily compute absorption spectra for spheres and sphere dimers, we implement our equations using the \textrm{2.5D technique}, which significantly reduces the computational time for axisymmetric structures \cite{Ciraci:2013jt,ciraci02,ciraciQHT}.
A detailed explanation of the FEM implementation can be found in Appendix \ref{appfem}. 
\my{In particular, we used Dirichlet boundary conditions without making any assumption of the asymptotic decay}.
%\textbf{HENRICK DESCRIBED BOUNDARY CONDITIONS}
A completely independent implementation has also been carried out using a finite-difference method for spherical systems in the quasistatic approximation: The results obtained with the two methods are numerically the same, \my{ and details of
the finite-difference implementation will be published elsewhere.}
%{\bf In particular the simulation domain-size is set to XXXX}.

In order to solve the system of Eqs.~\eqref{fq01}, an expression for the ground-state density function $n_0\left(\textbf{r}\right)$ is required. Throughout the article, we consider the following two ground-state density functions: (i) the exact KS density $n_0^\textrm{KS}\left(\textbf{r}\right)$ calculated using a DFT in-house code \cite{ciraciQHT}, and (ii) a model density defined as \cite{banerjee2008,ciraciQHT}
\begin{equation}
    n_0^{\textrm{Mod}}\left({r}\right) =\frac{1}{1 + \textrm{exp}{\left(\kappa^{\textrm{Mod}}\left(r - R\right)\right)}}
\label{fq10}
\end{equation}
normalized with a condition $\int n_0^{\textrm{Mod}}\textit{d}V = N_e$, where $N_e$ is the number of electrons. For the $\kappa^{\textrm{Mod}}$ coefficient, the $\kappa^{\textrm{Mod}} = 1.05/a_0$ value is fixed and fitted with asymptotic decay of the KS electronic density decay \cite{ciraciQHT}. Figure~\ref{f02} shows $n_0^{\textrm{KS}}$ and $n_0^\textrm{Mod}$ densities for a Na (Wigner-Seitz radius $r_s = 4$ a.u.) jellium nanosphere with $N_e = 1074$ electrons (nanosphere radius $R=2.168$~nm). Note that $n_0^{\textrm{Mod}}$ does not display Friedel oscillations inside the nanosphere volume (surface marked with the vertical line), which are instead present in $n_0^\textrm{KS}$. The inset shows that the asymptotic decay is the same for both cases.

%%%%%%%%%%%%%%%%%%%%%%%%%%%%%%%%%%%%%%%%%%%%%%%%%%%%%%%%%%%%%%%%%%%%%
\section{Absorption spectra}\label{sec05}
\begin{figure}[hbt]
\centering
\resizebox{0.95\columnwidth}{!}{\includegraphics[width=0.3\textwidth,angle=0]{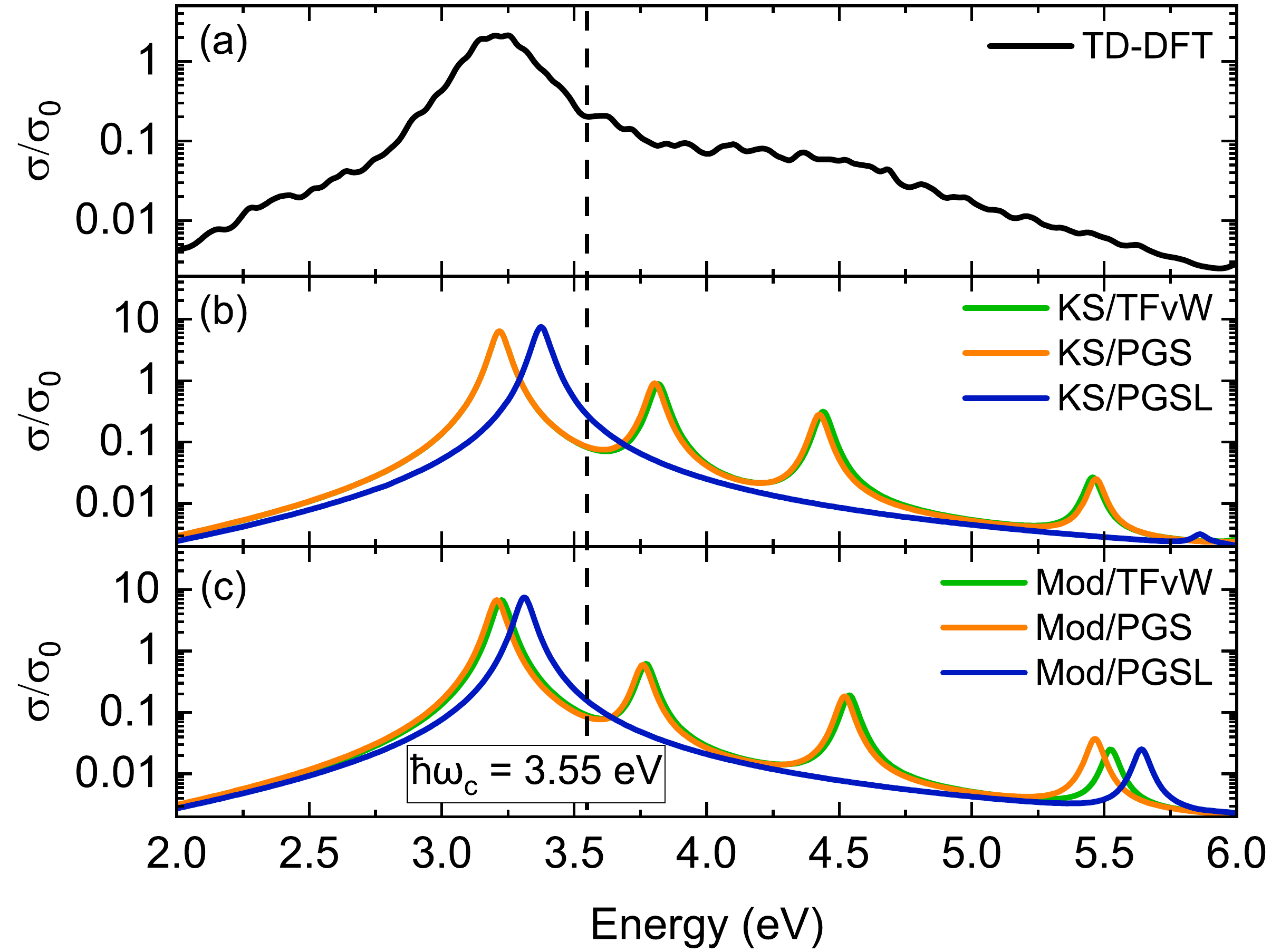}}
\caption{\small Normalized absorption cross section $\sigma/\sigma_0$ (see the Appendix A for definitions) in logarithmic scale for a Na jellium sphere with $N_e = 1074$
electrons as obtained from TD-DFT, QHT-TFvW, QHT-PGS, and QHT-PGSL using KS and model ground-state densities.
$\hbar\omega_c$ is the critical frequency: see Eq. \eqref{eq:crit}.}
\label{f03}
\end{figure}

%%%%%%%%%%%%%%%%%%%%%%%%%%%%%%%%%%%%%%%%%%%%%%%%%%%%%%%%%%%%%%%%%%%%%%%
%
In Fig.~\ref{f03}, we report the comparison of the normalized absorption cross section for a Na jellium nanosphere with $N_e=1074$ electrons as obtained using \my{QHT with three different KE functionals (TFvW, PGS, PGSL)} as well as the TD-DFT approaches (see Appendix \ref{appabs} for definitions and details). 
\my{The QHT results with a given KE functional F will be indicated in the following as KS/F or Mod/F, if the KS density or the model density is used as input density, respectively. When it is not relevant for the discussion to specify the input density the shorthand QHT-F will be used.
}

Figure \ref{f03} shows that the energy of the LSP resonance (first main peak) for QHT-TFvW and QHT-PGS is in good agreement (within 10~meV) with TD-DFT (approximately $3.22$~eV), which is broader due to  quantum-size effects, while KS/PGSL and Mod/PGSL give the LSP peaks at apprimately $ 3.37$~eV and $3.31$~eV, respectively, which are blueshifted with respect to TD-DFT results (for further analysis of the position of LSP peak, see Sec.~\ref{sec07}).
As we discuss in the Introduction, QHT-TFvW gives accurate energy of the LSP and predicts additional peaks at higher energies, which are not present in the TD-DFT. Almost the same situation is obtained for QHT-PGS, meaning that even the more general gradient approximation in Eq.~\eqref{fq05d} does not solve the problem of additional peaks. 
On the other hand, the QHT-PGSL absorption spectrum is quite different.
The main difference between QHT-TFvW and QHT-PGSL is not the energy shift of the LSP but the absence of additional resonances in the latter. \my{Actually, a second small peak is present in the QHT-PGSL spectrum at high energy, namely $\approx 5.85$~eV (hardly visible in Fig.~\ref{f03}-(b) for KS/PGSL and $\approx 5.7$~eV for Mod/PGSL. This peak is a Bennett state (\myy{which can be identified as shown in} Fig.~S6 of the Supplemental Material \cite{SM}) \myy{and it} will be further discussed in Section \ref{sec:pgsln}}.

%In addition, 
%In addition, the LSP (first peak) for Mod/QHT-PGS ($\approx$ 3.208~eV) and KS/QHT-PGS ($\approx$ 3.218~eV) are in good agreement with
%and have a small shift from Mod/QHT ($\approx$ 3.225~eV) and KS/QHT ($\approx$ 3.218~eV). It is also interesting is that KS/QHT and KS/QHT-PGS give very close values for LSP with a difference of less than 1 meV and Mod/QHT-PGS differs from Mod/QHT by $\approx$ 0.02~eV. KS/ (TD-DFT LSP peak is much broader because of the quantum size effects). 
%%%%%%%%%%%%%%%%%%%%%%%%%%%%%%%%%%%%%%%%%%%%%%%%%%%%%%%%
%\subsection{Limits of the conventional QHT}\label{sec02}
%%%%%%%%%%%%%%%%%%%%%%%%%%%%%%%%%%%%%%%%%%%%%%%%%%%%%%%%https://www.overleaf.com/project/5e3d7920017c950001954321

Although the QHT-TFvW predicts very well the LSP resonance when compared to more sophisticated TD-DFT approaches, the presence of additional peaks is a 
major shortcoming.
\mys
These peaks are, in fact, very sensitive to the details in the tail of the density (see Fig.~S4 in the Supplemental Material \cite{SM}, where a model density with different $\kappa^{\rm Mod}$ are considered). 
A small modification of the tail of the input density should not change the absorption spectrum significantly. 
This is the case for the Mod/PGSL absorption spectra, which are thus robust with respect to the input density. 
On the other hand, the Mod/TFvW absorption spectrum is instead very sensitive, and it results in being largely affected by additional peaks. 
%This is the case for the absoprtion spectrum of Mod/PGSL, which  is quite stable for all values of $\kappa_{mod}$. The LSP is always present and stable; only the Bennett state shows, as expected, some dependence on $\kappa_{mod}$.
%On the other hand for several peaks appears at high frequency 
%and moreover the LSP is completely destroyed for $\kappa_mod\lt 0.95$.
%The black line in Fig.  \ref{fig:modk}a represent the critical frequency, see Eq. %\ref{eq:crit}: below this line many states appears.
%Only for $\kappa_mod\lt 1.3$ the spectrum is quite stable.
\mye
These peaks have an energy higher than $\hbar\omega_c$
%\approx 3.55$ eV (see Ref. \cite{ciraciQHT}) and 
and can hardly be treated in an efficient numerical scheme. 
%it nevertheless has a major shortcoming.
%It can be demonstrated that QHT asymptotic solutions in the exponentially decaying density tail become propagating after a critical energy  \cite{ciraciQHT}, causing the appearance of spurious resonances and which can hardly be treated in a simple or efficient numerical scheme. 

\begin{figure}[ht]
\centering
\includegraphics[width=1\columnwidth]{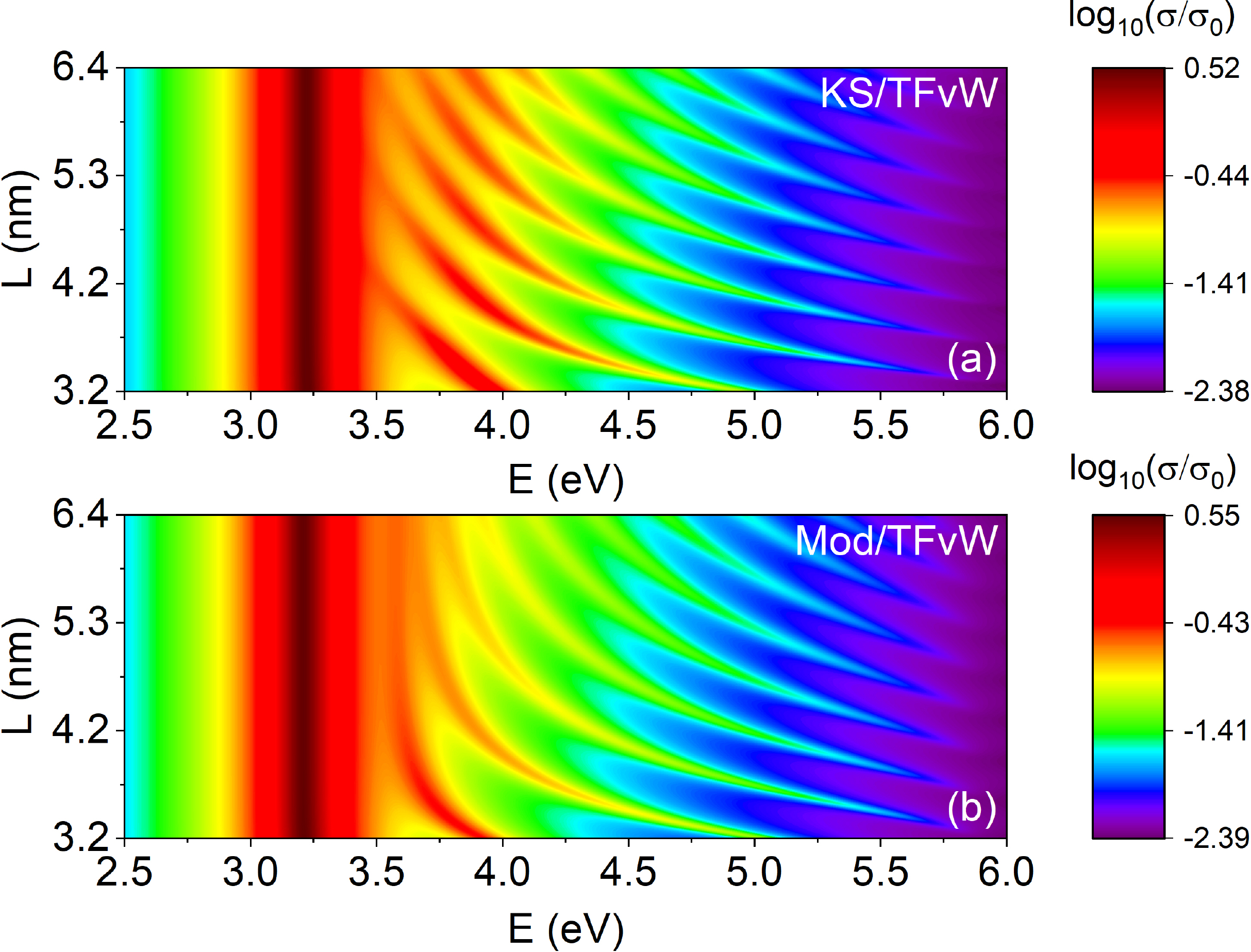}
\caption{\small The effect of computational domain size (L) on the normalized absorption spectra as obtained from KS/QHT (a) and Mod/QHT (b) for a Na jellium nanosphere with $N_e = 1074$ electrons.}
\label{f01}
\end{figure}

This behavior is shown in Fig.~\ref{f01}, where QHT-TFvW normalized absorption cross sections ($\sigma/\sigma_0$) for the same jellium nanosphere are calculated for increasing size of the simulation domain. These calculations have been done with an \my{in-house developed finite-difference code for spherical systems (see Section \ref{sec04})%(details will be published elsewhere)
}, which reproduces exactly the FEM results reported in this work but is more accurate in the asymptotic region\my{ (since it requires only a one-dimensional discretization)}.
The results are obtained with KS (upper panel) and model (lower panel) ground-state densities.
Clearly, as the domain size increases, more and more modes appear (and with reduced intensities) in the spectrum without any limit. Thus, the absorption spectrum is very sensitive to the domain size. 
We note that no previous report in the literature has considered the numerical convergence of those states in QHT calculations.
With an {\it infinite} computational domain size, there should be an infinite number of states with infinitely small peak intensity; i.e., {\it no peaks can be distinguished anymore}, and  only an unstructured shoulder could be present.

\begin{figure}[ht]
\centering
\includegraphics[width=0.9\columnwidth]{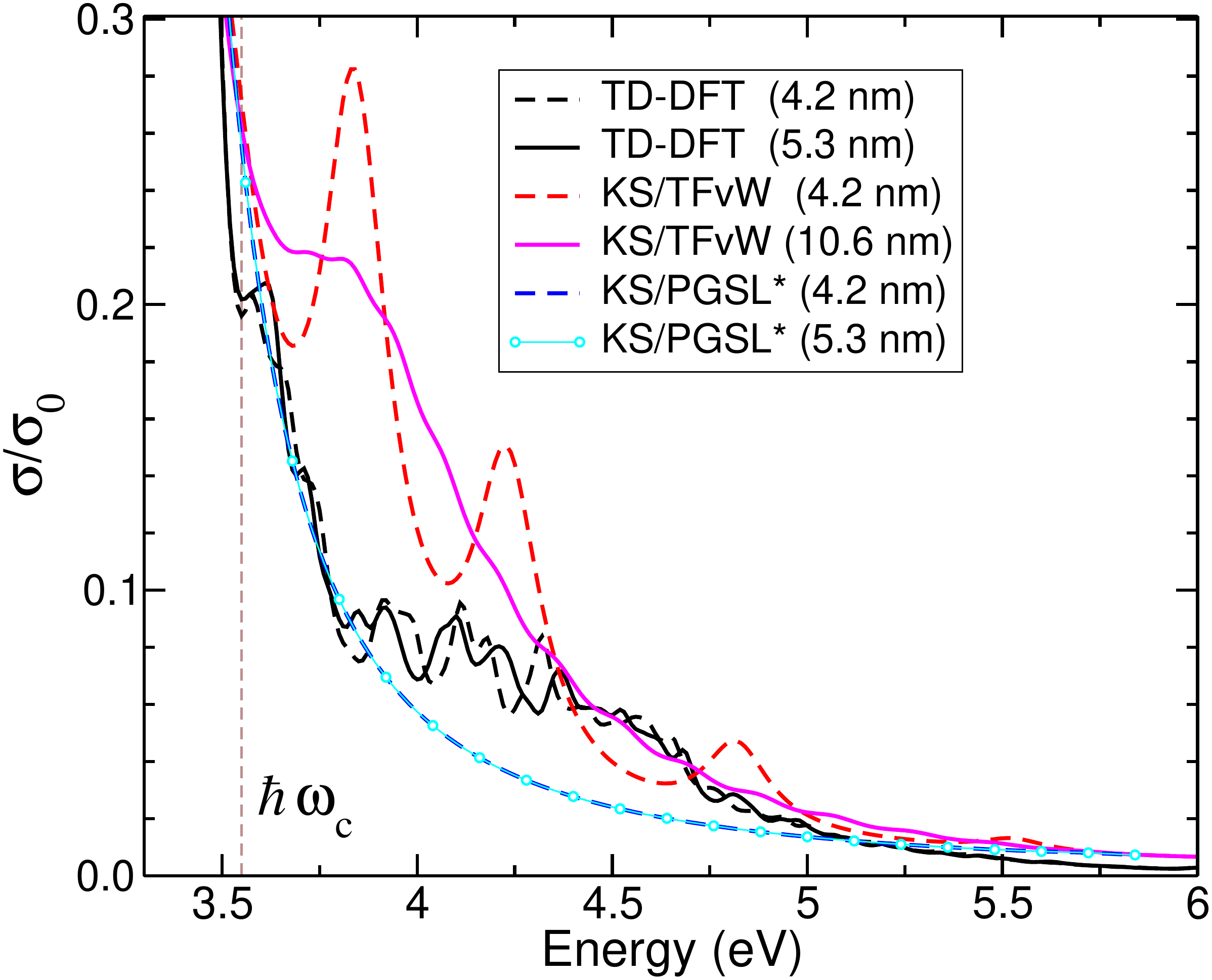}
\caption{\small High-energy part of the normalized absorption cross section for $N_e$=1074 electron Na jellium nanosphere considering different computational domain sizes \my{ for QHT with different functionals and TD-DFT. KS/PGSL* means that the spectrum have been red-shifted in order to have the same energy position of KS/TFvW. The damping (see Eq. \eqref{fq01b} and Appendix A) is $\gamma=0.2$ eV ($\gamma=0.234$ eV) for KS/TFvW (KS/PGSL*).
$\hbar\omega_c$ is the critical frequency, see Eq. \eqref{eq:crit}.}}
\label{fig:into}
\end{figure}

This behavior is indeed shown in Fig.~\ref{fig:into} where we report the QHT and TD-DFT results for two different computational domain sizes. We use a larger damping for KS/TFvW (namely, $\gamma$=0.2 eV) so that it will give the same intensity at the LSP peak as TD-DFT. 
While the TD-DFT results are converged with standard domain size, convergence seems to appear for KS/TFvW only with a domain size of 10.6nm (200 a.u.), where no more Bennett peaks can be distinguished and only a shoulder is present. However, this shoulder, which starts at $\hbar\omega_c$, is significantly higher  (about a factor of 2.5 in intensity) than the TD-DFT one, which starts later at about 3.7~eV.
Clearly, a domain size of 200 a.u. to obtain a converged absorption spectrum is not reasonable for any application in plasmonics, and it is obtained only with a specialized code for reference calculations.

In Fig.~\ref{fig:into}, we also report the KS/PGSL* results, where the * indicates that the spectra are red-shifted by 0.15~eV in order to have the same LSP energy position as QHT; the damping is fixed to $\gamma=0.234$ eV so that the peak intensity is also the same.
The plot shows that the KS/PGSL* does not change at all with the computational domain size (see also Fig S5 in the Supplemental Material \cite{SM}) , and overall it is much closer to TD-DFT than QHT.
%Although in this case the LSP energy $\hbar\omega_{lsp}$ is not affected by the presence of the extra peaks, its amplitude is
%, since now the total oscillator strength has to be split among several modes. 

A more quantitative comparison of methods can be done by considering the integrated absorption cross section
\begin{equation}
I(\omega)=\int_0^\omega \sigma(\omega')  d\omega',
\end{equation}
which converges to $(\pi e^2)/(2 \epsilon_0 m_e c)N_e$  for $\omega \to \infty$,  where $N_e$ is the number of electrons \cite{boh79,brack93,sum15}.

\begin{figure}[ht]
\centering
\includegraphics[width=0.9\columnwidth]{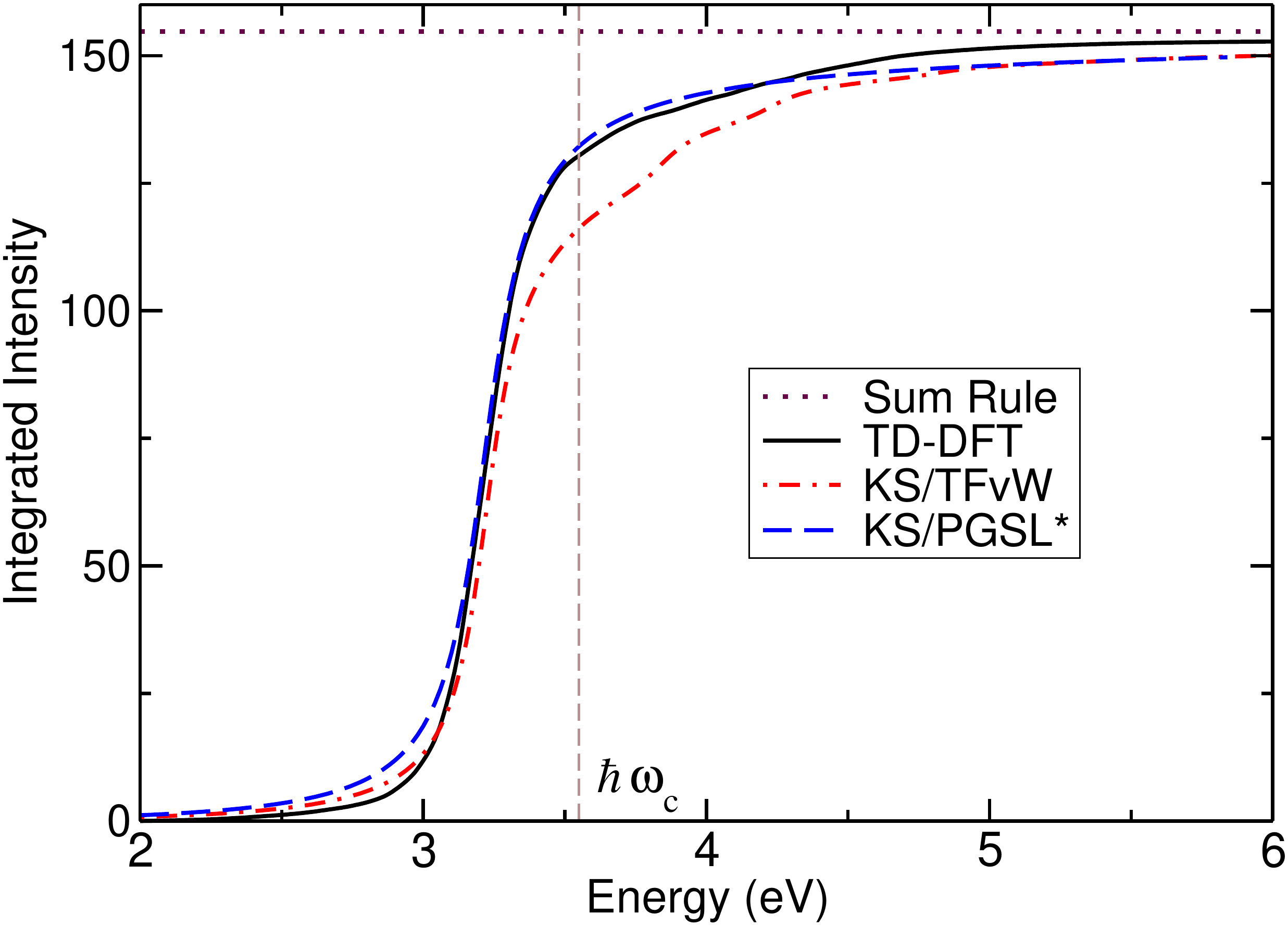}
\caption{Integrated intensity from KS/TFvW, KS/PGSL*, and TD-DFT for a Na jellium nanosphere with $N_e = 1074$ electrons. The classical limit is also
reported.}
\label{fosc}
\end{figure}
The integrated absorption is plotted in Fig.~\ref{fosc}, and it shows that $I(\omega)$ for QHT-TFvW and QHT-PGSL* converge to the same value for high energies. However, while the integrated absorption curve for TD-DFT and QHT-PGSL* are very close to each other, the growth in QHT-TFvW is much slower, meaning that the oscillator strength (i.e., the energy-integrated intensity) in QHT-TFvW is split into several Bennett modes, whereas the single peak in QHT-PGSL* contains it all. In fact, the integrated absorption for QHT-TFvW at $\hbar\omega=\hbar\omega_c$  is about 15\% smaller than  QHT-PGSL* and TD-DFT.
%In addition Fig. \ref{fosc} shows that QHT-PGSL* is much closer to the
%reference TD-DFT results, than QHT (

In Sec.~\ref{sec07}, a more detailed analysis of the oscillator strength and absorption cross section for different numbers of electrons is presented.
Here, we remark that these features are not limited to spherical NPs but could happen in other geometries or materials. In fact, for $\omega_c$ the identical expression was obtained for a jellium sphere \cite{ciraciQHT} and slab \cite{yan15prb}. Thus, in general, one could have for LSP $\omega_{\textrm{LSP}}\simeq \omega_c$ or even  $\omega_{\textrm{LSP}}>\omega_c$. In such cases, the QHT cannot describe the LSP peak (see also Fig.~S4 in the Supplemental Material \cite{SM}).

%The amplitude is decreasing with the increment of domain size as shown in the Supplementary Material (SM) Fig. S1 \cite{SM}. 
%\textbf{NEW FIGURE FOR THE MAIN MANUSCRIPT} 

%The reason behind this behaviour is fundamentally bond to the non-exactness of the KE functional used in QHT.

%In a Physical Review Letter article %\textbf{PRL\_Ref} submitted at the %same time as the current, we give a %robust proof that in the framework of %QHT in Laplacian-level there is no %dependence from $\omega$ of the %linear induced density decay at a %jellium surface.
%

%%%%%%%%%%%%%%%%%%%%%%%%%%%%%%%%%%%%%%%%%%%%%%%%%%%%%%%%%%%%%%
\section{Induced charge density}\label{sec06}
%%%%%%%%%%%%%%%%%%%%%%%%%%%%%%%%%%%%%%%%%%%%%%%%%%%%%%%%%%%%%
As we discuss in Sec.~\ref{sec:asy}, the decay of QHT-TFvW induced densities is frequency dependent, and solutions are pure exponentially decaying at the metal surface only if the incident plane-wave energy is lower than $\hbar\omega_c$, whereas using the PGSL functional, a fixed exponential decay should be obtained.
%
%$\kappa$ being the decay coefficient for the ground-state \cite{ciraciQHT}. 
%

\begin{figure}[ht]
\centering
\resizebox{0.85\columnwidth}{!}{\includegraphics[width=0.3\textwidth,angle=0]{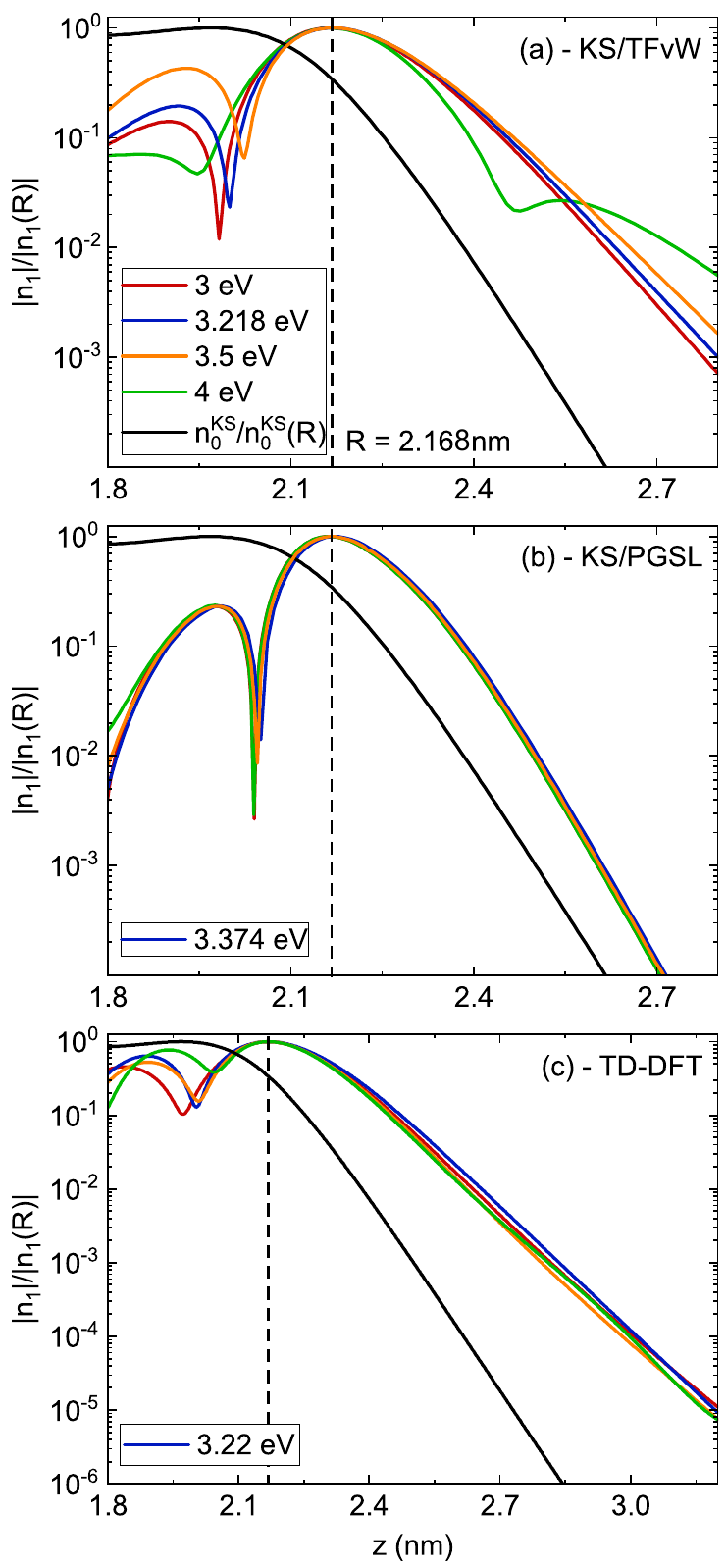}}
\caption{\small Modulus of induced charged density at different energies for a Na jellium sphere with $N_e = 1074$ electrons as calculated from KS/TFvW (a), KS/PGSL (b), and TD-DFT (c). The blue curves correspond to the densities associated with the LSP excitation energy. The critical frequency is $\hbar \omega_c = 3.55$ eV; see Ref. \cite{ciraciQHT}.}
\label{f04}
\end{figure}

%
This fact can be verified numerically by plotting the computed induced charge density $n_1$ (associated with the absorption).
%For QHT (panel a) the asymptotic decay changes with $\omega$ and %for $\omega>\omega_c$ it becomes oscillating. 
%It can be shown \cite{ciraciQHT}, in fact, that  %$n_1\left(\textbf{r}\right) \propto \textrm{exp}\left(-\beta %r\right)$ where $\kappa$ (which can be complex) depends on the %excitation energy $\hbar\omega$. 
%For QHT-PGSL (panel b) the scenario is completely different. 
In Fig.~\ref{f04}, we plot $\left|n_1\right|$ (in logarithmic scale) as obtained from the KS/TFvW and KS/PGSL for a Na jellium nanosphere with $N_e = 1074$. %(similar results, not shown, are obtained for the model density).
To have a clear comparison of decay rates, the curves for $\left|n_1\right|$ are shifted to have the maximum at $z=R$ and normalized to $\left|n_1\left(R\right)\right|$, while $n_0$ density is normalized only to $n_0\left(R\right)$.

For the KS/TFvW induced density, the decay slope shows a clear dependence on the incident energy $\hbar\omega$, becoming oscillatory for $\hbar\omega > \hbar\omega_c=3.55$ eV (note that KS/TFvW induced densities are not converged with respect to the computational domain size, as we discuss in Sec.~\ref{sec05}). 

On the other hand, the KS/PGSL calculations yield the same slope for all excitation energies, as we analytically demonstrate in Sec.~\ref{sec:asy}. A numerical fit of the decay gives a value of $\nu$ close to +1.12, i.e., the slowest from asymptotically decaying solutions (with $\nu > 2/3$); see Eq. \eqref{eq:nusol}.

It is important to note that the TD-DFT calculations [Fig.~\ref{f04}-(c)] give qualitatively similar results to the QHT-PGSL. In fact, for TD-DFT we get the same decay slope for the induced density (at least for $\hbar\omega < \hbar\omega_{p}$). However, as we discuss in Sec.~\ref{sec:asy}, in QHT-PGSL we have $\nu>1$ while $\nu<1$ in the QHT-TFvW, meaning that spill-out effects are somehow smaller in QHT-PSGL. 

Nonetheless, we need to point out that this feature is peculiar to PGSL, which is one of the few Laplacian-level KE functionals, and PGSL has not been developed for QHT calculations. Thus, another Laplacian-level KE functional can be developed with \my{different}  features. \my{ In Sec. \ref{sec07}, the induced charge density is further analyzed in terms of Feibelman $d$ parameters.}

Another important aspect is the numerical stability of the QHT-PGSL approach: Not only the absorption spectra do not depend on the domain size, but the fact that the decay constant is fixed and independent of the frequency allows the use of the mixed boundary condition for an exponential decay (i.e., $\hat{r}\cdot\nabla n_1 + \nu \kappa n_1=0$), allowing converged results even with a very small domain size (see Fig.~S5 in the Supplemental Material \cite{SM}).

\mys
%%%%%%%%%%%%%%%%%%%%%%%%%%%%%%%%%%%%%%%%%%%%%%%%%%%%%%%%%%%%%%
%%%%%%%%%%%%%%%%%%%%%%%%%%%%%%%%%%%%%%%%%%%%%%%%%%%%%%%%%%%%%% 
\section{Toward an accurate kinetic energy functional for QHT}\label{sec:pgsln}
%%%%%%%%%%%%%%%%%%%%%%%%%%%%%%%%%%%%%%%%%%%%%%%%%%%%%%%%%%%%%%
%%%%%%%%%%%%%%%%%%%%%%%%%%%%%%%%%%%%%%%%%%%%%%%%%%%%%%%%%%%%%%
In the previous sections, we show that the QHT results with the PGSL functional are distinctively different from the ones obtained with the more conventional TFvW and PGS functionals. In particular, the PGSL functional modifies the description of the density tail, removes all the additional high-energy peaks, and improves the oscillator strength of the LSP peak,  but it overestimates its energy. We recall that the PGSL functional has not been developed for QHT linear response but for ground-state OF-DFT calculations of bulk properties of metal and semiconductors \cite{fabio01}.
Nevertheless, we show in the previous sections that the Laplacian term (i.e., $\beta q^2$) present in the PGSL functional is of fundamental importance also for QHT.
In this section, we propose a modification of the PGSL functional to describe accurately the QHT linear-response properties.
We find that a modification of the $\beta$ parameter does not lead to any relevant modification of the results. This can be understood considering that the asymptotic solutions [i.e., Eq. \eqref{eq:nusol}] do not depend on $\beta$. 

Here, we consider the following kinetic energy density (named PGSLN):

\begin{equation}
\tau=\tau^{\textrm{vW}} +\tau^{\textrm{PGS}} + \tau^{\textrm{TF}}\left[ \beta {q_r}^2  +   2 \beta {q}_0^2 
 \ln(1+{q_r}/{q}_0)  \right], \label{eq:pgsln}
\end{equation}
where $q_0$ is a parameter. In this way, for large ${q_r}$ 
(${q_r}\gg {q}_0$), i.e., in the density tail, the functional will be equivalent to PGSL, whereas for  
${q_r}\ll {q}_0$
(i.e., inside the nanoparticle, where $|{q_r}|<0.2$; see Fig.~S7 of the Supplemental Material \cite{SM}), we have that
\begin{equation}
\tau \approx \tau^{\rm vW} +\tau^{\rm PGS} + 2 {q}_0 \beta {q_r} + O(  {q_r}^3 ) \, , 
\end{equation}
thus removing the quadratic term ${q_r}^2$.
For  small ${q_r}$, the PGSLN functional will be thus equivalent to PGS because a linear term in ${q_r}$ does not contribute to the kinetic energy or to the kinetic potential \cite{szmiga17}.
The PGSLN is an accurate total kinetic energy functional yielding also
accurate total energies of jellium nanospheres; see Table S2 of the Supplemental Material. \myy{The parameter $q_0$ has a well defined physical meaning, as it defines how rapidly the PGSL behavior is recovered at the density tail: the larger $q_0$, the farther the  quadratic term ${q_r}^2$ is recovered. The PGSLN is thus an interpolation between two density regimes where exact conditions are known; the asymptotic region where the quadratic term, ${q_r}^2$, will render the induced density decay independent from the frequency (see Section \ref{sec:asy}) and the region inside the nanoparticle where the density is slowly varying and the PGS functional satisfy the second-order gradient expansion \cite{fabio01}. The transition between these two density regimes is described by the $q_0$ parameter, whose actual value will be defined in the following.}

\begin{figure}[hbt]
    \centering
    \includegraphics[width=0.95\columnwidth]{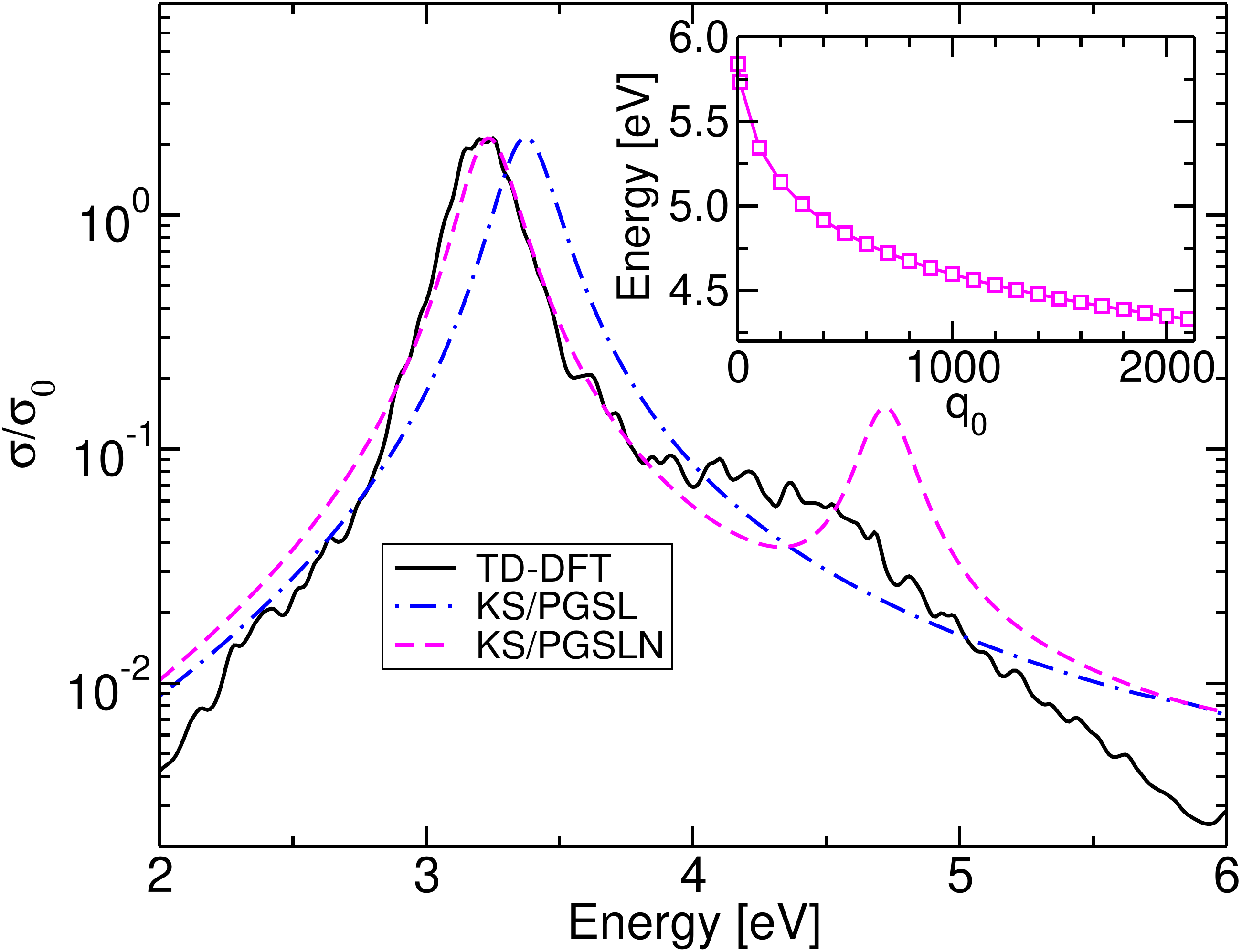}
    \caption{\my{Normalized absorption cross section (in logarithmic scale) for a Na jellium nanosphere with $N_e$=1074 electrons as computed
    from TD-DFT, KS/PGSL and KS/PGSLN with $q_0=700$.  The damping is $\gamma=0.234$ eV ($\gamma=0.224$ eV) for KS/PGSL (KS/PGSLN) so that all spectra have the same intensity at the LSP peak. }
    In the inset is the position of the Bennett peak for different
    values of $q_0$ for the PGSLN functional.}
    \label{fig:q0}
\end{figure}

In Fig.~\ref{fig:q0}, we report the absorption spectrum as computed
from TD-DFT, KS/PGSL, and KS/PGSLN using a larger damping for PGSL and PGSLN in order
to have the same intensity for the LSP peak.

The first main difference between KS/PGSL and KS/PGSLN is the presence of a well-defined second (Bennett) peak at $4.7$ eV \myy{(see also Fig.~S6 in the Supplemental Material \cite{SM})}. For KS/PGSL the Bennett peak (at 5.85 eV) cannot be distinguished at all when a large damping is used. 
The KS/PGSLN spectra are stable with respect to the computational domain size 
(see Fig.~S5 in the Supplemental Material \cite{SM}),
but \myy{the energy position of the Bennett state} changes with the values of $q_0$, as shown in the inset.
When $q_0\approx0$, the PGSLN functional is close to PGSL and indeed the position
of the Bennett peak is at very high energy (close to the volume plasmons, as also shown in Fig.~\ref{f03}).
Larger $q_0$ gives smaller energy of the Bennett peak.
We define the PGSLN functional with $q_0=700$ a.u.  in order to have the Bennett peak at 4.7 eV, as obtained from TD-DFT calculations for Na metal surfaces \cite{tsuei90,tsuei91}.
Fixing $q_0=700$ a.u. means that the PGSLN functional recovers the PGSL behavior only quite far outside the nanoparticle (see Fig.~S7 of the Supplemental Material \cite{SM}).
\myy{Note that fixing parameters from reference calculations of model systems is a standard procedure for DFT functional development since the known exact conditions are usually insufficient to build the full functional \cite{perdew2007,fabioijqc,umgga}}.

As stated in the Introduction, in TD-DFT calculations of nanoparticles, a well-defined Bennett state is not present, because
it is strongly damped and broadened due to the interaction with single-particle
transitions (not included, by definition, in any hydrodynamical approach).
Thus, the overall agreement between the PGSLN and TD-DFT spectra is not very good in the high-energy part (a sharp peak is present in QHT-PGSLN, whereas TD-DFT shows only a broad shoulder). A possible solution to be investigated in the future is to use in QHT a frequency-dependent damping (in contrast to a fixed value employed here) so that the QHT Bennett peak could be made broader (as it is in TD-DFT).

The second important difference is that the position of the LSP in PGSLN is redshifted to the correct (i.e., TD-DFT) position. Thus, the PGSLN functional not only predicts a correct and numerically stable Bennett peak, but it also corrects the overestimation of the LSP energy, peculiar to the PGSL functional. 
\myy{Thus the selected value of $q_0$, defined from the position of the Bennett state, also yields an LSP energy in excellent agreement with TD-DFT results. This result can be seen as an independent check of the reliability of the $q_0$ parameter, and it is an important result as the $q_0$ parameter could be also defined to reproduce the LSP peak energy position: in this case, as a direct consequence, the Bennett state will be at the right energy. Thus $q_0$ parameter is not a bare empirical parameter, but it describes the interplay between the Bennett state, which is related to the density decay far away from the system, and the LSP peak, which is instead related to density behaviour inside the nanoparticle.}
In Sec. \ref{sec07}, a detailed benchmark on the LSP energy position for Na jellium spheres of different dimensions will is presented.

\begin{figure}[hbt]
    \centering
    \includegraphics[width=0.95\columnwidth]{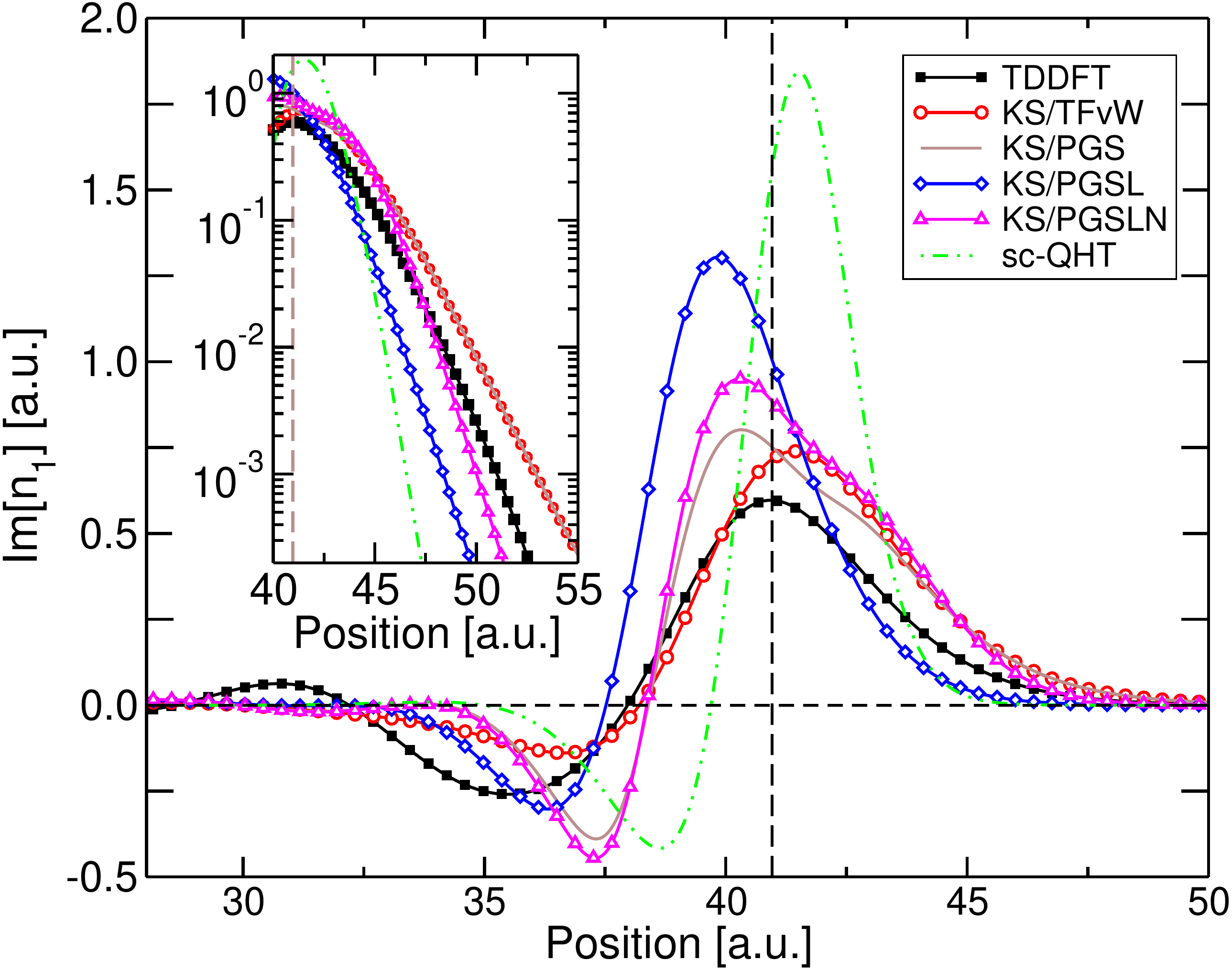}
    \caption{\my{Imaginary part of the induced density for  a Na jellium nanosphere with $N_e$=1074 as computed for TD-DFT, KS/TFvW, KS/PGS, KS/PGSL, KS/PGSLN, and SC-QHT. The inset shows the tail region in the logarithmic scale.}}
    \label{fig:ind}
\end{figure}

Finally, in Fig.~\ref{fig:ind}, we consider the induced density $n_1$ for different functionals.
Figure \ref{fig:ind} shows that KS/TFvW and KS/PGS give a quite accurate description of the induced density as compared to TD-DFT, but with an asymptotic tail (see the inset), which is slower than TD-DFT.
KS/PGSL is instead more confined inside the nanoparticles and decays faster (see also Fig.~\ref{f04}). The KS/PGSLN induced density is instead close to the KS/PGS one inside the nanoparticle, whereas in the tail, it approaches KS/PGSL.
In Fig.~\ref{fig:ind}, we also report the induced density from the SC-QHT approach: The shape of the $n_1$ is very different from all other QHT and TD-DFT results, as SC-QHT uses the self-consistent OF-DFT density as input density, which is very different
from the exact KS density.
More quantitative analysis with the Feibelman $d$ parameter is given in Sec. \ref{sec07}.
\mye
%%%%%%%%%%%%%%%%%%%%%%%%%%%%%%%%%%%%%%%%%%%%%%%%%%%%%%
%%%%%%%%%%%%%%%%%%%%%%%%%%%%%%%%%%%%%%%%%%%%%%%%%%%%%%%

\section{Benchmarking kinetic energy functionals for jellium nanospheres of different dimensions}\label{sec07}
%%%%%%%%%%%%%%%%%%%%%%%%%%%%%%%%%%%%%%%%%%%%%%%%%%%%%%
%%%%%%%%%%%%%%%%%%%%%%%%%%%%%%%%%%%%%%%%%%%%%%%%%%%%%%%
%
%\textbf{TOO MUCH DATA IN THIS SECTION. RESULTS ARE %NOT SOO GOD, SO WE DONT NEED TO DISCUSS TOO MUCH}

\begin{figure}[!hbt]
    \centering
    \resizebox{0.95\columnwidth}{!}{\includegraphics[width=0.3\textwidth,angle=0]{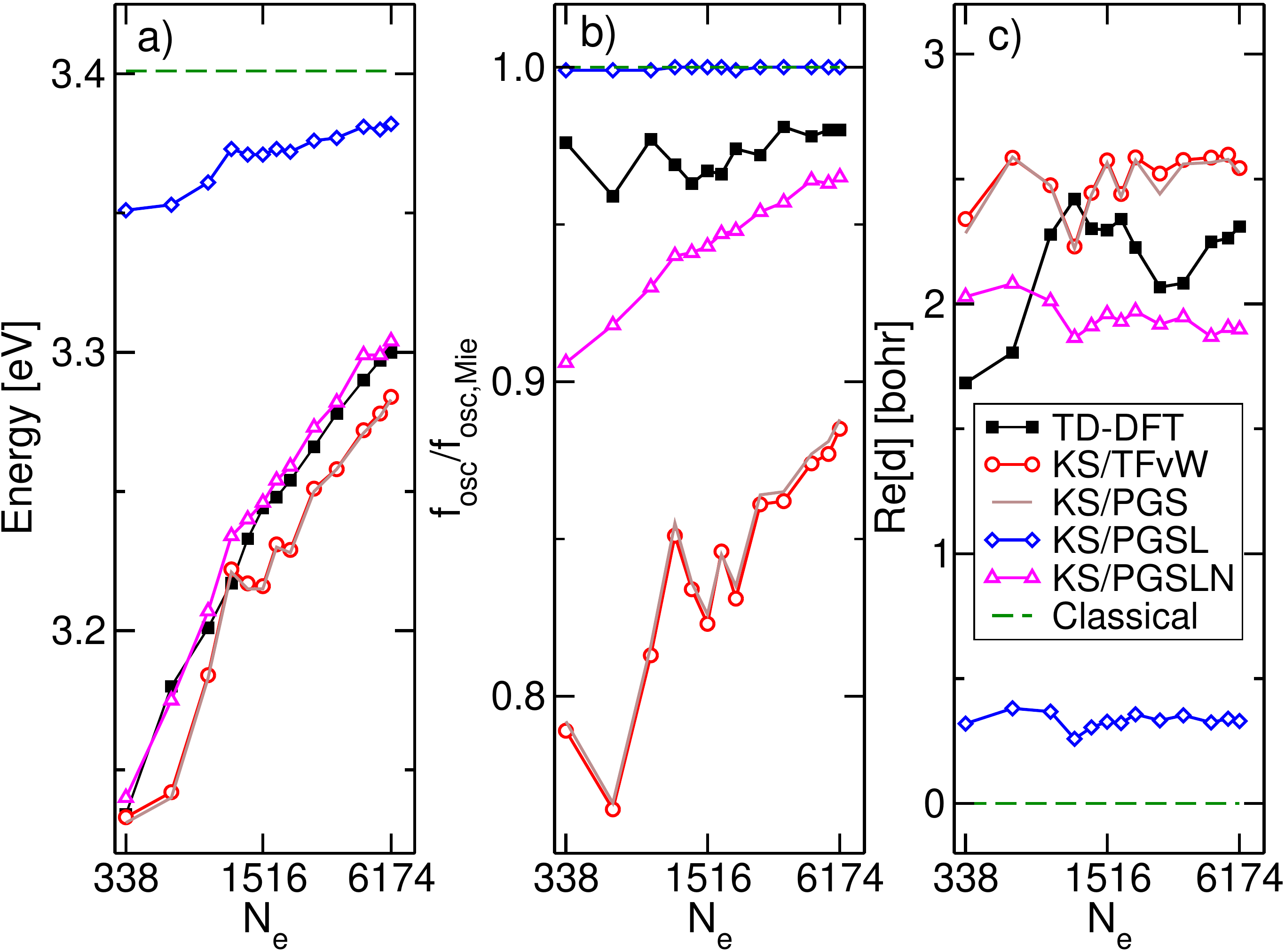}}
    \caption{\small (a) LSP energy,  (b) corresponding oscillator strength ($f_\textrm{osc}$) normalized  to the Mie one, and (c) $d$-parameter (real part) at the LSP energy for Na jellium spheres as a function of the number of electrons ($N_e$) as computed from TD-DFT, KS/TFvW, KS/PGS, KS/PGSL, and KS/PGSLN as well as, the classical results.
    \label{f06}} 
\end{figure}
An important aspect in nanoplasmonic systems is the LSP resonance dependence on the NP size \cite{scholl2012,reiners,li13}.
%As already mentioned in the introduction, LSP for Na jellium nanosphere must undergo red %shift with respect to Mie theory results.
In Fig.~\ref{f06}-(a) (horizontal axis is in logarithmic scale), we show the LSP resonance energy of various Na jellium nanospheres with the number of electrons $N_e$ varying from \my{338} to 6174 (the corresponding radius is $R = r_s N_e^{1/3}$) as computed from TD-DFT, KS/TFvW, KS/PGS, KS/PGSL, and KS/PGSLN. We see that for all approaches, LSP energy is lower than Mie theory value $\hbar\omega_{\textrm{Mie}} = 3.4$~eV (shown as a horizontal line)
\mys and approaching it for large $N_e$. 
The mean average errors (MAEs) with respect to reference TD-DFT are reported
in the first part of Table \ref{tab:all} for both KS and model density.
Note that a detailed comparison of QHT method vs TD-DFT 
can only be achieved using the KS density.
In fact, the model density is not the one used for the reference TD-DFT calculations. 
In any case, for applications involving large systems,  the model density 
is simpler to use; thus, it is relevant to verify (even if approximately) its accuracy.
\mye
%since the calculation of $n_0^{\textrm{KS}}$ via KS-DFT or OF-DFT would require additional computational effort. 
%$8,12,85$~meV,  for  Mod/QHT, Mod/PGS, and Mod/PGSL, respectively.

Figure ~\ref{f06}~(a) and Table \ref{tab:all} show that  the accuracy of KS/TFvW is very high (MAE=$18$ meV), which is somehow surprising, considering the shortcomings of the TFvW functional\mys{discussed in the Introduction.
For Mod/TFvW, the accuracy is even higher (MAE=$6$ meV, close to the numerical accuracy of our implementation).}
The PGS functional, which has some better properties than the 
TFvW functional \cite{fabio01}, yields similar accuracy.
\mys
On the other hand, PGSL overestimates the LSP peak by 80 meV for the model density and 129 meV for the KS density. This seems like quite a large error, but it is not if we consider that the widely used TF-HT has an error of 287 meV on a similar
test set \cite{ciraciQHT}.
The larger error of KS/PGSL with respect to Mod/PGSL can be traced back to the higher oscillating behavior of the KS density inside the NP (see Fig.~\ref{fq01}). Such quantum oscillations induce higher values of the Laplacian (see Fig.~S7 in the Supplemental Material \cite{SM}) and thus higher contributions to the energy and potential.
On the other hand, with the model density, both the gradient and the Laplacian are vanishing small inside the NP.
An ``exact'' KE functional should be able to describe both situations, but this is not the case of the PGSL functional, which has not been optimized for the jellium nanosphere or for the QHT approach.
Instead, a properly constructed functional like PGSLN has even better accuracy than KS/TFvW: The MAE of KS/PGSLN is, in fact, only $6$ meV.
\mye
%, but an overestimation of 80-129 ~meV is quite small, considering that that choice of the exchange-correlation functional can shift the results even more \cite{gisb01,aikens08,jacq13,baseggio16}.  

To describe the accuracy of a given theoretical method for the calculation of the absorption spectra, not only does the energy of the LSP peak has to be considered, but also the oscillator strength $f_{\textrm{osc}}$ associated with it.
The oscillator strength is readily available in an eigenvalue formulation of QHT \cite{banerjee2008,dingeig}. Our QHT implementation is frequency dependent, and, therefore, the oscillator strength is not directly computed, but it can be extracted from the absorption spectra using
the fitting procedure described in Sec.~IV of the Supplemental Material \cite{SM}.
The oscillator strength of the LSP peak can also be extracted from the TD-DFT spectra if the onset of the plasmon shoulder
is considered (Sec.~IV of the Supplemental Material \cite{SM}).
Previous attempts to compute the $f_{\textrm{osc}}$ of the LSP peak are based on the sum-rule approaches \cite{rein96}. 
In Fig.~\ref{f06}~(b), we report $f_{\textrm{osc}}$ of the LSP peak,  as obtained from TD-DFT and the same KE functionals.
Figure~\ref{f06}~(b) shows that for all methods, LSP converges to the classical Mie results for large $N_e$.
However, $f_{\textrm{osc}}$ for  KS/TFvW and KS/PGS is largely underestimated, as the main plasmon peak is subdivided into different peaks, as previously discussed. 
On the other hand, the main peak of KS/PGSL contains almost all
\mys the oscillator strength, as in the classical calculations.
The mean average error with respect to reference TD-DFT is reported
in the second part of Table \ref{tab:all}.
%The mean absolute relative error for these systems are 6.7\%, 6.4\%, and 2.9\% for 
%Mod/QHT, Mod/PGS, and Mod/PGSL. 
QHT-TFvW and QHT-PGS are thus quite inaccurate for the oscillator strength, whereas PGSL has an error of less than 3\%. In all cases, better accuracy is obtained using the model density.
Thus, while QHT-TFvW (QHT-PGS) and QHT-PGSL give either very good LSP energy or very good LSP oscillator strength, QHT-PGSLN is the only functional which gives very good accuracy for both properties.
\mye
%\myy{We also tested the fourth order gradient expansion\cite{ge4}, which is a %non-empirical Laplacian-level functional. Results are quite bad both for the %energy (MAE=116 meV) eV and for $f_osc$ (MARE 9\%)
\begin{table}
\begin{tabularx}{\columnwidth} { 
  | >{\raggedright\arraybackslash}X 
  | >{\raggedleft\arraybackslash}X 
  | >{\raggedleft\arraybackslash}X 
  |  >{\raggedleft\arraybackslash}X 
  | >{\raggedleft\arraybackslash}X |
  }
  
%    \begin{tabular*}{\columnwidth}{rrrrr}
\hline
 Density  &  TFvW       &  PGS  &     PGSL            &     PGSLN   \\
    \hline
\multicolumn{5}{|c|}{\my{LSP, MAE (meV)}}\\
\hline
\my{KS}  &   \my{18}        & \my{19} &    \my{\underline{129}}            &  \my{{\bf 6}}      \\ 
\my{Mod} &   \my{{\bf 6}}   & \my{12}  &    \my{\underline{80}}            &   \my{14}          \\
\hline
  \multicolumn{5}{|c|}{\my{$f_{osc}$, MARE \%}}\\
  \hline
\my{KS}  &  \my{\underline{13.7}}  &   \my{\underline{13.4}}    & \my{{\bf 2.8}}             &    \my{{\bf 2.9}}     \\
\my{Mod} &   \my{\underline{6.3}}  &   \my{\underline{6.1}}    & \my{2.4}              &   \my{{\bf 0.8}}    \\
\hline
  \multicolumn{5}{|c|}{\my{Re[d], MAE (bohr)}}\\
  \hline
\my{KS}  & \my{0.35}   &     \my{{\bf 0.33}}  &    \my{\underline{1.84}}         &     \my{{\bf 0.32}}\\
\my{Mod} & \my{0.19}   &      \my{{\bf 0.15}} &   \my{\underline{1.26}}          & \my{0.41} \\ 
  \hline
    \end{tabularx}
    \caption{\my{Performances of the QHT approach using different kinetic energy functionals (TFvW, PGS, PGSL, and PGSLN) with different input density (KS, model): the first  block reports
    the mean absolute error (MAE) in meV for the energy position of the main LSP peak;  the  second block reports the mean absolute relative error in percent (MARE\%) for the oscillator strength of the main LSP peak; the last block reports the MAE (in a.u.) for the Feibelman $d$ parameter (real part) as computed from the induced density at the LSP energy}. \myy{Best results (or close to them) are in bold; worst results (or close to them) are underlined. }}
    \label{tab:all}
\end{table}

\mys
Finally, we consider the Feibelman $d$ parameter \cite{feibelmann82}, i.e.,
\begin{equation}
    d=\frac{\int 4\pi r^2 (r-R) n_1 dr }{\int 4\pi r^2 n_1 dr}
    \label{eq:fei}
\end{equation}
where $R$ is the radius of the jellium nanosphere, and $n_1$ is the radial part of the induced density.
Equation~\eqref{eq:fei} is valid for a spherical density, and the real part of $d$ describes  the position of the center of mass of $n_1$ with respect $R$. 
The results are reported in Fig.~\ref{f06}-(c) and in the last section of Table \ref{tab:all}.
 While for PGSL the $d$ parameter is underestimated (i.e., the induced density
 is more confined inside the nanoparticles), TFvW, PGS, and PGSLN give
 quite accurate results, as also shown in Fig.~\ref{fig:ind}.
 
\myy{The PGSLN is thus simultaneously very accurate for the LSP energy position, Bennett energy position, LSP oscillator strength and Feibelman $d$ parameter, for 
 all the systems considered. This is quite large test set of properties and systems, showing the reliability of the $q_0$ parameter and of the PGSLN functional form.}
%the
%If KS density is used (see Fig. \ref{f06} (d)), the LSP energy follows a much smoother trajectory. Mod/QHT reproduces TD-DFT results with the maximum error of only 10 meV for $N_e \geq 338$ \cite{ciraciQHT}. Mod/QHT-PGS gives slightly smaller values but its difference from Mod/QHT is $\approx$ 65 meV for $N_e = 8$ and get as small as 10 meV for $N_e = 4570$.

%Results using the KS-density (not reported) are similar, but in this case the %KS/QHT-PGSL overestimates the TD-DFT even more, as also shown in Fig.~\ref{f03}.
%{Mod/QHT-PGSL, on the other hand, is again blue-shifted with respect to %Mod/QHT and Mod/QHT-PGS, but gives results much closer to TD-DFT. 

%Thus, results in Fig. \ref{f06} (d) can be consider very promising for %further development of Laplacian-level kinetic-energy functionals.
%In addition, we can state that results with $n_0^{\textrm{Mod}}$ density perform better than $n_0^{\textrm{KS}}$ density. This is a good finding considering that for 

%Finally, when considering the variation of $\alpha$ and $\beta$ in Figs. \ref{f06} (e) and (f), we observe the same shifts of LSP energy as already discussed in Figs. \ref{f06} (b) and (c). 
\mye

%%%%%%%%%%%%%%%%%%%%%%%%%%%%%%%%%%%%%%%%%%%%%%%%%%%%%%%%%%%%%%55
\section{Application to spherical dimer}\label{sec08}
%%%%%%%%%%%%%%%%%%%%%%%%%%%%%%%%%%%%%%%%55
\begin{figure*}[!hbt]
    \centering
    \includegraphics[width=0.9\textwidth,angle=0]{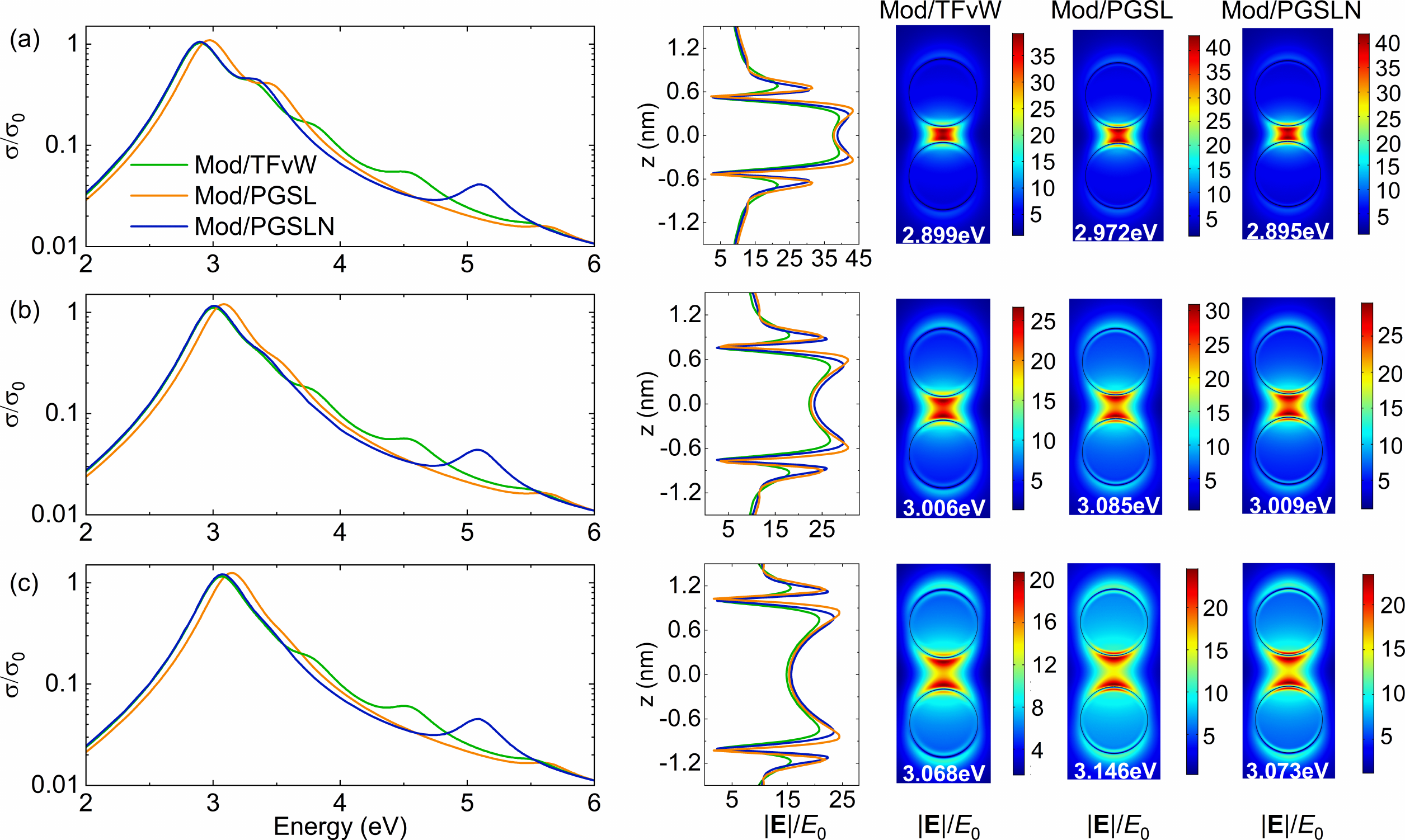}
  \caption{Normalized absorption cross section $\sigma/\sigma_0$ (in the logarithmic scale) and norm of the total field $\left|\textbf{E}\right|/E_0$ for dimers of Na jellium spheres with $N_e = 1074$ electrons as obtained from the Mod/TFvW, Mod/PGS and Mod/PGSL. From top to bottom: panel (a) refers to gap = 1\,nm, (b) to gap = 1.5\,nm, and (c) refers to gap = 2\,nm.}
  \label{f07} 
\end{figure*}
Our FEM implementation allows us to calculate absorption spectra for axisymmetric structures. An important example of such a system is a nanosphere dimer. The NP dimer has been widely studied in the literature since it supports gap plasmons that can squeeze light down to subnanometer volumes, making it an ideal system for exploring the quantum and nonlocal phenomena \cite{perez-gonzalez2010,barbry2015,ciraci2017visc,jeong2019,schumacher2019}. Here we consider a dimer of Na jellium spheres with 1074 electrons each. In Fig.~\ref{f07}-(a), we present a comparison of the absorption cross section as calculated from the Mod/\my{TFvW}, Mod/\my{PGSL}, and Mod/\my{PGSLN} (the cross section is normalized to the $2\sigma_0 = \pi R^2$ with $R$ being the radius of a single sphere). The plane wave that excites the structure is polarized along the $z$ axis, and the input ground-state density is the sum of model densities \my{(see Eq. \ref{fq10})} of two spheres.
\my{As we can see, the Mod/\my{TFvW} give oscillations in the spectrum (at $\approx 3.9$~eV and $\approx 4.6$~eV, which are absent in Mod/PGSL, and Mod/PGSLN approaches.} Our convergence analysis shows that these oscillations, as in the case of the sphere (see Fig.~\ref{f03}), persist \my{and depend on}  the computational domain. These oscillations should not be confused with the small undulation next to the main plasmon peak that is more clearly visible in the gap = 1~nm case [Fig.~\ref{f07}-(a)]. This undulation comes from higher-order plasmon resonances and gets higher for smaller sizes of the gap \cite{barbry2015}.
\begin{table}[!hbt]
\begin{tabular}{|c|l|c|}
\hline
Method                        & \multicolumn{1}{c|}{gap} & $\left|\textbf{E}\right|/E_0$ \\ \hline
\multirow{3}{*}{Mod/\my{TFvW}}      & 1.0\,nm             & 33.9     \\ \cline{2-3} 
                              & 1.5\,nm                   & 22.5     \\ \cline{2-3} 
                              & 2.0\,nm                   & 16.5     \\ \hline
\multirow{3}{*}{Mod/\my{PGS}}  & 1.0\,nm                   & 35.4     \\ \cline{2-3} 
                              & 1.5\,nm                   & 23.2     \\ \cline{2-3} 
                              & 2.0\,nm                   & 16.9     \\ \hline
\multirow{3}{*}{Mod/\my{PGSL}} & 1.0\,nm                   & 38.6     \\ \cline{2-3} 
                              & 1.5\,nm                   & 25.4     \\ \cline{2-3} 
                              & 2.0\,nm                   & 18.7     \\ \hline
\multirow{3}{*}{\my{Mod/PGSLN}} & 1.0\,nm                   & \my{36.8}     \\ \cline{2-3} 
                              & 1.5\,nm                   & \my{24.6}     \\ \cline{2-3} 
                              & 2.0\,nm                   & \my{18.2}     \\ \hline
\end{tabular}
\caption{The average value of $\left|\textbf{E}\right|/E_0$ in the dimer gap as calculated from \my{QHT with different KE functionals (TFvW, PGS, PGSL and PGSLN) and the model density}.}
\label{tab}
\end{table}
For all considered cases of gap size, Mod/PGSL gives blueshifted plasmon resonance energy with respect to other methods. \my{On the other hand, Mod/TFvW and Mod/PGSLN match very well at the plasmon resonance, with the maximum difference of 0.005~eV, but, as stated before, Mod/PGSLN does not show the oscillations. Also, the Bennett peak is observed in Mod/PGSL ($\approx$ 5.7 eV), and Mod/PGSLN ($\approx$ 5.1 eV)  approaches that is stable to change in the computational domain size.} The respective values of the plasmon resonance are shown in the map plots of the total field enhancement in Fig.~\ref{f07}. \my{There we also see that the field gets more enhanced for Mod/QHT-PGSL as for other approaches, that is more clearly observed in the cut lines of the field distribution around the z-axis. This behavior is expected since Mod/PGSL does not result in additional peaks of the absorption spectra as opposed to TFvW and now more energy is moved to the main plasmon peak. For Mod/PGSLN, the Bennett peak is more pronounced as opposed to Mod/PGSL, and, consequently, we have less enhancement PGSLN calculations at the main plasmon peak. Also, as Table~\ref{tab} shows, more field is concentrated in the gap for Mod/QHT-PGSL.} 

%%%%%%%%%%%%%%%%%%%%%%%%%%%%%%%%%%%%%55
\section{Conclusions and Future Perspectives}\label{sec09}
%%%%%%%%%%%%%%%%%%%%%%%%%%%%%%%%%%%%%555

We extended the quantum hydrodynamic theory to Laplacian-level kinetic energy functionals.
In particular, we \my{started our investigation considering} the PGSL functional, which is shown to be accurate for OF-DFT calculations \my{of metals and semiconductors \cite{fabio01}.}
%and well reproduced the linear response of the homogeneous electron gas %\cite{fabio01}. 
We \my{analyze} in detail Na jellium nanospheres, and the results are compared to \my{gradient-level kinetic functional and} reference TD-DFT calculations. 
The key results obtained are focused on two main findings:
\begin{enumerate}
  \item QHT-TFvW and QHT-PGS that are based on gradient-level KE functionals of electron density, together with an LSP resonance, give additional resonances in the absorption spectrum of Na jellium nanospheres. 
  \my{These resonances have usually an energy higher than the critical frequency, and thus they are very
  sensitive to the computational domain-size}.
  Well-defined additional resonances are not present in TD-DFT or in QHT-TFvW with an {\it infinite} computational domain size. In both cases, only a shoulder is present at the high-energy side to the plasmon peak, with the TD-DFT result being much smaller and at higher energy than in QHT-TFvW.  
  On the other hand, QHT-PGSL yields only the LSP peak in the absorption spectrum, with an overall spectrum and oscillator strength closer to TD-DFT. %This result indicates the necessity to consider second order terms in the Taylor expansion of density functionals. 
  \item The theoretical and numerical asymptotic analysis of the induced charge density as obtained from QHT-TFvW and QHT-PGS shows that the decay slope is changing at different energies of incident radiation. Contrarily, QHT-PGSL shows the same decay slope for all energies, \my{and thus, no critical frequency exists anymore}. This result strongly simplifies the boundary conditions so that converged a calculation can be obtained with a very small computational domain-size.
%  \item We also extend our theory to nanosphere dimers and demonstrate that QHT-PGSL correction results in the absorption spectra that is free of oscillations that are present in QHT and QHT-PGS approaches.
\end{enumerate}
Our results thus demonstrate that the convergence of the QHT absorption spectra is problematic, and most of the QHT-TFvW results reported so far are thus not accurate enough for energies above LSP resonance. 
The QHT-PGSL, on the other hand, does not suffer from these problems.
%and can be successfully applied to investigate quantum and spill-out effects in different systems %as shown in Sec.~\ref{sec09}, where we report the results for  NP dimers.

\mys
The PGSL functional, which is characterized by a term proportional to $q^2$,
solves some fundamental limitation of the QHT-TFvW approaches: (i) the presence of the critical frequency, (ii) the sensitivity to input density and the computational domain size, and (iii) the underestimation of the oscillator strength for the LSP peak.
On the other hand, the PGSL results are not very accurate when the LSP energy position and LSP $d$ parameter are considered. Moreover, PGSL predicts a Bennett state too close to the volume plasmon.

We find that all these shortcomings can be removed if the  $q^2$ term is kept only outside the nanoparticles. We thus develop a new functional, PGSLN, which combines only the good features of QHT-TFvW (or QHT-PGS) and QHT-PGSL.
Thus, QHT-PGSLN is very accurate for all properties that are of interest in plasmonics, allowing an efficient and numerically converged computation of collective excitations in quantum systems.
%retains  offers several and distinctive advantage with respect QHT-TFvW, namely: i) 
%, namely
%\begin{itemize}
%\item[-] results doens't depend on the computational domain size, which can also be %very small;
%\item[-] it is more accurate for LSP oscillator strength;
%\item[-] results are almost independent from the input density;
%\end{itemize}
\mye

%From a computational point-of-view, QHT-PSGL is close to the widely used TF-HT with %hard-wall: in fact, the size of the computational domain can be set very close to the %size of the nanoparticles, as discussed in Sec. \ref{sec06}. However, QHT-PGSL solves %the two main drawbacks of the TF-HF approach: the neglect of spill-out effects and the %blueshift of LSP energies.

%QHT-PGSL has thus 

%: the main drawbacks of QHT-PGSL are related to the too fast decay of the induced %density, which then causes reduced spill-out effects and  increased plasmon energy. 
%However, we point out that the QHT-PGSL functional, used in this work, has not been %developed for QHT, but for bulk properties of bulk metal and semiconductors.
%We believe the Laplacian-level QHT results can be improved considered more sophisticated %functional that could depend, for example, on a different function of $q$ and/or on a %product of $w$ and $q$. 
\myy{Clearly, the PGSLN functional is a very simple functional introduced here to show the power of the Laplacian-level QHT, but further tests and developments will be required to verify and extend its applicability.} The Laplacian-level QHT is thus a new platform, very promising for the future, as the Laplacian ingredient includes many more degrees of freedom in developing
accurate KE functionals than a more conventional functional based on density gradient.
So far, however, the development of a semilocal KE functional focused only on ground-state properties, considering only the total KE and the KE potential (i.e.,
the first functional derivatives).
Instead, for the QHT response properties, the KE kernel (i.e., the second functional derivative) is required, but, so far, it has not been considered at all in the semilocal KE functional development \cite{perdew2007,kara09,cancio16,fabio01,seino18,golub18,const19}.
%Despite QHT and QHT-PGS allow to calculate the plasmon energy with a high accuracy (MAE of about 10 meV) if compared with TD-DFT, they lacks in describing correctly the oscillator strength and the high-energy regions.
%ptimization On the other hand, QHT-PG$\alpha$L$\beta$ gives energies about blue shifted with respect to QHT and QHT-PGS as well as reduced spill-out effects.
%On the other hand QHT-PGSL given better oscillator strength for the main peak.
%This drawback is related to the fact that the KE functional is never %known exactly \cite{hohenberg1964}. 
%We have chosen a Laplacian term dependence by Exp. \eqref{fq05c} which can be %modified by considering

In addition, it is crucial to understand the role of static and dynamic corrections to the energy functional.
Although here we consider only static corrections at the second-order gradient and Laplacian level, the analysis of dynamic correction represents another important route to explore. \my{In particular a frequency dependent damping can be important
to further improve the accuracy of QHT-PGSLN with respect to TD-DFT.} 
Overall, we believe that our current results will help to better understand the role of functional dependence on electron density in plasmonic systems. 

%
% Specify following sections are appendices. Use \appendix* if there
% only one appendix.

% If you have acknowledgments, this puts in the proper section head.
\section*{Author contributions}
H. B. and C. C. conceived the idea of applying the PGSL functional to QHT.
H. B. derived the first-order potential and implemented all the equations
in FEM. H. B. performed calculations for the dimer. H. B. and F. D. S. did the calculations for the nanospheres with different numbers of electrons.
F. D. S. derived the proof for the asymptotic decay and the PGSLN functional.
F. D. S. and C. C. supervised the research. 

All authors contributed to writing the paper. H. M. B. and F. D. S. contributed equally to this work.
%\end{acknowledgments}
\appendix
%%%%%%%%%%%%%%%%%%%%%%%%%%%%%%%%%%%%%%%%%%%%%%%%%%%%%%%%%%%%%%%%%
\section{Absorption spectrum}\label{appabs}
%%%%%%%%%%%%%%%%%%%%%%%%%%%%%%%%%%%%%%%%%%%%%%%%%%%%%%%%%%%%%%%%%5
In QHT, the absorption cross section is calculated as
\begin{equation}
    \sigma\left(\omega\right) = \frac{\omega}{2 I_0}\int\textrm{Im}\left\{\textrm{\textbf{E}}\cdot\textrm{\textbf{P}}^{*}\right\}\textit{d}V,
\label{fq09}
\end{equation}
with $I_0$ being the intensity for the incident plane wave with frequency $\omega$. The electric field ${\bf E}$ and the polarization vector ${\bf P}$
are obtained solving Eqs.~\eqref{fq01a} and \eqref{fq01b}.
Considering the very small size \my{of the}  investigated nanoparticles, only dipole modes are excited (for spherical nanoparticles).
An important parameter for the shape of the absorption spectra is the damping parameter ($\gamma)$; see Eq.~\eqref{fq01b}. If not stated differently, in all QHT calculations, we use $\gamma=66$~meV.

The normalized absorption cross section (absorption efficiency) is then obtained by normalizing $\sigma$ to the geometric cross section of a nanosphere $\sigma_0 = \pi R^2$ with $R$ being the radius of the nanosphere. 

The TD-DFT absorption spectra are computed with a finite-difference in-house code (with spherical symmetry) introduced in Ref. \citenum{ciraciQHT}; a radial uniform grid is used to represent KS orbitals and densities. In TD-DFT, no retardation effects are included, and only longitudinal electric \my{fields} are considered \cite{ullrich}.
The absorption cross section \cite{zangwill1980,ekardt1985,bertsch1990,prodan2002} is calculated as
\begin{equation}
    \sigma \left(\omega\right) = \frac{\omega}{c \epsilon_0}\textrm{Im}\left\{\alpha_{zz}\left(\omega\right)\right\},
\label{fq11}
\end{equation}
where the polarizability is given by
\begin{equation}
    \alpha_{zz}\left(\omega\right) = -e^2\int\textit{d}\textbf{r}\textit{d}\textbf{r}^\prime z\chi\left(\textbf{r},\textbf{r}^\prime ,\omega\right)z^\prime ,
\label{fq12}
\end{equation}
with $\chi\left(\textbf{r},\textbf{r}^\prime ,\omega\right) = \delta n\left(\textbf{r}\right)/\delta\left(eV_{\textrm{ext}}\left(\textbf{r}^\prime \right)\right)$ being the interacting density-density response function \cite{ullrich},  which is obtained solving the Dyson equation
\begin{equation}
\chi = \chi_0 +\chi_0 (v_{coul}+f_{\textrm{XC}}^{\textrm{LDA}}) \chi \, .
\label{eq:chi}
\end{equation}

In Eq.~\eqref{eq:chi}, $v_{\textrm{Coul}}$, is the Coulomb interaction, $f_{\textrm{XC}}^{\textrm{LDA}}$ is the adiabatic LDA XC kernel, and $\chi_0$ is the noninteracting density-density response function, which is computed using the Green's function \cite{ekardt1985} using occupied KS orbitals from the ground-state calculation (again using LDA).
The broadening parameter for the Green's function calculations is, if not stated differently, $\Gamma_0=33$~meV. 

%We note also that all QHT and TD-DFT absorption spectra reported in this %work are dipole-spectradipole %spectra in the quasi-static approximation, %thus no higher multipoles peaks %are present, despite they also occurs in %the same energy range of the %Bennett states\cite{cris17} 

%In all TDDFT calculation we use the LDA functional.

%%%%%%%%%%%%%%%%%%%%%%%%%%%%%%%%%%%%%%%%%%%%%%%%
\section{FEM implementation}\label{appfem}
%%%%%%%%%%%%%%%%%%%%%%%%%%%%%%%%%%%%%%%%%%%%
In order to lower the order of derivatives, we multiply Eq.~\eqref{fq01b} by test function $\textrm{\textbf{\~{P}}}$ and integrate by parts, which give us
\begin{widetext}
\begin{equation}
\int\left\{-\frac{e}{m_e}\left(\frac{\delta G\left[n\right]}{\delta n}\right)_1 \left(\nabla \cdot \textrm{\textbf{\~{P}}} \right) + \frac{1}{n_0}\left[\left(\omega^2 + i\gamma\omega\right)\textrm{\textbf{P}} + \epsilon_0\omega_p^2\textrm{\textbf{E}}\right]\cdot \textrm{\textbf{\~{P}}}\right\}dV = 0,
\label{aeq01}
\end{equation}
\end{widetext}
\noindent where we assume that the integral on the boundary goes to zero. Even after integration by parts, the $\left(\frac{\delta G\left[n\right]}{\delta n}\right)_1$ potential contains derivatives up to the fourth order of $n_1$ [see the Exps.~\eqref{fq08}], so auxiliary variables should be added to lower the order of differentiation. By introducing two variables $\textrm{\textbf{F}} = \nabla n_1$ and $\textrm{\textbf{O}} = \nabla\left(\nabla^2 n_1\right) = \nabla^2\textrm{\textbf{F}}$, we have only first-order derivatives. Considering axisymmetry of considered structures, we adopt \textrm{2.5D technique} \cite{Ciraci:2013jt,ciraci02,ciraciQHT}, and the dependence of $\textrm{\textbf{E}}, \textrm{\textbf{P}}, \textrm{\textbf{F}}$, and $\textrm{\textbf{O}}$ on the azimuthal coordinate is taken in $e^{-im\phi}$ form with $m \in \mathbb{Z}$. The dependence on $m$ for test functions $\textrm{\textbf{\~{E}}}, \textrm{\textbf{\~{P}}}, \textrm{\textbf{\~{F}}}$, and $\textrm{\textbf{\~{O}}}$ is of $e^{im\phi}$ form. Thus, instead of a three-dimensional problem, we can have $2m_{\textrm{max}} + 1$ problems (with $m_{\textrm{max}}$ being the maximum value for $m$). Moreover, for the dimensions considered in the current work $m_{\textrm{max}} = 0$ is enough for the convergence of results. Finally, only one two-dimensional problem needs to be solved. Hence, we come to the following system of equations:
\begin{widetext}
\begin{subequations}
    \begin{gather}
        2\pi \int\left\{\left(\nabla\times\textrm{\textbf{E}}^{\left(0\right)}\right)\cdot\left(\nabla\times\mathrm{\mathbf{\tilde{E}}}^{\left(0\right)}\right)-\left(k_0^2\textrm{\textbf{E}}^{(0)} + \mu_0\omega^2\textrm{\textbf{P}}^{(0)}\right)\cdot\mathrm{\mathbf{\tilde{E}}}^{\left(0\right)}\right\}\rho d\rho dz = 0, \label{aeq02a} \\
        2\pi \int\left\{-\frac{e}{m_e}\left(\frac{\delta G\left[n\right]}{\delta n} \right)_1^{\left(0\right)}\left(\nabla \cdot \mathrm{\mathbf{\tilde{P}}}^{\left(0\right)} \right) + \frac{1}{n_0}\left[\left(\omega^2 + i\gamma\omega\right)\textrm{\textbf{P}}^{(0)} + \epsilon_0\omega_p^2\left(\textrm{\textbf{E}}^{(0)} + \textrm{\textbf{E}}^{(0)}_\textrm{inc}\right)\right]\cdot \mathrm{\mathbf{\tilde{P}}}^{\left(0\right)}\right\}\rho d\rho dz = 0, \label{aeq02b} \\
        2\pi \int\left\{\left(\nabla\cdot\textrm{\textbf{P}}^{\left(0\right)}\right)\left(\nabla\cdot\mathrm{\mathbf{\tilde{F}}}^{\left(0\right)}\right) + e\textrm{\textbf{F}}^{\left(0\right)}\cdot\mathrm{\mathbf{\tilde{F}}}^{\left(0\right)}\right\}\rho d\rho dz = 0, \label{aeq02c}\\
        2\pi \int\left\{\left(\nabla\cdot\textrm{\textbf{F}}^{\left(0\right)}\right)\left(\nabla\cdot\mathrm{\mathbf{\tilde{O}}}^{\left(0\right)}\right) + \textrm{\textbf{O}}^{\left(0\right)}\cdot\mathrm{\mathbf{\tilde{O}}}^{\left(0\right)}\right\}\rho d\rho dz = 0, \label{aeq02d}
    \end{gather}
    \label{aeq02}
\end{subequations}
\end{widetext}
\noindent where the $\left(0\right)$ superscript denotes the zero-order coefficients of the $\textrm{\textbf{v}}\left(\rho,\phi,z\right) = \sum_{m \in \mathbb{Z}} \mathrm{\mathbf{v}^{\left(\mathit{m}\right)}}\left(\rho,z\right)e^{-im\phi}$ vector field expansion of cylindrical harmonics. We find that \textit{curl} elements for Eq.~\eqref{aeq02a} and \textit{divergence} elements \cite{comsol} for other equations of the system \eqref{aeq02} are the best choices for stable solutions. 

For the wave equation \eqref{aeq02a}, simulation domain radius $R_{\textrm{dom}}$ is defined via the $R_{\textrm{dom}} = R + 500a_0$ condition. $R = r_sN_e^{1/3}$ is the radius of the nanosphere, and, for dimers, it is the radius of one of the spheres. Perfectly matched layers (PMLs) are used in order to emulate an infinite domain and avoid unwanted reflections. The PML thickness is set to $t_{\rm PML} = 200a_0$ for all the considered systems. Also, a zero flux boundary condition is imposed on the electric field at the outer boundary of the PML.
For Eqs.~\eqref{aeq02b}~-~\eqref{aeq02d}, simulations are done in a smaller domain, considering faster decay of variables $\textrm{\textbf{P}}, \textrm{\textbf{F}}$, and $\textrm{\textbf{O}}$ compared to the electric field. The domain, as depicted in Fig.~S1-(a) of Supplemental Material \cite{SM}, is a semicircle (consider the axial symmetry) for the nanospheres, and for the dimers it is the union of two circles centered at the centers of the nanospheres. Moreover, to facilitate the calculations, only the ``quarter'' of the dimer is simulated with a corresponding perfect electric conductor condition at the intersection segment of two circles, as shown in Fig.~S1-(b) of Supplemental Material \cite{SM}. The radius for the circles is $r_{\textrm{dom}} = R + 25a_0$ for QHT and QHT-PGS, but for QHT-PGSL, it is in the range $r_{\textrm{dom}} \approx R + 12a_0$. The simulation domain is smaller for QHT-PGSL because $\nu \approx 1.12$ decay slope is bigger in this case (see Sec.~\ref{sec06}). Dirichlet boundary conditions $\textrm{\textbf{P}} = 0, \textrm{\textbf{F}} = 0$, and $\textrm{\textbf{O}} = 0$ are set on the simulation domain boundary.
%\textcolor{red}{Fabio and Cristian, please pay attention that the indicated domain size is not fixed for PGSL. I had to use different %sizes for different particles. But, in principle, the same domain size can be set, if the we play with the density of the mesh.}
%
As we state in Sec. \ref{sec06}, mixed boundary condition with a fixed exponential decay can be used for QHT-PGSL so that a very small simulation domain is enough for converged calculations.

\bibliography{bib_pgsl_qht}

\end{document}

\title{Supplementary Material for\\ ``Laplacian-Level Quantum Hydrodynamic Theory for Plasmonics''}

\author{Henrikh M. Baghramyan$^{1}$}
\author{Fabio Della Sala$^{1,2}$}
\author{Cristian Cirac\`{i}$^{1}$}
 \email{cristian.ciraci@iit.it}

\affiliation{$^{1}$Center for Biomolecular Nanotechnologies, Istituto Italiano di Tecnologia, Via Barsanti 14, 73010 Arnesano (LE), Italy}
\affiliation{$^{2}$Institute for Microelectronics and Microsystems (CNR-IMM), Via Monteroni, Campus Unisalento, 73100 Lecce, Italy}

\date{\today}

\maketitle
\def\mys{\color{black}}
\def\mye{\color{black}}
\newcommand{\my}[1]{\textcolor{black}{#1}}
\renewcommand{\thepage}{S\arabic{page}} 
\renewcommand{\thesection}{S\arabic{section}}  
\renewcommand{\thetable}{S\arabic{table}}  
\renewcommand{\thefigure}{S\arabic{figure}}
\renewcommand{\theequation}{S\arabic{equation}}

%\begin{figure}[!htb]
%    \centering
%    \includegraphics[width=0.4\textwidth,angle=0]{suppl_04_mod%_imag}
%  \caption{\small Na jellium sphere with $N_e = 1074$ %electrons. Imaginary part of induced charge density $n_1$ and %norm of the total field $\textrm{\textbf{E}}$ normalized to %incident field amplitude $E_0$ at LSP resonance as calculated %from Mod/QHT, Mod/QHT-PGS, and Mod/QHT-PGSL.}
%  \label{sup04} 
%\end{figure}

%\clearpage
%\newpage

\begin{figure}[!hbt]
    \centering
    \resizebox{0.4\columnwidth}{!}{\includegraphics[width=0.3\textwidth,angle=0]{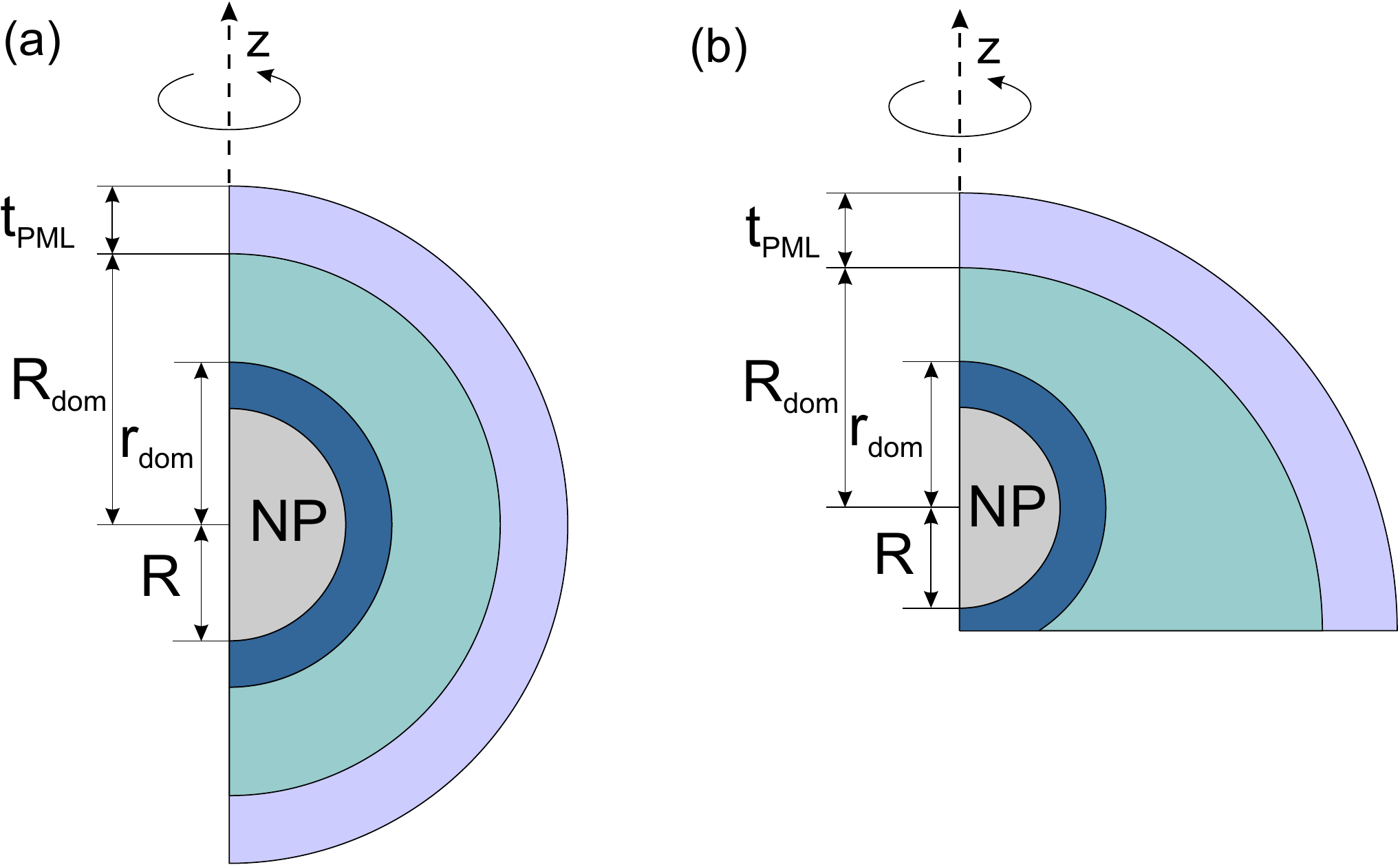}}
  \caption{\small Schematic image of the simulation domain for a spherical nanoparticle (NP) - (a) and NP dimer - (b). $\textrm{R}$ denotes the radius of the NP, $\textrm{r}_{\textrm{dom}}$ - the radius of the simulation domain for $\textrm{\textbf{P}}, \textrm{\textbf{F}}$, and $\textrm{\textbf{O}}$ variables, $\textrm{R}_{\textrm{dom}}$ - the simulation domain for $\textrm{\textbf{E}}$, and ${\rm t}_{\rm PML}$ is the thickness of the perfectly matched layer (PML).}
  \label{fnum_sch} 
\end{figure}

\section{Functional derivative}

Here we derive the functional derivative of $T_s\left[n\right]$,  i.e., Eq.~(10) of the main manuscript. 
Following Ref. \citenum{fabio_01}, we have

\begin{equation}\frac{\delta T_s}{\delta n}=\frac{\partial \tau}{\partial n}-\nabla \cdot \frac{\partial \tau}{\partial(\nabla n)}+\nabla^{2} \frac{\partial \tau}{\partial\left(\nabla^{2} n\right)},
\label{eq_func_01}
\end{equation}

where $\frac{\partial \tau}{\partial i} \equiv \tau_{i}(i=n, w, q)$. In Exp.~\eqref{eq_func_01}

\begin{equation}
\begin{aligned}
\nabla \cdot \frac{\partial \tau}{\partial(\nabla n)}&=\nabla \cdot \left(\tau_{w} \frac{\partial w}{\partial(\nabla n)}\right)=2 \nabla \cdot\left(\tau_{w} \nabla n\right)=2\left(\tau_{w} \nabla^{2} n+\nabla \tau_{w} \cdot \nabla n\right) \\
&=2\left(\tau_{w} \nabla^{2} n+\left(\tau_{w n} \nabla n+\tau_{w w} \nabla w+\tau_{w q} \nabla q\right) \cdot \nabla n\right) \\
&=2\left(\tau_{w} \nabla^{2} n+\tau_{w n} \nabla n \cdot \nabla n+\tau_{w w} \left(\nabla n \cdot \nabla w\right)+\tau_{w q} \left( \nabla n \cdot \nabla q \right) \right),
\end{aligned}
\label{eq_func_02}
\end{equation}

\begin{equation}
\begin{aligned}
\nabla^{2} \frac{\partial \tau}{\partial\left(\nabla^{2} n\right)}&=\nabla \cdot \left(\nabla \frac{\partial \tau}{\partial\left(\nabla^{2} n\right)}\right)=\nabla \cdot\left(\nabla \tau_{q}\right)=\nabla \cdot \left(\tau_{q n} \nabla n+\tau_{q w} \nabla w+\tau_{q q} \nabla q\right) \\
&=\nabla \cdot\left(\tau_{q n} \nabla n\right)+\nabla \cdot\left(\tau_{q w} \nabla w\right)+\nabla \cdot\left(\tau_{q q} \nabla q\right) \\
&=\tau_{q n} \nabla^{2} n+\nabla n \cdot \left(\tau_{q n n} \nabla n+\tau_{q n w} \nabla w+\tau_{q n q} \nabla q\right)+\tau_{q w} \nabla^{2} w \\
&+\nabla w \cdot \left(\tau_{q w n} \nabla n+\tau_{q w w} \nabla w+\tau_{q w q} \nabla q\right)+\tau_{q q} \nabla^{2} q+\nabla q \cdot \left(\tau_{q q n} \nabla n+\tau_{q q w} \nabla w+\tau_{q q q} \nabla q\right) \\
&=\tau_{q n} \nabla^{2} n+\tau_{q n n} w+\tau_{q n w} \left(\nabla n \cdot \nabla w\right)+\tau_{q n q} \left(\nabla n \cdot \nabla q\right)+\tau_{q w} \nabla^{2} w+\tau_{q w n} \left(\nabla n \cdot \nabla w\right) \\
&+\tau_{q w w}|\nabla w|^{2}+\tau_{q w q} \left(\nabla n \cdot \nabla w\right)+\tau_{q q} \nabla^{2} q+\tau_{q q n} \left(\nabla n \cdot \nabla q\right)+\tau_{q q w} \left(\nabla w \cdot \nabla q\right)+\tau_{q q q}|\nabla q|^{2}.
\end{aligned}
\label{eq_func_03}
\end{equation}

Thus, Exp.~\eqref{eq_func_01} simplifies to

\mys

\begin{equation}
\begin{aligned}
\frac{\delta T_s}{\delta n}&=\tau_{n}-2\left(\tau_{w} \nabla^{2} n+\tau_{w n} w+\tau_{w w} \left(\nabla n \cdot \nabla w\right)+\tau_{w q} \left(\nabla n \cdot \nabla q\right)\right) \\
&+\tau_{q n} q+\tau_{q n n} w+\tau_{q n w} \left(\nabla n \cdot \nabla w\right)+\tau_{q n q} \left(\nabla n \cdot \nabla q\right)+\tau_{q w} \nabla^{2} w\\
&+\tau_{q w n} \left(\nabla n \cdot \nabla w\right)+\tau_{q w w}|\nabla w|^{2}+\tau_{q w q} \left(\nabla w \cdot \nabla q\right)+\tau_{q q} \nabla^{2} q+\tau_{q q n} \left(\nabla n \cdot \nabla q\right) \\
&+\tau_{q q w} \left(\nabla w \cdot \nabla q\right)+\tau_{q q q}|\nabla q|^{2}\\
&\, = \tau_{n}+w\left(\tau_{n n q}-2 \tau_{n w}\right)+\left(\tau_{q n}-2 \tau_{w}\right) q + 2\left(\tau_{n q q}-\tau_{w q}\right) \left(\nabla n \cdot \nabla q\right) \\
& + 2\left(\tau_{n w q}-\tau_{w w}\right) \left(\nabla n \cdot \nabla w\right) +2\tau_{w q q} \left(\nabla w \cdot \nabla q\right)+\tau_{w w q}|\nabla w|^{2} +\tau_{w q} \nabla^{2} w +\tau_{q q} \nabla^{2} q +\tau_{q q q}|\nabla q|^{2}, \\
\end{aligned}
\label{eq_func_04}
\end{equation}

\noindent that is the Eq.~(10) of the main manuscript.
\mye

% Considering that our choice for $\tau\left(n,w,q\right)$ (Exps.~(5) and (6) of the main manuscript),

% \begin{equation}
%     \tau\left(n,w,q\right) = A n^{-1}w + B n^{5/3}e^{\alpha C n^{-8/3}w} + D \beta n^{-5/3}q^2
% \label{eq_func_05}
% \end{equation}

% \noindent is quadratic with respect to $q$, and mixed derivatives of it that involve both $q$ and $w$ are zero, the final expression of $\frac{\delta T_s}{\delta n}$ is (as Exp. (7) of the main manuscript):

% \begin{equation}
% \frac{\delta T_s}{\delta n} = \tau_n + w\tau_{nnq} - 2w\tau_{nw} + \tau_{nq}q - 2q\tau_w 
% + 2\tau_{nqq} \left(\nabla n \cdot \nabla q\right) - 2\tau_{ww}\left(\nabla n \cdot \nabla w\right)  + \tau_{qq}\nabla^2q.
% \label{eq_func_06}
% \end{equation}

\section{Linearization}

\my{Here we derive the Exps.~(12) of the main manuscript, that is for a general kinetic energy density $\tau(n,w,q)$, but without products of $w$ and $q$}. For the linearization, $\frac{\delta T_s}{\delta n}$ is expressed in Taylor series around $n_0$ ground state density (unperturbed), such that $n=n_0+n_1$:

\begin{equation}
\begin{aligned}
\left(\frac{\delta T}{\delta n}\right)_{1}&=\left(\tau_{n}\right)_{1}+\left(w \tau_{n n q}\right)_{1}-2\left(w \tau_{n w}\right)_{1}+\left(q \tau_{n q}\right)_{1}-\left(q \tau_{w}\right)_{1} \\
&+2\left(\tau_{n q q} \nabla n \cdot \nabla q\right)_{1}-2\left(\tau_{w w} \nabla n \cdot \nabla w\right)_{1}+\left(\tau_{q q} \nabla^{2} q\right)_{1}+\left(\tau_{q q q}|\nabla q|^{2}\right)_{1}.
\label{eq_func_07}
\end{aligned}
\end{equation}

Keeping only terms linear with respect to $n_1$, we have for $\left(\tau_{n}\right)_{1}$:

\begin{equation}
\begin{aligned}
\left(\tau_{n}\right)_{1}&=\tau_{n n}^{(0)}\left(n-n_{0}\right)+\tau_{n w}^{(0)}\left(w-w_{0}\right)+\tau_{n q}^{(0)}\left(q-q_{0}\right)= \\
&=\tau_{n n}^{(0)} n_{1}+\tau_{n w}^{(0)}\left[\nabla\left(n_{0}+n_{1}\right) \cdot \nabla\left(n_{0}+n_{1}\right)-\left|\nabla n_{0}\right|^{2}\right]+\tau_{n q}^{(0)}\left[\nabla^{2}\left(n_{0}+n_{1}\right)-\nabla^{2} n_{0}\right] \\
&=n_{1}\tau_{n n}^{(0)}+2\tau_{n w}^{(0)} \left(\nabla n_{0} \cdot \nabla n_{1}\right)+\tau_{n q}^{(0)} \nabla^{2} n_{1},
\end{aligned}
\label{eq_func_08}
\end{equation}

\noindent where the superscript (0) denotes that the function is evaluated at $n=n_0$. Similarly, for other terms in Exp.~\eqref{eq_func_07} (and canceling out partial derivatives that are zero) we get:

\begin{equation}
\begin{aligned}
\left(w \tau_{n n q}\right)_{1}&=\left(w \tau_{n n n q}\right)^{(0)} n_{1}+2 \left(\nabla n_{0} \cdot \nabla n_{1}\right)\left(\tau_{n n q}+w \tau_{n n q w}\right)^{(0)}+\left(w \tau_{n n q q}\right)^{(0)} \nabla^{2} n_{1}=\\
&=\left|\nabla n_{0}\right|^{2}\tau_{n n n q}^{(0)} n_{1}+2 \left(\nabla n_{0} \cdot \nabla n_{1}\right)\tau_{n n q}^{(0)}+\left|\nabla n_{0}\right|^{2}\tau_{n n q q}^{(0)} \nabla^{2} n_{1},
\end{aligned}
\label{eq_func_09}
\end{equation}

\begin{equation}
\begin{aligned}
-2\left(w \tau_{n w}\right)_{1}&=-2\left[w \tau_{n w n}^{(0)} n_{1}+2 \left(\nabla n_{0} \cdot \nabla n_{1}\right)\left(\tau_{n w}+w \tau_{n w w}\right)^{(0)}+\left(w \tau_{n w q}\right)^{(0)} \nabla^{2} n_{1}\right] \\
&=-2\left|\nabla n_{0}\right|^{2}\tau_{n w w}^{(0)} n_{1}-4 \left(\nabla n_{0} \cdot \nabla n_{1}\right)\tau_{n w}^{(0)}-4 \left(\nabla n_{0} \cdot \nabla n_{1}\right)\left|\nabla n_{0}\right|^{2}\tau_{n w w}^{(0)},
\end{aligned}
\label{eq_func_10}
\end{equation}

\begin{equation}
\begin{aligned}
\left(q \tau_{n q}\right)_{1}&=\nabla^{2} n_{0}\tau_{n q n}^{(0)} n_{1}+2 \left(\nabla n_{0} \cdot \nabla n_{1}\right) \nabla^{2} n_{0}\tau_{n q w}^{(0)}+\tau_{n q}^{(0)} \nabla^{2} n_{1}+\nabla^{2} n_{0}\tau_{n q q}^{(0)} \nabla^{2} n_{1} \\
&=\nabla^{2} n_{0}\tau_{n n q}^{(0)} n_{1}+\tau_{n q}^{(0)} \nabla^{2} n_{1}+\nabla^{2} n_{0}\tau_{n q q}^{(0)} \nabla^{2} n_{1},
\end{aligned}
\label{eq_func_11}
\end{equation}

\begin{equation}
\begin{aligned}
-2\left(q \tau_{w}\right)_{1}&=-2 \nabla^{2} n_{0}\tau_{w n}^{(0)} n_{1}-4 \left(\nabla n_{0} \cdot \nabla n_{1}\right) \nabla^{2} n_{0}\tau_{w w}^{(0)}-2\left[\nabla^{2} n_{0}\tau_{w q}^{(0)} \nabla^{2} n_{1}+\tau_{w}^{(0)} \nabla^{2} n_{1}\right] \\
&=-2 \nabla^{2} n_{0} \tau_{w n}^{(0)} n_{1}-4 \left(\nabla n_{0} \cdot \nabla n_{1}\right) \nabla^{2} n_{0}\tau_{w w}^{(0)}-2\tau_{w}^{(0)} \nabla^{2} n_{1},
\end{aligned}
\label{eq_func_12}
\end{equation}

\begin{equation}
\begin{aligned}
2\left(\tau_{n q q} \left(\nabla n \cdot \nabla q\right)\right)_{1}
&=2\Biggl[\tau_{n q q n}^{(0)} \left(\nabla n_{0} \cdot \nabla\left(\nabla^{2} n_{0}\right)\right) n_{1}+\tau_{n q q}^{(0)}\left[\Bigl(\left(\nabla n \cdot \nabla q\right)_{\nabla n}\Bigr)^{(0)} \cdot \nabla n_{1}\right]\\ 
&+\tau_{n q q q}^{(0)}  \left(\nabla n_{0} \cdot \nabla\left(\nabla^{2} n_{0}\right)\right)\nabla^{2} n_{1}+\tau_{n q q}^{(0)}\left[\left(\left(\nabla n \cdot \nabla q\right)_{\nabla q}\right)^{(0)} \cdot \nabla\left(\nabla^{2} n_{1}\right)\right]\Biggr] \\
&=2\tau_{n n q q}^{(0)} \left(\nabla n_{0} \cdot \nabla\left(\nabla^{2} n_{0}\right)\right) n_{1}+2\tau_{n q q}^{(0)} \left(\nabla n_{1} \cdot \nabla\left(\nabla^{2} n_{0}\right)\right) \\
&\my{ +2\tau_{n q q q}^{(0)}  \left(\nabla n_{0} \cdot \nabla\left(\nabla^{2} n_{0}\right)\right)\nabla^{2} n_{1} }+2\tau_{n q q}^{(0)} \left(\nabla n_{0} \cdot \nabla\left(\nabla^{2} n_{1}\right)\right),
\end{aligned}
\label{eq_func_13}
\end{equation}

\begin{equation}
\begin{aligned}
-2\left(\tau_{w w}(\nabla n \cdot \nabla w)\right)_{1} &=-2\Biggl[\tau_{w w n}^{(0)}\left(\nabla n_{0} \cdot \nabla\left(\left|\nabla n_{0}\right|^{2}\right)\right) n_{1}+\Bigl(\left(\tau_{w w}(\nabla n \cdot \nabla w)\right)_{\nabla n}\Bigr)^{(0)} \cdot \nabla n_{1}\\
&+\left(\left(\tau_{w w}(\nabla n \cdot \nabla w)\right)_{q}\right)^{(0)} \nabla^{2} n_{1}+\Bigl(\left(\tau_{w w}(\nabla n \cdot \nabla w)\right)_{\nabla w}\Bigr)^{(0)} \cdot\left(\nabla w-\nabla w_{0}\right)\Biggr] \\
&=-2\Biggl[\tau_{w w n}^{(0)}\left(\nabla n_{0} \cdot \nabla\left(\left|\nabla n_{0}\right|^{2}\right)\right) n_{1}+2\tau_{w w w}^{(0)}\left(\nabla n_{0} \cdot \nabla n_{1}\right)\left(\nabla n_{0} \cdot \nabla\left(\left|\nabla n_{0}\right|^{2}\right)\right) \\
&+\tau_{w w}^{(0)}\left(\nabla\left(\left|\nabla n_{0}\right|^{2}\right) \cdot \nabla n_{1}\right)+\tau_{w w}^{(0)}\left(\nabla n_{0} \cdot \nabla\left(\nabla\left(n_{0}+n_{1}\right) \cdot \nabla\left(n_{0}+n_{1}\right)-\left|\nabla n_{0}\right|^{2}\right)\right)\Biggr] \\
&=-2\tau_{w w n}^{(0)}\left(\nabla n_{0} \cdot \nabla\left(\left|\nabla n_{0}\right|^{2}\right)\right) n_{1}-4\tau_{w w w}^{(0)}\left(\nabla n_{0} \cdot \nabla n_{1}\right)\left(\nabla n_{0} \cdot \nabla\left(\left|\nabla n_{0}\right|^{2}\right)\right) \\
&-2\tau_{w w}^{(0)}\left(\nabla\left(\left|\nabla n_{0}\right|^{2}\right) \cdot \nabla n_{1}\right)-2\tau_{w w}^{(0)}\Bigl(\nabla n_{0} \cdot \nabla\left(2\left(\nabla n_{0} \cdot \nabla n_{1}\right)\right)\Bigr) \\
&=-2\tau_{n w w}^{(0)}\left(\nabla n_{0} \cdot \nabla\left(\left|\nabla n_{0}\right|^{2}\right)\right) n_{1}-4\tau_{w w w}^{(0)}\left(\nabla n_{0} \cdot \nabla n_{1}\right)\left(\nabla n_{0} \cdot \nabla\left(\left|\nabla n_{0}\right|^{2}\right)\right) \\
&-2\tau_{w w}^{(0)}\left(\nabla\left(\left|\nabla n_{0}\right|^{2}\right) \cdot \nabla n_{1}\right)-4\tau_{w w}^{(0)}\left(\nabla n_{0} \cdot \nabla\left(\nabla n_{0} \cdot \nabla n_{1}\right)\right),
\label{eq_func_14}
\end{aligned}
\end{equation}
%{\color{red}  }
\begin{equation}
\begin{aligned}
\left(\tau_{q q} \nabla^{2} q\right)_{1}&=\tau_{q q n}^{(0)} \nabla^{2}\left(\nabla^{2} n_{0}\right) n_{1}+\left(\left(\tau_{q q} \nabla^{2} q\right)_{w}\right)^{(0)} 2\left(\nabla n_{0} \cdot \nabla n_{1}\right) \\
&+\tau_{q q}^{(0)}\left(\nabla^{2}\left(\nabla^{2}\left(n_{0}+n_{1}\right)\right)-\nabla^{2}\left(\nabla^{2} n_{0}\right)\right)+\tau_{q q q}^{(0)} \nabla^{2}\left(\nabla^{2} n_{0}\right) \nabla^{2} n_{1} \\
&=\tau_{n q q}^{(0)} \nabla^{2}\left(\nabla^{2} n_{0}\right) n_{1}+\tau_{q q}^{(0)} \nabla^{2}\left(\nabla^{2} n_{1}\right)\my{ + \tau_{q q q}^{(0)} \nabla^{2}\left(\nabla^{2} n_{0}\right) \nabla^{2} n_{1} }.
\end{aligned}
\label{eq_func_15}
\end{equation}

\my{ \begin{equation}
\begin{aligned}
\left(\tau_{q q q}|\nabla q|^{2}\right)_{1}&=\left|\nabla\left(\nabla^{2} n_{0}\right)\right|^{2} n_{1} \tau_{q q q n}^{(0)}+2\left|\nabla\left(\nabla^{2} n_{0}\right)\right|^{2}\left(\nabla n_{0} \cdot \nabla n_{1}\right) \tau_{q q q w}^{(0)}+\tau_{q q q}^{(0)}\left(\left(|\nabla q|^{2}\right)_{\nabla q}^{(0)} \cdot\left(\nabla q-\nabla q_{0}\right)\right) \\
&+\tau_{q q q q}^{(0)}\left|\nabla\left(\nabla^{2} n_{0}\right)\right|^{2} \nabla^{2} n_{1} \\
&=\left|\nabla\left(\nabla^{2} n_{0}\right)\right|^{2} n_{1} \tau_{n q q q}^{(0)}+2 \tau_{q q q}^{(0)}\left(\nabla\left(\nabla^{2} n_{0}\right) \cdot \nabla\left(\nabla^{2} n_{1}\right)\right)+\tau_{q q q q}^{(0)}\left|\nabla\left(\nabla^{2} n_{0}\right)\right|^{2} \nabla^{2} n_{1}
\end{aligned}
\label{eq_func_16}
\end{equation}  }

Substituting Exps.~\eqref{eq_func_08}~-~\eqref{eq_func_16} into Exp.~\eqref{eq_func_07} we get:

\begin{equation}
\begin{aligned}
\left(\frac{\delta T}{\delta n}\right)_{1}&=n_{1}\tau_{n n}^{(0)}+2\tau_{n w}^{(0)} \left(\nabla n_{0} \cdot \nabla n_{1}\right)+\tau_{n q}^{(0)} \nabla^{2} n_{1}\\
&+\left|\nabla n_{0}\right|^{2}\tau_{n n n q}^{(0)} n_{1}+2 \left(\nabla n_{0} \cdot \nabla n_{1}\right)\tau_{n n q}^{(0)}+\left|\nabla n_{0}\right|^{2}\tau_{n n q q}^{(0)} \nabla^{2} n_{1} \\
&-2\left|\nabla n_{0}\right|^{2}\tau_{n n w}^{(0)} n_{1}-4 \left(\nabla n_{0} \cdot \nabla n_{1}\right)\tau_{n w}^{(0)}-4 \left(\nabla n_{0} \cdot \nabla n_{1}\right)\left|\nabla n_{0}\right|^{2}\tau_{n w w}^{(0)} \\
&+\nabla^{2} n_{0}\tau_{n n q}^{(0)} n_{1}+\tau_{n q}^{(0)} \nabla^{2} n_{1}+\nabla^{2} n_{0}\tau_{n q q}^{(0)} \nabla^{2} n_{1} \\
&-2 \nabla^{2} n_{0} \tau_{w n}^{(0)} n_{1}-4 \left(\nabla n_{0} \cdot \nabla n_{1}\right) \nabla^{2} n_{0}^{(0)}\tau_{w w}^{(0)}-2\tau_{w}^{(0)} \nabla^{2} n_{1} \\
&+2\tau_{n n q q}^{(0)} \left(\nabla n_{0} \cdot \nabla\left(\nabla^{2} n_{0}\right)\right) n_{1}+2\tau_{n q q}^{(0)} \left(\nabla n_{1} \cdot \nabla\left(\nabla^{2} n_{0}\right)\right) \my{ +2\left(\nabla n_{0} \cdot \nabla\left(\nabla^{2} n_{0}\right)\right) \nabla^{2} n_{1}\tau_{n q q q}^{(0)} } \\
&+2\tau_{n q q}^{(0)} \left(\nabla n_{0} \cdot \nabla\left(\nabla^{2} n_{1}\right)\right)\my{ - 2\tau_{n w w}^{(0)}\left(\nabla n_{0} \cdot \nabla\left(\left|\nabla n_{0}\right|^{2}\right)\right) n_{1}-4\tau_{w w w}^{(0)}\left(\nabla n_{0} \cdot \nabla n_{1}\right)\left(\nabla n_{0} \cdot \nabla\left(\left|\nabla n_{0}\right|^{2}\right)\right) }\\
&-2\tau_{w w}^{(0)}\left(\nabla\left(\left|\nabla n_{0}\right|^{2}\right) \cdot \nabla n_{1}\right)-4\tau_{w w}^{(0)}\left(\nabla n_{0} \cdot \nabla\left(\nabla n_{0} \cdot \nabla n_{1}\right)\right) \\
&+\tau_{n q q}^{(0)} \nabla^{2}\left(\nabla^{2} n_{0}\right) n_{1} + \tau_{q q}^{(0)} \nabla^{2}\left(\nabla^{2} n_{1}\right)
\my{ +\, \tau_{q q q}^{(0)} \nabla^{2}\left(\nabla^{2} n_{0}\right) \nabla^{2} n_{1} } \\
&\my{ + \left|\nabla\left(\nabla^{2} n_{0}\right)\right|^{2} n_{1} \tau_{n q q q}^{(0)}+2 \tau_{q q q}^{(0)}\left(\nabla\left(\nabla^{2} n_{0}\right) \cdot \nabla\left(\nabla^{2} n_{1}\right)\right)+\tau_{q q q q}^{(0)}\left|\nabla\left(\nabla^{2} n_{0}\right)\right|^{2} \nabla^{2} n_{1} }.
\end{aligned}
\label{eq_func_total}
\end{equation}

Term with partial derivative factor $\tau_{n n}^{(0)}$ determines the potential $\left(\frac{\delta T_s^{\textrm{I}}}{\delta n}\right)_1$, $\tau_{w}^{(0)}, \tau_{n w}^{(0)}$, $\tau_{w w}^{(0)}, \tau_{n w w}^{(0)}, \tau_{n n w}^{(0)}$ and $\tau_{w w w}^{(0)}$ - the potential $\left(\frac{\delta T_s^{\textrm{II}}}{\delta n}\right)_1$, and $\tau_{n q}^{(0)}, \tau_{q q}^{(0)}, \tau_{n n q}^{(0)}, \tau_{n q q}^{(0)}, \tau_{q q q}^{(0)}, \tau_{n n n q}^{(0)}, \tau_{n n q q}^{(0)}, \tau_{n q q q}^{(0)}$ and $\tau_{q q q q}^{(0)}$ - the potential $\left(\frac{\delta T_s^{\textrm{III}}}{\delta n}\right)_1$.

%\clearpage

\section{Asymptotic form of the Laplacian term for spherical system}
%For spherical systems we have that
%\begin{equation}
%  \nabla n_0(r)  \nabla  \left(\frac{\delta T_s^{L\beta}}{\delta n}\right)_1=\sum_{n=0}^6 %F_k[r,n_0(r)] \frac{d^k n_1(r)}{dr^k},     
%\end{equation}
%where $n_1({\bf r})=n_1(r)\cos(\theta)$ and
\my{The functions defined by Eq. (17) in the main manuscript are for the spherical ground-state density $n_0(r)$}:
\begin{eqnarray}
F_6&=&{\frac {1}{80}}\,{\frac {\sqrt [3]{3}\beta}{{\pi }^{4/3} \left( {\it 
n_\textrm{0}} \left( r \right)  \right) ^{2/3}}}, \\
F_5&=&{\frac {1}{240}}\,{\frac {\sqrt [3]{3}\beta\, \left( -17\, \left( {
\frac {d}{dr}}{\it n_\textrm{0}} \left( r \right)  \right) r+18\,{\it n_\textrm{0}}
 \left( r \right)  \right) }{r{\pi }^{4/3} \left( {\it n_\textrm{0}} \left( r
 \right)  \right) ^{5/3}}}, \\
 F_4&=& -{\frac {1}{240}}\,{\frac {\sqrt [3]{3}\beta\, \left( 18\, \left( {
\it n_\textrm{0}} \left( r \right)  \right) ^{2}-65\, \left( {\frac {d}{dr}}{
\it n_\textrm{0}} \left( r \right)  \right) ^{2}{r}^{2}+40\, \left( {\frac {d^{2
}}{d{r}^{2}}}{\it n_\textrm{0}} \left( r \right)  \right) {r}^{2}{\it n_\textrm{0}}
 \left( r \right) +108\, \left( {\frac {d}{dr}}{\it n_\textrm{0}} \left( r
 \right)  \right) r{\it n_\textrm{0}} \left( r \right)  \right) }{{r}^{2}{\pi }^
{4/3} \left( {\it n_\textrm{0}} \left( r \right)  \right) ^{8/3}}}, 
\end{eqnarray}

\begin{eqnarray}
 F_3&=& {\frac {1}{2160}}\,
 { {\sqrt [3]{3}\beta\, 
    {r}^{-2}{\pi }^{-4/3} \left( {\it n_\textrm{0}} \left( r \right)  \right) ^{-11/3}                   
    } 
 }    
  \Bigg(
 -1400\, \left( {\frac {d}{dr}}          {\it n_\textrm{0}} \left( r \right)  \right) ^{3}{r}^{2}
 -450\,  \left( \frac {d^{3}}{d{r}^{3}} {\it n_\textrm{0}} \left( r \right)  \right) {r}^{2} 
 \left( {\it n_\textrm{0}} \left( r \right)  \right) ^{2} \nonumber \\
&+&1935\, \left( \frac {d}{dr}{\it n_\textrm{0}} \left( r \right)  \right)  \left( {\frac {d^{2}
}{d{r}^{2}}}{\it n_\textrm{0}} \left( r \right)  \right) {r}^{2}{\it n_\textrm{0}} \left( 
r \right) +324\, \left( {\frac {d}{dr}}{\it n_\textrm{0}} \left( r \right) 
 \right)  \left( {\it n_\textrm{0}} \left( r \right)  \right) ^{2} \nonumber \\
 &-&1980\, \left( {\frac {d^{2}}{d{r}^{2}}}{\it n_\textrm{0}} \left( r \right)  \right) r
 \left( {\it n_\textrm{0}} \left( r \right)  \right) ^{2}
 +3690\,   \left( {\frac {d}{dr}}{\it n_\textrm{0}} \left( r \right)  \right) ^{2}r{\it n_\textrm{0}} \left( r \right)  
 \Bigg), 
 \end{eqnarray}
 
 \begin{eqnarray}
 F_2&=& 
 {\frac {1}{6480}}\,
 {\sqrt [3]{3}\beta\, {{\pi }^{-4/3} \left( {\it n_\textrm{0}} \left( r
 \right)  \right) ^{-14/3}{r}^{-3}}}
 \Bigg( 
 4840\, 
 \left( {\frac {d}{dr}}{\it n_\textrm{0}} \left( r \right)  \right) ^{4}{r}^{3}
 -1080\,
 \left( {\frac {d^{4}}{d{r}^{4}}}{\it n_\textrm{0}} \left( r \right)  \right) 
 \left( {\it n_\textrm{0}} \left( r \right)  \right) ^{3}{r}^{3} \nonumber \\
 &+&3780\, \left( {
\frac {d^{2}}{d{r}^{2}}}{\it n_\textrm{0}} \left( r \right)  \right) ^{2}
 \left( {\it n_\textrm{0}} \left( r \right)  \right) ^{2}{r}^{3}-28560\, \left( 
{\frac {d}{dr}}{\it n_\textrm{0}} \left( r \right)  \right) ^{3}{\it n_\textrm{0}} \left( 
r \right) {r}^{2}+2808\, \left( {\frac {d}{dr}}{\it n_\textrm{0}} \left( r
 \right)  \right)  \left( {\it n_\textrm{0}} \left( r \right)  \right) ^{3} \nonumber \\
 &+&2160
\, \left( {\frac {d^{2}}{d{r}^{2}}}{\it n_\textrm{0}} \left( r \right)  \right) 
 \left( {\it n_\textrm{0}} \left( r \right)  \right) ^{3}r-14640\, \left( {
\frac {d}{dr}}{\it n_\textrm{0}} \left( r \right)  \right) ^{2} \left( {\frac {d
^{2}}{d{r}^{2}}}{\it n_\textrm{0}} \left( r \right)  \right) {\it n_\textrm{0}} \left( r
 \right) {r}^{3} \nonumber \\
 &+&5805\, \left( {\frac {d}{dr}}{\it n_\textrm{0}} \left( r
 \right)  \right)  \left( {\frac {d^{3}}{d{r}^{3}}}{\it n_\textrm{0}} \left( r
 \right)  \right)  \left( {\it n_\textrm{0}} \left( r \right)  \right) ^{2}{r}^{
3}-5940\, \left( {\frac {d^{3}}{d{r}^{3}}}{\it n_\textrm{0}} \left( r \right) 
 \right)  \left( {\it n_\textrm{0}} \left( r \right)  \right) ^{3}{r}^{2} \nonumber \\
 &-&270\,
 \left( {\frac {d}{dr}}{\it n_\textrm{0}} \left( r \right)  \right) ^{2} \left( 
{\it n_\textrm{0}} \left( r \right)  \right) ^{2}r+32580\, \left( {\frac {d}{dr}
}{\it n_\textrm{0}} \left( r \right)  \right)  \left( {\frac {d^{2}}{d{r}^{2}}}{
\it n_\textrm{0}} \left( r \right)  \right)  \left( {\it n_\textrm{0}} \left( r \right) 
 \right) ^{2}{r}^{2} 
\Bigg), 
 \end{eqnarray}
 \begin{eqnarray}
 F_1&=&  -{\frac {1}{1296}}\,  \sqrt [3]{3}\beta\, 
 {{\pi} ^{-4/3} \left( {\it n_\textrm{0}} \left( r \right)  \right) ^{-14/3}{r}^{-4}}
 \Bigg ( 
 108\, \left( {\frac {d^{5}}{d{r}^{5}}}{\it n_\textrm{0}} \left( r \right)  \right) 
 \left( {\it n_\textrm{0}} \left( r \right)  \right) ^{3}{r}^{4}  \nonumber \\
 &+&3180\, \left( {\frac {d^
{2}}{d{r}^{2}}}{\it n_\textrm{0}} \left( r \right)  \right) ^{2} \left( {\frac {
d}{dr}}{\it n_\textrm{0}} \left( r \right)  \right) {\it n_\textrm{0}} \left( r \right) {r
}^{4}+432\, \left( {\frac {d}{dr}}{\it n_\textrm{0}} \left( r \right)  \right) 
 \left( {\it n_\textrm{0}} \left( r \right)  \right) ^{3} \nonumber \\
 &-&783\, \left( {\frac {d
}{dr}}{\it n_\textrm{0}} \left( r \right)  \right)  \left( {\frac {d^{4}}{d{r}^{
4}}}{\it n_\textrm{0}} \left( r \right)  \right)  \left( {\it n_\textrm{0}} \left( r
 \right)  \right) ^{2}{r}^{4}-432\, \left( {\frac {d^{2}}{d{r}^{2}}}{
\it n_\textrm{0}} \left( r \right)  \right)  \left( {\it n_\textrm{0}} \left( r \right) 
 \right) ^{3}r \nonumber \\
 &-&3528\, \left( {\frac {d^{2}}{d{r}^{2}}}{\it n_\textrm{0}} \left( 
r \right)  \right) ^{2} \left( {\it n_\textrm{0}} \left( r \right)  \right) ^{2}
{r}^{3}-1296\, \left( {\frac {d^{2}}{d{r}^{2}}}{\it n_\textrm{0}} \left( r
 \right)  \right)  \left( {\frac {d^{3}}{d{r}^{3}}}{\it n_\textrm{0}} \left( r
 \right)  \right)  \left( {\it n_\textrm{0}} \left( r \right)  \right) ^{2}{r}^{
4} \nonumber \\
&+&648\, \left( {\frac {d^{4}}{d{r}^{4}}}{\it n_\textrm{0}} \left( r \right) 
 \right)  \left( {\it n_\textrm{0}} \left( r \right)  \right) ^{3}{r}^{3}-4136\,
 \left( {\frac {d}{dr}}{\it n_\textrm{0}} \left( r \right)  \right) ^{3} \left( 
{\frac {d^{2}}{d{r}^{2}}}{\it n_\textrm{0}} \left( r \right)  \right) {r}^{4}-
10208\, \left( {\frac {d}{dr}}{\it n_\textrm{0}} \left( r \right)  \right) ^{4}{
r}^{3} \nonumber \\
&+&1332\, \left( {\frac {d}{dr}}{\it n_\textrm{0}} \left( r \right) 
 \right) ^{2} \left( {\it n_\textrm{0}} \left( r \right)  \right) ^{2}r+2736\,
 \left( {\frac {d}{dr}}{\it n_\textrm{0}} \left( r \right)  \right) ^{2} \left( 
{\frac {d^{3}}{d{r}^{3}}}{\it n_\textrm{0}} \left( r \right)  \right) {\it n_\textrm{0}}
 \left( r \right) {r}^{4} \nonumber \\
 &-&432\, \left( {\frac {d^{3}}{d{r}^{3}}}{\it 
n_\textrm{0}} \left( r \right)  \right)  \left( {\it n_\textrm{0}} \left( r \right) 
 \right) ^{3}{r}^{2}+432\, \left( {\frac {d}{dr}}{\it n_\textrm{0}} \left( r
 \right)  \right) ^{3}{\it n_\textrm{0}} \left( r \right) {r}^{2}+720\, \left( {
\frac {d}{dr}}{\it n_\textrm{0}} \left( r \right)  \right)  \left( {\frac {d^{2}
}{d{r}^{2}}}{\it n_\textrm{0}} \left( r \right)  \right)  \left( {\it n_\textrm{0}}
 \left( r \right)  \right) ^{2}{r}^{2} \nonumber \\
 &+&18288\, \left( {\frac {d}{dr}}{
\it n_\textrm{0}} \left( r \right)  \right) ^{2} \left( {\frac {d^{2}}{d{r}^{2}}
}{\it n_\textrm{0}} \left( r \right)  \right) {\it n_\textrm{0}} \left( r \right) {r}^{3}-
4806\, \left( {\frac {d}{dr}}{\it n_\textrm{0}} \left( r \right)  \right) 
 \left( {\frac {d^{3}}{d{r}^{3}}}{\it n_\textrm{0}} \left( r \right)  \right) 
 \left( {\it n_\textrm{0}} \left( r \right)  \right) ^{2}{r}^{3} 
 \Bigg ), 
\end{eqnarray}
\begin{eqnarray}
 F_\textrm{0}&=&-{\frac {1}{3888}}\,\sqrt [3]{3}\beta\, {\pi }^{-4/3}
 \left( {\it n_\textrm{0}} \left( r \right)  \right) ^{-{\frac {17}{3}}}{r}^{-5} 
 \Bigg (
 13176\, \left( {\frac {
d}{dr}}{\it n_\textrm{0}} \left( r \right)  \right)  \left( {\frac {d^{2}}{d{r}^
{2}}}{\it n_\textrm{0}} \left( r \right)  \right)  \left( {\frac {d^{3}}{d{r}^{3
}}}{\it n_\textrm{0}} \left( r \right)  \right)  \left( {\it n_\textrm{0}} \left( r
 \right)  \right) ^{2}{r}^{5} \nonumber \\
 &+&2646\, \left( {\frac {d}{dr}}{\it n_\textrm{0}}
 \left( r \right)  \right)  \left( {\frac {d^{3}}{d{r}^{3}}}{\it n_\textrm{0}}
 \left( r \right)  \right)  \left( {\it n_\textrm{0}} \left( r \right)  \right) 
^{3}{r}^{3} -83424\, \left( {\frac {d}{dr}}{\it n_\textrm{0}} \left( r \right) 
 \right) ^{3} \left( {\frac {d^{2}}{d{r}^{2}}}{\it n_\textrm{0}} \left( r
 \right)  \right) {\it n_\textrm{0}} \left( r \right) {r}^{4} \nonumber \\
 &-&4860\, \left( {
\frac {d^{4}}{d{r}^{4}}}{\it n_\textrm{0}} \left( r \right)  \right)  \left( {
\frac {d}{dr}}{\it n_\textrm{0}} \left( r \right)  \right)  \left( {\it n_\textrm{0}}
 \left( r \right)  \right) ^{3}{r}^{4} +4968\, \left( {\frac {d}{dr}}{
\it n_\textrm{0}} \left( r \right)  \right)  \left( {\frac {d^{2}}{d{r}^{2}}}{
\it n_\textrm{0}} \left( r \right)  \right)  \left( {\it n_\textrm{0}} \left( r \right) 
 \right) ^{3}{r}^{2} \nonumber \\
 &-&2880\, \left( {\frac {d}{dr}}{\it n_\textrm{0}} \left( r
 \right)  \right) ^{2} \left( {\frac {d^{2}}{d{r}^{2}}}{\it n_\textrm{0}}
 \left( r \right)  \right)  \left( {\it n_\textrm{0}} \left( r \right)  \right) 
^{2}{r}^{3} +25488\, \left( {\frac {d}{dr}}{\it n_\textrm{0}} \left( r \right) 
 \right) ^{2} \left( {\frac {d^{3}}{d{r}^{3}}}{\it n_\textrm{0}} \left( r
 \right)  \right)  \left( {\it n_\textrm{0}} \left( r \right)  \right) ^{2}{r}^{
4} \nonumber \\
&+&36648\, \left( {\frac {d^{2}}{d{r}^{2}}}{\it n_\textrm{0}} \left( r \right) 
 \right) ^{2} \left( {\frac {d}{dr}}{\it n_\textrm{0}} \left( r \right) 
 \right)  \left( {\it n_\textrm{0}} \left( r \right)  \right) ^{2}{r}^{4}-783\,
 \left( {\frac {d^{5}}{d{r}^{5}}}{\it n_\textrm{0}} \left( r \right)  \right) 
 \left( {\frac {d}{dr}}{\it n_\textrm{0}} \left( r \right)  \right)  \left( {
\it n_\textrm{0}} \left( r \right)  \right) ^{3}{r}^{5} \nonumber \\
&-&12408\, \left( {\frac {d
}{dr}}{\it n_\textrm{0}} \left( r \right)  \right) ^{3} \left( {\frac {d^{3}}{d{
r}^{3}}}{\it n_\textrm{0}} \left( r \right)  \right) {\it n_\textrm{0}} \left( r \right) {
r}^{5} 
-21252\, \left( {\frac {d^{2}}{d{r}^{2}}}{\it n_\textrm{0}} \left( r
 \right)  \right) ^{2} \left( {\frac {d}{dr}}{\it n_\textrm{0}} \left( r
 \right)  \right) ^{2}{\it n_\textrm{0}} \left( r \right) {r}^{5} \nonumber \\
 &-&1728\, \left( 
{\frac {d^{2}}{d{r}^{2}}}{\it n_\textrm{0}} \left( r \right)  \right)  \left( {
\frac {d^{4}}{d{r}^{4}}}{\it n_\textrm{0}} \left( r \right)  \right)  \left( {
\it n_\textrm{0}} \left( r \right)  \right) ^{3}{r}^{5} 
+ 4104\, \left( {\frac {d^{4}}{d{r}^{4}}}{\it n_\textrm{0}} \left( r \right)  \right)  \left( {\frac {d}{d
r}}{\it n_\textrm{0}} \left( r \right)  \right) ^{2} \left( {\it n_\textrm{0}} \left( r
 \right)  \right) ^{2}{r}^{5} \nonumber \\
 &-&9504\, \left( {\frac {d^{2}}{d{r}^{2}}}{
\it n_\textrm{0}} \left( r \right)  \right)  \left( {\frac {d^{3}}{d{r}^{3}}}{
\it n_\textrm{0}} \left( r \right)  \right)  \left( {\it n_\textrm{0}} \left( r \right) 
 \right) ^{3}{r}^{4} 
 -1080\, \left( {\frac {d^{3}}{d{r}^{3}}}{\it n_\textrm{0}}
 \left( r \right)  \right) ^{2} \left( {\it n_\textrm{0}} \left( r \right) 
 \right) ^{3}{r}^{5} \nonumber \\
 &+&2772\, \left( {\frac {d^{2}}{d{r}^{2}}}{\it n_\textrm{0}}
 \left( r \right)  \right) ^{3} \left( {\it n_\textrm{0}} \left( r \right) 
 \right) ^{2}{r}^{5}+81\, \left( {\frac {d^{6}}{d{r}^{6}}}{\it n_\textrm{0}}
 \left( r \right)  \right)  \left( {\it n_\textrm{0}} \left( r \right)  \right) 
^{4}{r}^{5} \nonumber \\
&-&2484\, \left( {\frac {d}{dr}}{\it n_\textrm{0}} \left( r \right) 
 \right) ^{2} \left( {\it n_\textrm{0}} \left( r \right)  \right) ^{3}r+1944\,
 \left( {\frac {d^{2}}{d{r}^{2}}}{\it n_\textrm{0}} \left( r \right)  \right) ^{
2} \left( {\it n_\textrm{0}} \left( r \right)  \right) ^{3}{r}^{3}+1296\,
 \left( {\frac {d^{2}}{d{r}^{2}}}{\it n_\textrm{0}} \left( r \right)  \right) 
 \left( {\it n_\textrm{0}} \left( r \right)  \right) ^{4}r \nonumber \\
 &-&648\, \left( {\frac {
d^{4}}{d{r}^{4}}}{\it n_\textrm{0}} \left( r \right)  \right)  \left( {\it n_\textrm{0}}
 \left( r \right)  \right) ^{4}{r}^{3} + 486\, \left( {\frac {d^{5}}{d{r
}^{5}}}{\it n_\textrm{0}} \left( r \right)  \right)  \left( {\it n_\textrm{0}} \left( r
 \right)  \right) ^{4}{r}^{4} \nonumber \\
 &+& 17248\, \left( {\frac {d}{dr}}{\it n_\textrm{0}}
 \left( r \right)  \right) ^{4} \left( {\frac {d^{2}}{d{r}^{2}}}{\it 
n_\textrm{0}} \left( r \right)  \right) {r}^{5}-5472\, \left( {\frac {d}{dr}}{
\it n_\textrm{0}} \left( r \right)  \right) ^{3} \left( {\it n_\textrm{0}} \left( r
 \right)  \right) ^{2}{r}^{2} \nonumber \\
 &-&1584\, \left( {\frac {d}{dr}}{\it n_\textrm{0}}
 \left( r \right)  \right) ^{4}{\it n_\textrm{0}} \left( r \right) {r}^{3}+34496
\, \left( {\frac {d}{dr}}{\it n_\textrm{0}} \left( r \right)  \right) ^{5}{r}^{4
} -1296\, \left( {\frac {d}{dr}}{\it n_\textrm{0}} \left( r \right)  \right) 
 \left( {\it n_\textrm{0}} \left( r \right)  \right) ^{4}
 \Bigg ),  
\end{eqnarray}
The above expressions are valid everywhere (for spherical systems). In the asymptotic region,
we can use $n_0(r)=A_0 \exp(-\kappa r)$, obtaining:
\begin{eqnarray}
F_6&=& {\frac {1}{80}}\,{\frac {\sqrt [3]{3}\beta}{{\pi }^{4/3} \left( A_0\,{{\rm e}^{-\kappa\,r}} \right) ^{2/3}}}
\\
F_5&=& {\frac {1}{240}}\,{\frac {\sqrt [3]{3}\beta\, \left( 18+17\,\kappa\,r
 \right) }{{\pi }^{4/3} \left( A_0\,{{\rm e}^{-\kappa\,r}}
 \right) ^{2/3}r}}\\
F_4&=& {\frac {1}{240}}\,{\frac {\sqrt [3]{3}\beta\, \left( -18+25\,{\kappa}^
{2}{r}^{2}+108\,\kappa\,r \right) }{{\pi }^{4/3} \left( A_0\,{
{\rm e}^{-\kappa\,r}} \right) ^{2/3}{r}^{2}}} \\
F_3&=& -{\frac {1}{2160}}\,{\frac {\sqrt [3]{3}\beta\,\kappa\, \left( 85\,{
\kappa}^{2}{r}^{2}+324-1710\,\kappa\,r \right) }{{\pi }^{4/3} \left( A_0\,{{\rm e}^{-\kappa\,r}} \right) ^{2/3}{r}^{2}}}\\
F_2&=& {\frac {1}{240}}\,{\frac {\sqrt [3]{3}\beta\, \left( -18+25\,{\kappa}^
{2}{r}^{2}+108\,\kappa\,r \right) }{{\pi }^{4/3} \left( A_0\,{
{\rm e}^{-\kappa\,r}} \right) ^{2/3}{r}^{2}}}\\
F_1&=& -{\frac {1}{1296}}\,{\frac {\sqrt [3]{3}\beta\,\kappa\, \left( 191\,{
\kappa}^{4}{r}^{4}-432+900\,\kappa\,r+394\,{\kappa}^{3}{r}^{3}-720\,{
\kappa}^{2}{r}^{2} \right) }{{\pi }^{4/3} \left( A_0\,{{\rm e}^{-
\kappa\,r}} \right) ^{2/3}{r}^{4}}}\\
F_\textrm{0}&=& -{\frac {1}{1944}}\,{\frac {\sqrt [3]{3}\beta\,\kappa\, \left( -594\,
\kappa\,r+648+335\,{\kappa}^{4}{r}^{4}+252\,{\kappa}^{2}{r}^{2}-261\,{
\kappa}^{3}{r}^{3}+65\,{\kappa}^{5}{r}^{5} \right) }{{\pi }^{4/3}
 \left( A_0\,{{\rm e}^{-\kappa\,r}} \right) ^{2/3}{r}^{5}}}.
\end{eqnarray}

In the asymptotic region, we have also to consider only the lowest $1/r$ power, thus obtaining the Exp.~(19) in the main manuscript.

%\clearpage
%%%%%%%%%%%%%%%%%%%%%%%%%%%%%%%%%%%%%%%%%%%%%%%%%%%%%%%%%%%%%%%%%%%%%%%%5
\section{Fitting procedure to extract the oscillator strength}
\begin{figure}[!hbt]
    \centering
    \includegraphics[width=0.5\textwidth]{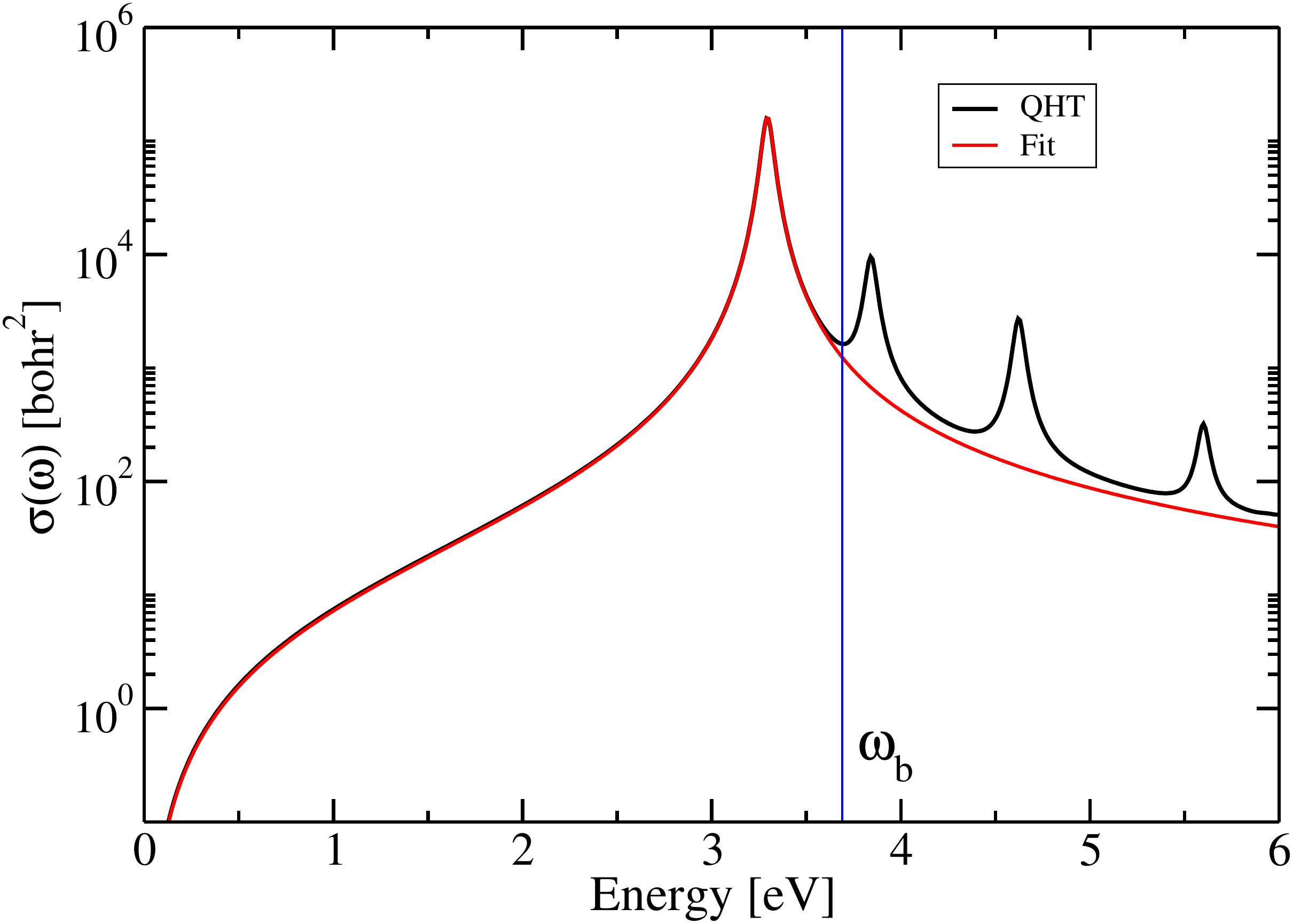}
    \caption{Photoabsorption cross-section (log scale) for a Na jellium nanosphere with $N_b$=1074 electrons, as obtained for Mod/QHT and the resulting fit using the procedure described in the text. The vertical line shows the energy position of $\omega_b$.}
    \label{fig:wawb}
\end{figure}

%%%%%%%%%%%%%%%%%%%%%%%%%%%%%%%%%%%%%%%%%%%%%%%%%%%%%%%%%%%%%%%%%%%%%%%%%%%5
We fit the photoabsorption cross-section (of all QHT approaches and reference TD-DFT) with the function
\begin{equation}
f(\omega)= \frac{2}{\pi} \frac{\omega^2 \gamma}
{(\omega^2 -\omega_0^2)^2+ \omega^2 \gamma^2 },
\label{eq:fit}
\end{equation}
where $\gamma$ is the broadening parameter and $\omega_0$ the peak position.   Note that $f(\omega)$ has a maximum at $\omega=\omega_0$ and $f(\omega_0)=2/(\pi \gamma)$.
This function integrates to 1 in the range $0<\omega<+\infty$ and originates from the Drude classical solution of a nanosphere, whose photoabsorption cross-section in CGS units is \cite{kreib}:
\begin{equation}
    \sigma(\omega)=\frac{4\pi}{c} {\rm Im}\left [ R^3\frac{\epsilon(\omega)-1}{\epsilon(\omega)+2} \right] = \frac{2 \pi^2 R^3 \omega_p^2}{3 c} f(\omega) =\frac{2\pi^2}{c}\frac{e^2}{m_e} N_e f(\omega),
\end{equation}
where we used that $w_0=w_p/\sqrt{3}$, $\omega_p=\sqrt{4\pi e^2/m_e (N/V)}$ and $V=(4\pi/3) R^3$.
For a jellium nanosphere of very large $R$,  QHT, and TD-DFT will reach the classical limit \cite{brack93}, thus, the function $f(\omega)$ is the right function to be used in the fitting procedure, where we minimize:
\begin{equation}
    E= \frac{ \int_{\omega_a}^{\omega_b} |f_{osc} f(w) -\sigma_m(\omega)|d\omega} {\int_{\omega_a}^{\omega_b} \sigma_m(\omega)d\omega},
\end{equation}
where $\sigma_m(\omega)$ is the input absorption spectra (QHT or TDDFT).
The error depends on three parameters  $f_{osc},\omega_0$, and $\gamma$.
%is the fitting parameter. 
For the range, we consider $\omega_a=1$ eV, whereas $\omega_b$ is the energy position where other peaks/shoulders starts after the LSP resonance. 
In this way, only the LSP peak is included in the fit, whereas 
other high-energy peaks or shoulder are not considered.
As an example, we show in Fig. \ref{fig:wawb} the QHT spectra for $N_e=1074$. 
The resulting fit is very accurate for all energies less than $\omega_b$. Other peaks above $\omega_b$ are not included in the peaks, so that the oscillator strength of the first main peak can be obtained. 
For QHT, one can also do a fit with several $f(\omega)$, one for each peak, obtaining the same results.

\begin{figure}[!hbt]
    \centering
    \includegraphics[width=0.6\textwidth]{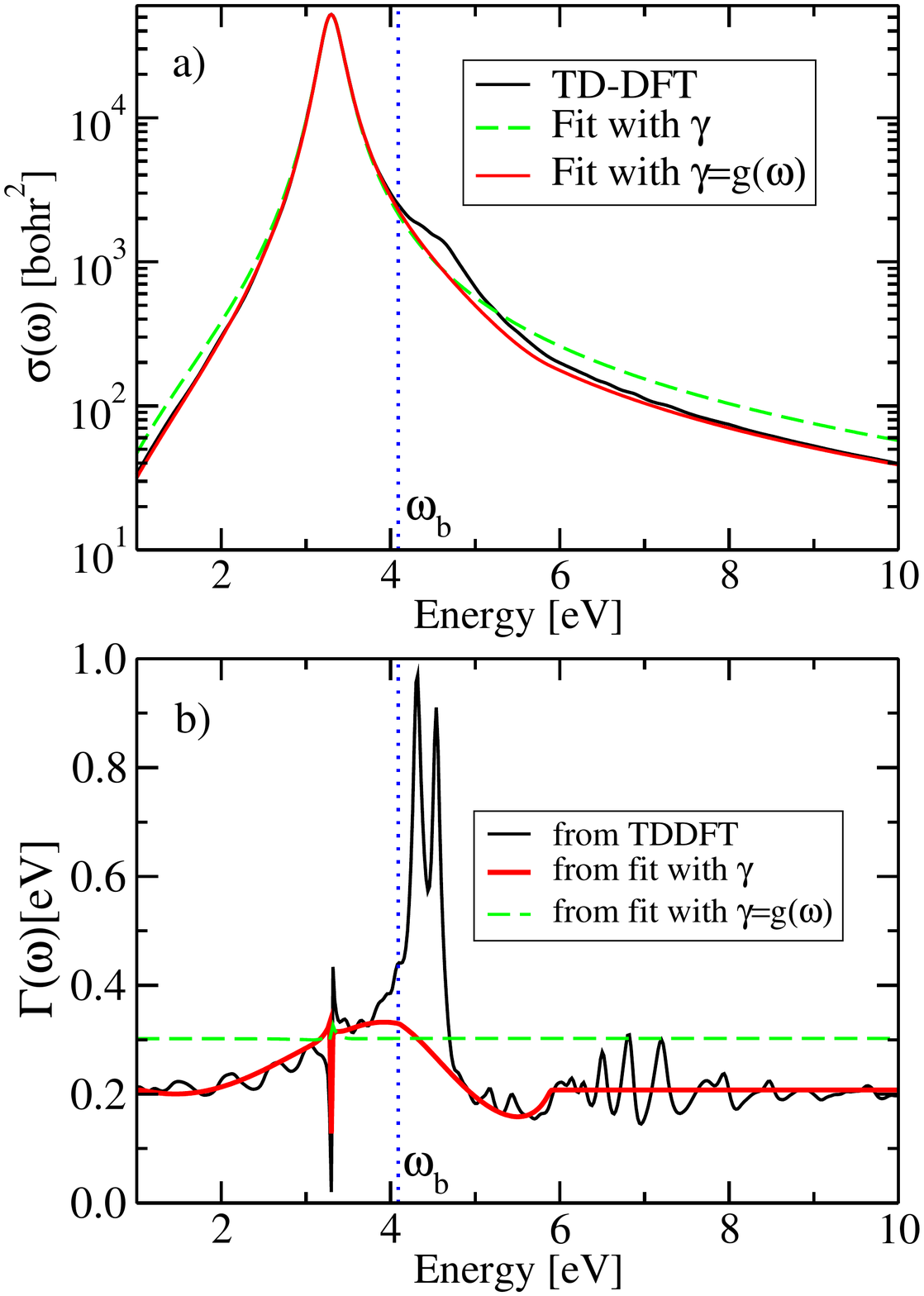}
    \caption{a) TD-DFT photoabsorption cross-section (black line) for $N_e$=6174 and using $\Gamma_0$=0.1eV and its fits using a constant broadening (dashed-green line) and an energy-dependent broadening (red line). The fits consider only the region from 1 eV up to $\omega_b$=4.1eV, that represent the onset of the plasmon shoulder. b) Function $\Gamma(\omega)$ obtained from the three curves in panel a).}
    \label{fig:pp}
\end{figure}

The situation complicates for the TD-DFT spectrum.
In Fig. \ref{fig:pp}~a), we report the TD-DFT spectrum for $N_e=6174$ (i.e., the largest system considered) and the corresponding fit (green dashed line) with Eq. (\ref{eq:fit}). The error in the fit is 2.1\% in the fitting range and fitted $f_{osc}$ is 890.
The fitting curve is not very accurate as the TD-DFT values (for low and high energies) are not well reproduced. 

In order to improve the fit, we considered an energy-dependent broadening \cite{zaremba}, 
i.e., $\gamma=g(\omega)$ and to model the function $g(\omega)$, we inverted Eq. (\ref{eq:fit}) as a function on $\omega$. 
For an arbitrary spectrum $\sigma(\omega)$ with $f_{osc}=\int_0^\infty \sigma(\omega) d\omega$
we define the function: 

\begin{equation}
\Gamma(\omega)=\frac{\pi (\omega_0^4-\omega^4)}{\omega^3} \frac{\sigma(\omega)^2 }{\displaystyle 
f_{osc} \frac{d \sigma(\omega)}{d\omega}}  \; .
\label{eq:invfit}
\end{equation}

It can be easily verified that if $\sigma(\omega)=f_{osc} f(\omega)$, then   we obtain a constant function $\Gamma(\omega)=\gamma$.
In Fig. \ref{fig:pp}~b), we report the function $\Gamma(\omega)$ for the TD-DFT absorption spectra
using Eq. (\ref{eq:invfit}). There are some spikes at $\omega=\omega_0$ due to the presence of vanishing the
first derivative at the main plasmon peak. For high- and low-energy, we see that $\Gamma(\omega)$ 
approaches exactly $2\Gamma_0$, where  $\Gamma_0$ is the broadening used in TD-DFT calculation. 
Thus, we define a model ${g}(\omega)$ with the following limits
 ${g}(\omega\le \omega_{low})=2 \Gamma_0$, where $\omega_{low}$ is close to $\omega_a$ 
 and ${g}(\omega > \omega_p)=2 \Gamma_0$, where $\omega_p$ is the plasma frequency.
In between these limits, the function ${g}(\omega)$ grows up to the 
maximum values of $g(\omega_1)=\gamma$, where $\omega_1$ is fixed to be $\omega_b$ i.e., the energy
where the shoulder in the spectrum starts.
If the TD-DFT absorption spectrum is fitted with such function, we obtain the red-curve in Fig. \ref{fig:pp}a) with an error reduced by a factor of 4 ($0.6\%$) and also a reduced oscillator strength ($f_{osc}=872$).
The non-linear  fit is done using four parameters: $f_{osc},\omega_0,\gamma$, and $\omega_{low}$. Note also that the function $f(\omega)$ has to be renormalized to integrate to 1 when a frequency-dependent broadening is used. 
In Fig. \ref{fig:pp}~b) we report the function $\Gamma(\omega)$ obtained from
the fitted curve. The agreement with the TD-DFT results is very good, for all energies before and after the plasmon shoulder.

In Table \ref{tab:fits}, we report all the parameters for the fits of the TD-DFT spectra.
Note that for the fits, we compute the TD-DFT with a broadening of $\Gamma_0=$ 0.1 eV, to have a smooth curve, which is required especially for small spheres.
The $f_{osc}$ of the LSP peak is thus very accurate (about $1\%$) for 
systems where the plasmon shoulder is clearly observed (i.e., for $N_e>1000$).

\begin{table}[!hbt]
    \centering
    \begin{tabular}{r|ccccc}
    $N_e$ & $\omega_0$ & $\gamma$ & $f_{osc}$ & $\omega_1$ & $\omega_b$ \\
    \hline
%254	& 3.08797 & 0.88216 & 35.94383 & 2.59131 & 3.83797  \\
338	& 3.13421 & 0.52539 & 47.51501 & 1.83333 & 3.60421  \\
556	& 3.18204 & 0.44196 & 76.59263 & 1.00001 & 3.95204  \\
832	& 3.20140 & 0.44291 & 117.08588 & 1.33059 & 3.97140  \\
1074 &	3.21699 & 0.37609 & 149.89713 & 1.56759 & 3.72699  \\
1284 &	3.23376 & 0.36143 & 177.38700 & 1.44677 & 3.72376  \\
1516 &	3.24390 & 0.36138 & 210.54419 & 1.00000 & 4.04390  \\
1760 &	3.24730 & 0.36607 & 244.62864 & 1.40239 & 4.04730  \\
2048 &	3.25425 & 0.36936 & 286.90874 & 1.25072 & 4.05425  \\
2654 &	3.26583 & 0.35402 & 371.71087 & 1.22778 & 3.89583  \\
3404 &	3.27786 & 0.35134 & 480.75394 & 1.07239 & 4.07786  \\
4570 &	3.28959 & 0.33304 & 643.47974 & 1.35421 & 4.08959  \\
5470 &	3.29687 & 0.32296 & 771.47150 & 0.99999 & 4.09687  \\ 
6174 &	3.30032 & 0.31742 & 871.93036 & 1.06694 & 4.10032  \\
\hline
    \end{tabular}
    \caption{Parameters from the fits of the TD-DFT photo-absorption cross-section. All values but $f_{osc}$ in eV.}
    \label{tab:fits}
\end{table}

\vspace{2cm}

%\section{Bennet states}

%\newpage
%\clearpage
\section{Additional details on the PGSL and  PGSLN functionals}

\begin{figure}[!hbt]
    \centering
    \includegraphics[width=0.6\textwidth,angle=0]{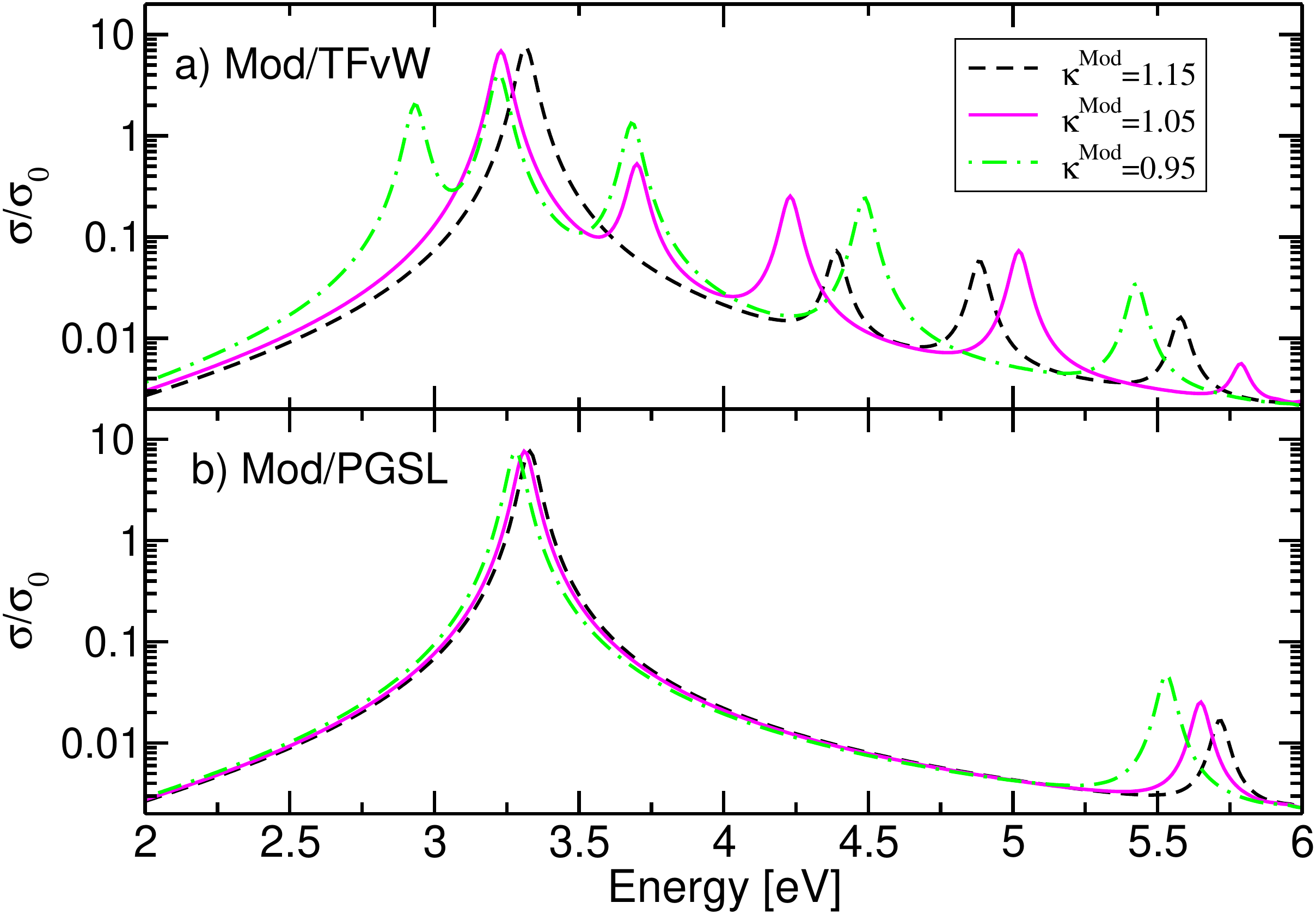}
    \caption{\my{Normalized absorption cross-section $\sigma/\sigma_0$ in a logarithmic scale for a Na jellium sphere with $N_e = 1074$ electrons as obtained from Mod/TFvW and Mod/PGSL using a model density with different values of $\kappa^{Mod}$. 
    For $\kappa^{Mod}=0.95$ the main LSP peak of Mod/TFvW is corrupted by the additional peaks, as the critical frequency is 3.06 eV, i.e., below the LSP energy.
    The absorption spectra of Mod/PGSL is instead almost independent from    the values of $\kappa^{Mod}$.}}
    \label{fig:my_label}
\end{figure}

\begin{figure}[!hbt]
    \centering
    \includegraphics[width=0.6\textwidth,angle=0]{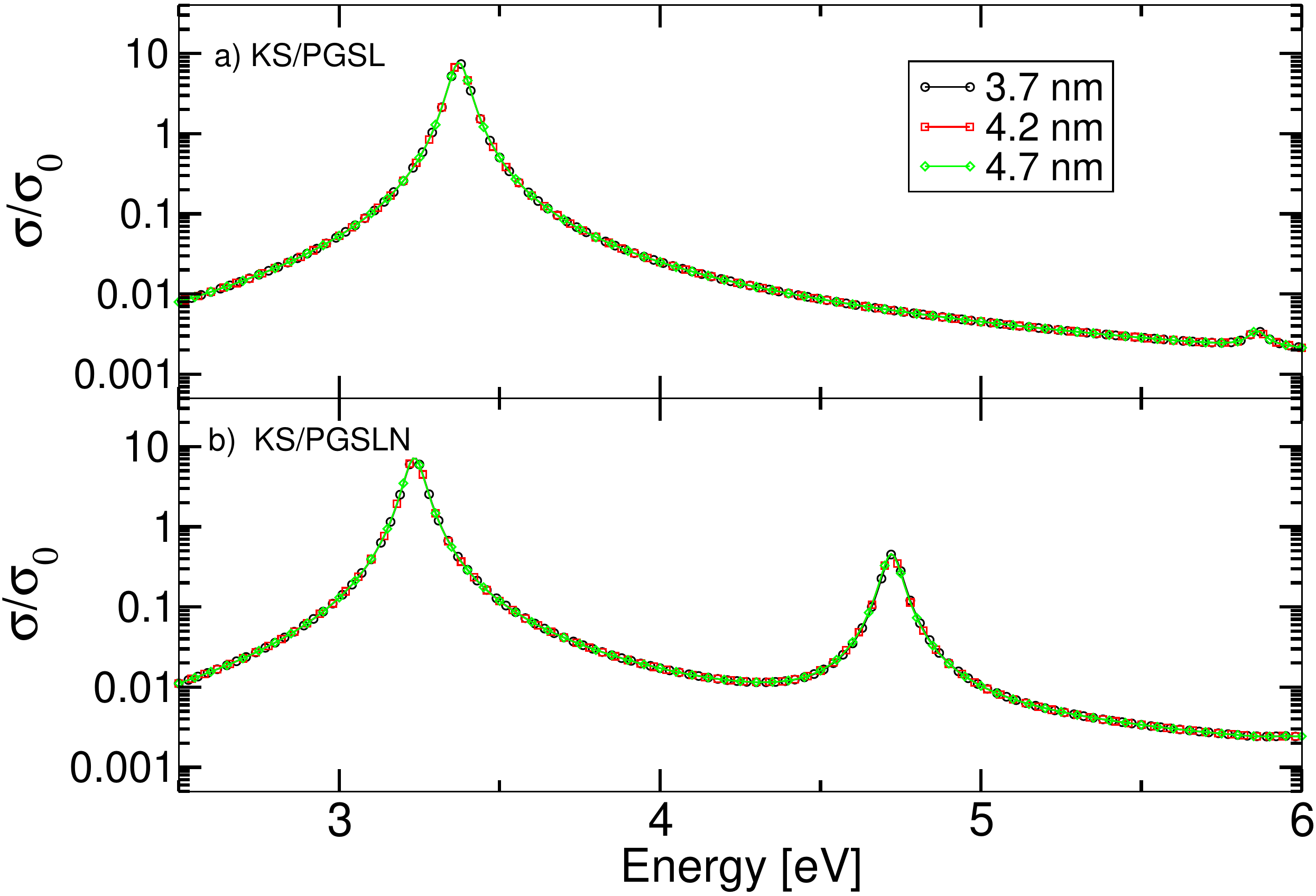}
    \caption{\my{Normalized absorption cross-section $\sigma/\sigma_0$ in a logarithmic scale for a Na jellium sphere with $N_e = 1074$ electrons as obtained from KS/PGSL and KS/PGSLN using different computational domain size. 
    In those calculations we used mixed boundary conditions for an exponential decay (i.e., $\hat{r}\cdot\nabla n_1 + \nu \kappa n_1=0$) with $\nu=1.123$.
    The absorption spectra is fully independent of the computational domain size.}}
    \label{fig:my_label}
\end{figure}

\begin{figure}[!hbt]
    \centering
    \includegraphics[width=0.8\textwidth,angle=0]{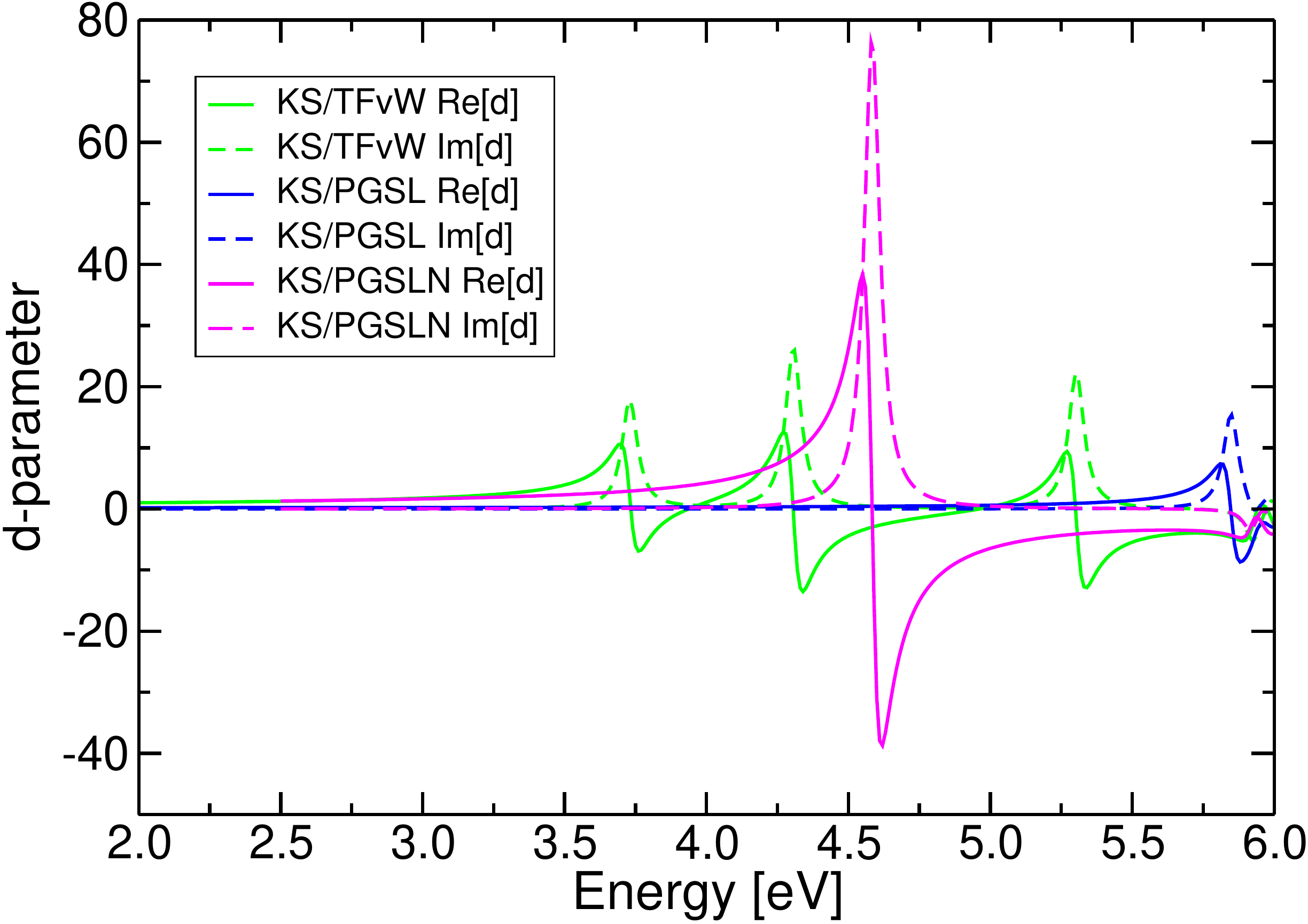}
  \caption{Feibelman d-parameter, see Eq. (25) in the manuscript, as a function of the energy for a Na jellium sphere with $N_e = 1074$ electrons as calculated from KS/TFvW, KS/PGSL, and KS/PGSLN. Real part solid-line, imaginary part dashed-line.
  %(a), KS/PGSL (b), Mod/PGSLN (c), and KS/PGSLN (d).
  %Note that for Bennett peaks, the following relation holds: 
  %$4\pi\int_0^\infty  r^2 n_1(r)dr=0$, where $n_1(r)$ is the radial part of the %induced density.
   The position of the Bennett peaks can be easily identified when Re[d] experiences an abrupt change from the positive to the negative value, while Im[d] shows a peak \cite{yan,tsuei}.
 Note, instead, that no structure is present at the LSP peak position (3.2-3.3 eV). 
 The absolute intensities of the peaks are inversely proportional to the damping parameter (here $\gamma=66$ meV).
  %The Bennett peak for KS/PGSL is close to volume plasmon and has a very small %intensity, so that induced density is also very small (and mixed with volume %plasmon). }}
  \label{Bennet_dens} 
  }
\end{figure}

\begin{table}[!hbt]
    \centering
    \begin{tabular}{l|c|c}
    method     &     Kinetic Energy (a.u.)  & Relative Error (\%) \\
    \hline
         Kohn-Sham & 72.4381866  &  - \\
         TFvW      & 77.2190644  &  +6.60\% \\
         PGS       & 74.5726180  &  +2.94\% \\
         PGSL      & 77.3438041  &  +6.77\%  \\
         PGSLN     & 74.9702980  &  +3.49\% \\
         \hline
    \end{tabular}
    \caption{\my{Kinetic energy for a Na jellium sphere with $N_e=1074$ electrons and relative error in percent with respect to the Kohn-Sham result, using different functionals; kinetic energies are computed using Kohn-Sham density.}}
    \label{tab:my_label}
\end{table}

\begin{figure}[!hbt]
    \centering
    \includegraphics[width=0.5\columnwidth]{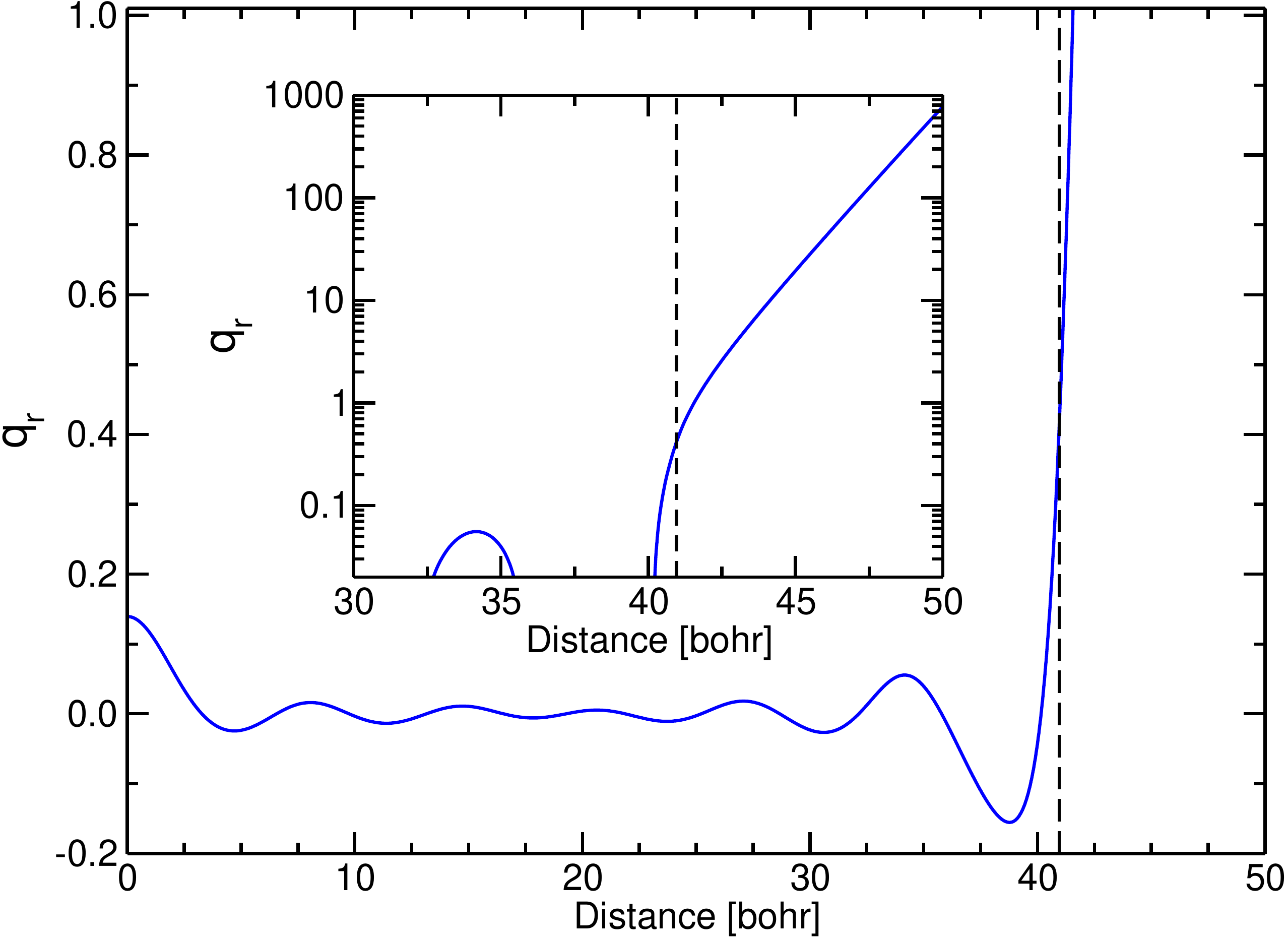}
    \caption{\my{Reduced Laplacian for a Na jellium sphere with $N_e=1074$ electrons; the inset shows the tail region in the logarithmic scale. The dashed vertical line represents the radius of the jellium nanosphere.}} 
    \label{fig:qplot}
\end{figure}

\clearpage
\bibliography{lap_sm}